\documentclass[12pt, rmp,aps,longbibliography]{revtex4-2}

 \usepackage{graphics,graphicx,subfigure,bm}
    \usepackage{amsmath,amssymb,pifont}
    \usepackage{caption}
    \usepackage[section]{placeins}
    \usepackage[usenames,dvipsnames]{color}
\usepackage{graphicx,color,xfrac}
\usepackage{MnSymbol}

\newcommand{\Rey}{\text{Re}}
\newcommand{\We}{\text{We}}
\newcommand{\Mo}{\text{Mo}}
\newcommand{\Bo}{\text{Bo}}
\newcommand{\Oh}{\text{Oh}}

\newcommand{\Ga}{\text{Ga}}
\newcommand{\Ca}{\text{Ca}}
\newcommand{\Fr}{\text{Fr}}
\newcommand{\Wi}{\text{Wi}}
\newcommand{\St}{\text{St}}

\newcommand{\R}[1]{\textcolor{red}{#1}}
\newcommand{\B}[1]{\textcolor{blue}{#1}}

\begin{document}


\title{Gas bubble dynamics}
\author{Dominique Legendre}
\affiliation{Institut de M\'ecanique des Fluides de Toulouse (IMFT) - Universit\'e de Toulouse, CNRS, INPT, UPS, 31400 Toulouse, France}
\author{Roberto Zenit}
\affiliation{Center for Fluid Mechanics, School of Engineering, Brown University, Providence RI 02912, USA}

\date{\today}

\begin{abstract}

The study of gas bubble dynamics in  liquids is justified by the numerous applications and natural phenomena where this two-phase flow is encountered. Gas bubbles move as forces are applied to them; their dynamics are full of nuances that need to be addressed carefully. Since the mass of gas bubbles is practically negligible, in comparison to that of the surrounding liquid,  their reaction to the fluid  is controlled by the added mass acceleration and is thus impacted by all the forces arising from the fluid action. Furthermore, since their surface can be deformed by the same forces acting on them, their shape may change leading to changes in their resistance to move, the drag force, and therefore affecting their speed and their interaction with the surrounding flow which is often  turbulent. The liquid rheology, as well as its surfactant content can also affect the bubble shape and motion as well. Understanding these issues, in addition to the effect of interactions with other bubbles, walls, and non-uniform flows, provides sufficient elements to model and predict bubble behavior through the solution of dynamic equations. In this review, we cover the key aspects of non-condensable gas bubble dynamics. We survey classical references on the subject and provide an overview of the main findings in the past 20 years. We conclude with a scope and suggestions for future research directions, with  special attention to the dynamics of bubble in turbulence, in non-Newtonian fluid and/or in the presence of electrolytes.
\end{abstract}


\maketitle
\newpage
\tableofcontents

\section{Introduction}

The motion of bubbles in liquids is relevant in many natural, industrial and everyday phenomena. Understanding how bubbles move is the  key ingredient to developing new technologies, improving industrial processes and coping with natural flow phenomena. {We present a few examples of current importance to motivate the need for this review.}

\subsection{Motivation}

In this introduction, our aim is to delineate the subjects to be covered in this paper. The examples illustrate what needs to be known if we wanted to study systems like these. we do not  delve deeply into the details of these specific problems,  we refer you to the cited publications in each case. 
\begin{figure}[ht!]
\includegraphics[width=0.9\textwidth]{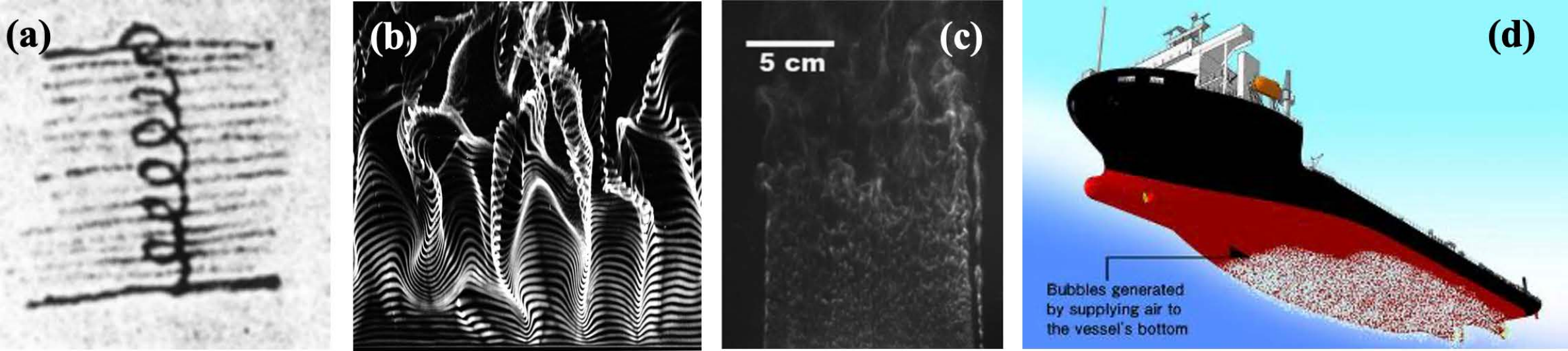}
\includegraphics[width=0.9\textwidth]{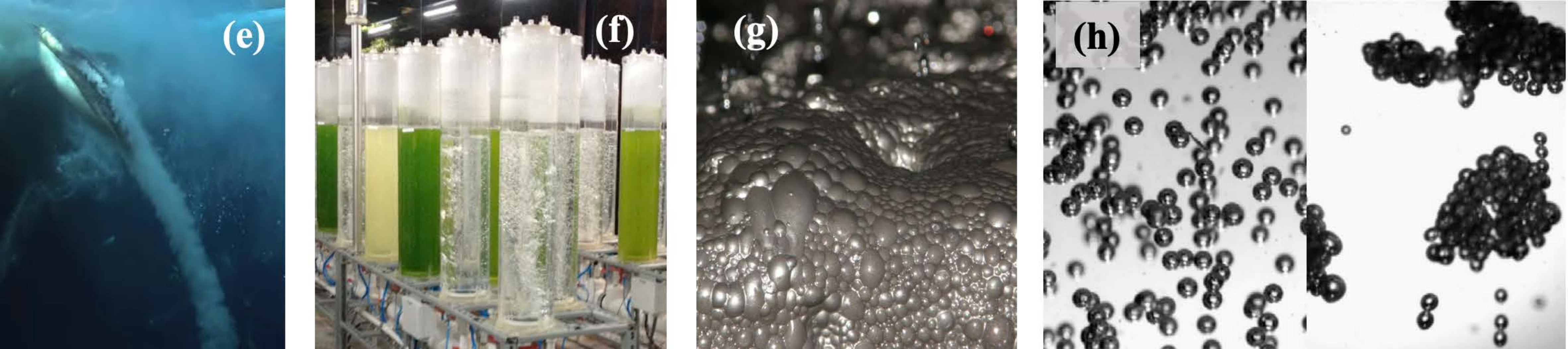}
\includegraphics[width=0.9\textwidth]{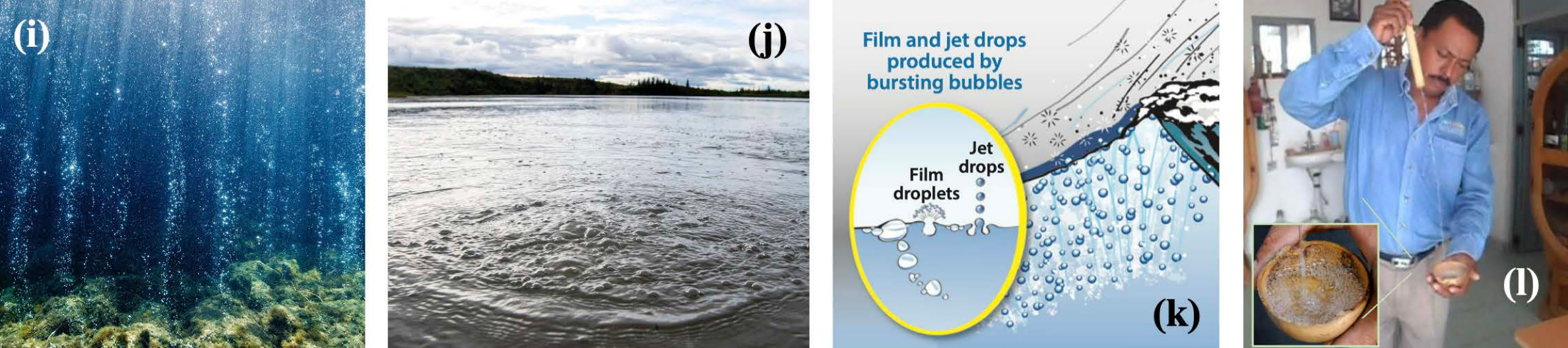}
\caption{{(a) Zigzagging bubble drawing from Leonardo da Vinci (circa 1510), image taken from \cite{prosperetti2004}. 
(b) Hydrogen bubble visualization of streak pattern in the near-wall region of a turbulent boundary layer \cite{Smith2012}.
(c) Hydrogen bubble plume formation from an electrode during water hydrolysis \cite{Chandran2015}.
(d) Bubble drag reduction for commercial ships \citep{kawabuchi2011}. 
(e) Bubble drag reduction in swimming penguins  \cite{davenport2011}.
(f) Bubble column bioreactors, used to harvest algae, image from Universidad EAFIT.
(g) Surface of a flotation tank showing copper sulfide particles floated by bubbles, image from Geomartin, CC BY-SA 4.0, via Wikimedia Commons.
(h) Comparison of a bubbly flow, (left) Newtonian, (right) non-Newtonian, adapted from \citet{velez2011b}.
 (i) CO$_2$ bubbles in ocean seeps, Image from Pasquale Vassallo, Stazione Zoologica Anton Dohrn. 
 (j) Methane gas released from seep holes at the bottom of Esieh Lake ripples the surface, image from \cite{Lecher2017}. 
 (k) Sea spray generation from bubbles bursting at the surface, which create films and jet drops \cite{Veron2015}.
(l) The alcohol content in Mezcal is determined using traditional techniques, which involve the lifetime of surface bubbles \cite{Rage2020}.}}
\label{fig_examples}
\end{figure}

{In  his delightful paper, \citet{prosperetti2004} quotes a classical Greek proverb ``Homo Bulla,'' or ``Man is a bubble'', in reference to the frailty of human life. He then uses this as a prompt to discuss bubble dynamics.  The apparent infirmity of a bubble arises from its lack of mass. Despite this fact, the motion of a  bubble carries a significant amount of inertia through its added mass, as discussed below. So, even if bubbles are `fragile' they {resist external flow effects thanks to the surface tension and  are able to} impose significant changes to the surrounding flow; therefore, their effect cannot be discarded or neglected. 
One example that clearly displays this feature is the zigzagging  motion of a rising bubble. It has been observed that when the size of an air bubble rising in a low viscosity liquid surpasses a certain critical size, its shape is no longer spherical, and its motion is no longer rectilinear.  Interestingly, the first scientist to make this observation was Leonardo da Vinci \cite{Leonardo}, as shown in Fig. \ref{fig_examples}(a). Prosperetti coined the term \emph{Leonardo's paradox} to describe this unexpected behavior \cite{prosperetti2004}: the bubble moves sideways even though buoyancy only points upwards. We discuss this in Section \ref{sec_zigzag}.}

{Very small bubbles have been used for a long time as tracers to visualize flows. An example of the streak lines in the near-wall region of a turbulent boundary layer is shown in Fig. \ref{fig_examples}(b), taken from \citet{Smith2012}. This visualization technique, conceived in the 1950s \cite{Clayton1967}, uses the hydrogen bubbles produced at the cathode of a water electrolysis process. Production of hydrogen has become significant due to the need to replace fossil fuels. One of the methods to produce hydrogen on an industrial scale is precisely electrolysis \cite{Chandran2015}, as shown in Fig. \ref{fig_examples}(c).  The main obstacle in increasing production results from the need for a better understanding of the process of bubble formation and detachment \cite{angulo2020}.}

The recent paper by \citet{lohse2018} provides a personal view of the importance of bubbles in modern technologies and natural phenomena extending from inkjet printing, ultrasound contrast agents, drag reduction, and surface cleaning, all the way to snapping shrimp and sonoluminescence.  
One of the subjects discussed by Lohse is bubble-induced drag reduction, see Fig. \ref{fig_examples}(d). {Due to its large economic implications and possible consequences for the naval shipping industry \cite{kawabuchi2011}, this subject is often cited to justify the study of bubble dynamics. Additionally, in nature, some animals are believed to benefit from the same principle to reduce their drag during swimming. For example,  \citet{davenport2011} argued that penguins trap bubbles in their plumage to increase their speed when leaving the water, as illustrated in Fig. \ref{fig_examples}(e). The review of \citet{ceccio2010friction} provides a good starting point for understanding bubble drag reduction. Some authors attribute the reduction of drag mainly to the presence of large bubbles  \cite{murai2014frictional,verschoof2016bubble}; however, small bubbles can also lead to a reduction in drag \cite{xu2002,song2018,pang2018,garcia2023}. Among the different effects that have been identified to be relevant for this problem are the interaction of bubbles with turbulence, the bubble-bubble interactions and bubble-wall interactions. Some of these issues are introduced in this review.}

{The dynamics at the bubble scale play an important role in  chemical and biochemical reactors \cite{schluter2021small}.}
A widely referenced bubble-based engineering application is the so-called bubble column reactor \cite{kantarci2005}{, shown in Fig. \ref{fig_examples}(f).} These devices are used in chemical, biochemical and petroleum industries to provide high rates of mass and heat transfer, while being compact and low-maintenance. Such bubble reactors can be used to improve microalgae harvesting \cite{Demir2023}. These microorganisms are able to convert light energy into biomass, using water and inorganic nutrients. However, using bubbles could also be inefficient since algae cells are, in general, hydrophilic, which results in poor bubble-cell interactions, thus requiring bubble functionalization. Flotation is a process that uses such bubble systems to collect solid particles in suspension using rising bubbles \cite{Huang2011, Saththasivam2016,Federle2024}, in applications as varied as wastewater treatment, metallurgy and bioreactors, see Fig. \ref{fig_examples} (g). In this case, bubbles of a certain size ascend in a fluid at a certain terminal speed, interacting with other bubbles and containing walls. Another important aspect of bubbly columns is the large mixing rates that are induced by rising bubbles \cite{risso2018agitation}. As they move agitation is induced and large-scale buoyancy driven instabilities can appear. The properties of bubble-induced agitation are different from classical shear-induced turbulence. Understanding mixing and large-scale motion can only be achieved if the local dynamics of individual bubbles and bubble-bubble interactions are known and understood.  In many biological applications, small traces of bio-proteins may cause a non-Newtonian behavior of the surrounding liquid. {When the surrounding fluid is non-Newtonian the nature of the bubbly flow changes drastically \cite{Zenit2018}, as shown in Fig. \ref{fig_examples}(h). Some of the issues that arise when considering such fluids are discussed in Section \ref{bubblesnonNewtonian}.}

Due to the large gas-liquid surface area available in a bubbly liquid, these flows are often very effective for mass transfer processes. Bubble columns are used to provide oxygen to cell cultures \cite{Henzler1993}. In such cases, oxygen bubbles injected at the bottom of a column would gradually diffuse into the liquid as they rise, reducing their size. A similar process with environmental importance is the formation of CO$_2$ and  methane bubbles released from the seabed to the surface \cite{leifer2002bubble,oceanseeps2016},  as seen in Fig. \ref{fig_examples}(i). The fraction of CH$_4$ that reaches the surface depends upon the release depth, bubble size, dissolved gas concentrations, temperature, surface-active substances, and bulk fluid turbulence in the upwelling flow. Carbon dioxide sources also result in the acidification of the ocean waters, which in turn affect biodiversity \cite{Teixido2018}. {A direct consequence of climate change is the development of lakes across the tundra in the Arctic. As the permafrost thaws, it releases carbon dioxide and methane gas, generating intense bubbly flows that rise to the surface \cite{Lecher2017}, as seen in Fig. \ref{fig_examples}(j). }

{When bubbles reach a free surface, they can remain floating briefly but eventually burst. The generation of ocean spray aerosols, which plays a crucial role in radiative and cloud processes, is the result of the bubble bursting process \cite{Veron2015}. When the bubble bursts, the jets and films shatter into fragments, creating small droplets. This process is depicted in Fig. \ref{fig_examples}(k). The time that a bubble remains at the surface depends on the properties of the fluid, as well as the level of surfactant content \cite{Atasi2020}. Curiously, this process is used by traditional mezcal manufacturers in Mexico to determine the ethanol content of their distilled spirits \cite{Rage2020}, as shown in Fig. \ref{fig_examples}(l).}

In this review paper, we will provide a general description of bubble dynamics: how bubbles move in response to forces. The key issues discussed here are a direct consequence of the challenges posed by the applications and natural phenomena briefly described above. Among them are bubble terminal velocity, surface deformation, bubble interaction with flow, bubble-bubble and bubble-wall interactions, and the effects of liquid rheology. Our paper does not attempt to cover all possible aspects of bubble dynamics. The literature in the subject is vast and varied, making an exhaustive review impractical. Instead, we have selected the  physical mechanisms that, in our opinion, are the most relevant for entering the field of bubble dynamics. We end our review with a set of general equations of motion for bubbles and conclude with some general remarks and ideas for future directions.

\subsection{Gas bubble dynamics}

In this review we will focus on the case of non-condensable gas bubbles, for which the size remains practically constant. In some cases, mass transfer occurs across the bubble surface but no phase change is considered. In this context, we use the term bubble dynamics to refer to the motion of bubbles when subjected to external forces, induced by the surrounding fluid. Note that the term bubble dynamics is often used to study the process of bubble size evolution when subjected to changes in pressure or temperature in the surrounding liquid \cite{brennen2014cavitation}. This growth rate or collapse of bubbles is of particular importance for vapor bubbles \cite{prosperetti2017vapor}. We do not discuss these issues here.

\begin{figure}[h]
\includegraphics[width=6.5in]{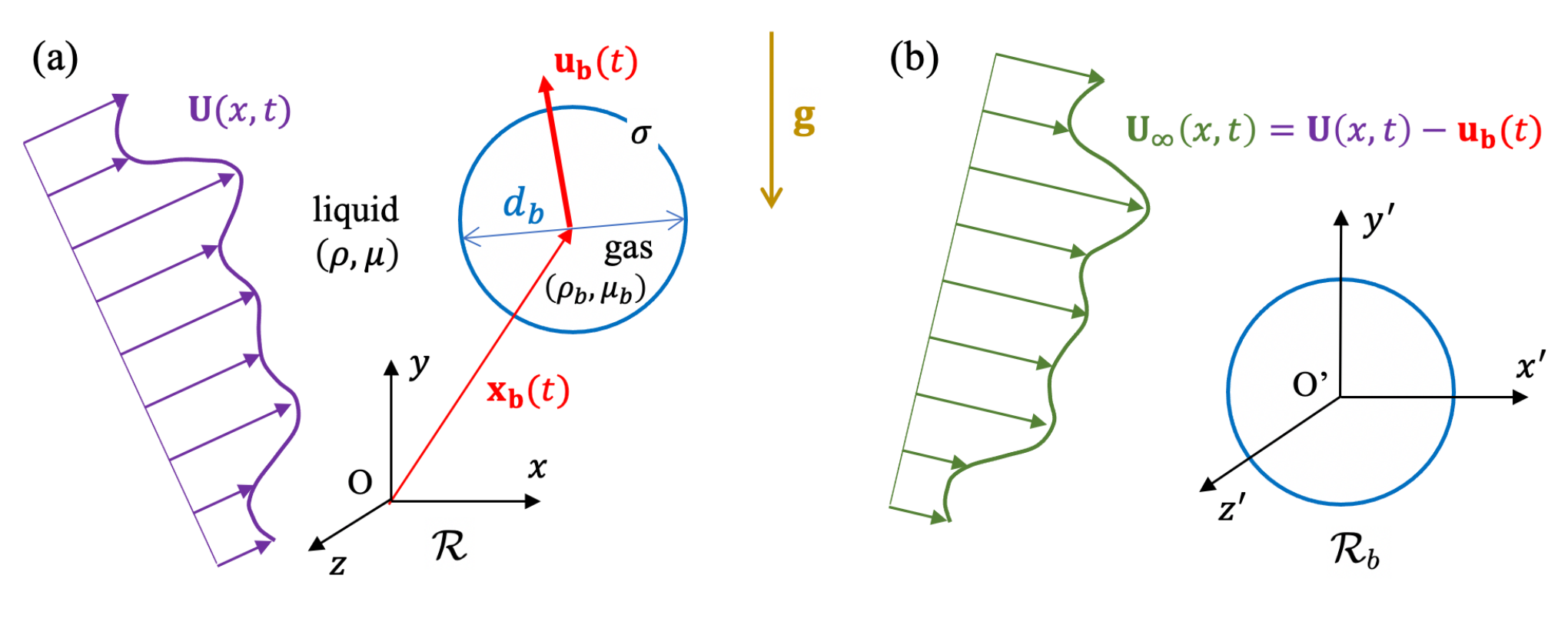}
\caption{{Bubble dynamics: problem statement and frames of reference. (a)  A bubble  shown in the laboratory frame of reference $\mathcal{R}$: a  bubble of diameter $d_b$, density $\rho_b$ and viscosity $\mu_b$ moves at velocity $\mathbf{u_b}$ in  a liquid of density $\rho$, viscosity $\mu$, surface tension $\sigma$, and velocity field $\mathbf{U}$ which is, in general,  non-uniform and unsteady. The gravitational field is represented by $\mathbf{g}$.  (b) The same problem viewed in the reference frame $\mathcal{R}_b$ of the moving bubble. The velocity field here is $\mathbf{U_\infty}= \mathbf{U}-\mathbf{u_b}$}.}
\label{fig_pb_statment}
\end{figure}

For a bubble of {mass, $m_b=\rho_b \vartheta_b$ ($\vartheta_b$ and $\rho_b$ are the bubble volume and gas density, respectively)}, moving at a velocity $\mathbf{u_b}$ {in a fluid of density $\rho$ and viscosity $\mu$,  as depicted in  Fig. \ref{fig_pb_statment}(a),}  a Lagrangian equation of motion can be written as:
\begin{equation}\label{eq_motion}
    m_b \frac{d \mathbf{u_b}}{dt}=m_b \mathbf{g} +\int_S \mathbf{T}\cdot \mathbf{n_b} \,dS 
\end{equation}
where $\mathbf{T}$ is the fluid stress tensor, $S$ is the bubble surface, $\mathbf{n_b}$ is the normal vector on the bubble surface and $\mathbf{g}$ is the gravitational acceleration. {The bubble dynamics are discussed in the laboratory frame of reference $\mathcal{R}$ or in the reference frame $\mathcal{R}_b$ moving with the bubble as illustrated in Fig. \ref{fig_pb_statment}}. According to Eq. \ref{eq_motion}, if $m_b$ is very small, changes in the stress field around the bubble (right-hand-side of Eq. \ref{eq_motion}), would cause large values of $d\mathbf{u_b}/dt$. In other words, bubbles would respond drastically to small changes in the forces on them. {However, it is important to emphasize that, {the amount of mass to be accelerated by the bubble is imposed by} one of these forces,  the added mass force contribution $- C_M m_f d \mathbf{u_b}/dt$ where $C_M$ is the so-called added mass coefficient and $m_f= \rho \vartheta_b$ is the displaced mass of fluid by the presence of the bubble {(see Section \ref{sec_added_mass})}.}

\subsection{Previous reviews on the subject}
There are numerous studies that address the dynamics of bubbles. The most cited source for the subject is the book by \citet{clift1978} and reprinted numerous times. This book remains an invaluable source of information, particularly for experimental results conducted up to the end of the 1970s. Another very important source of experimental data is the paper by \citet{maxworthy1996}, who conducted a vast set of experiments {on rising bubbles} in a wide range of parameters that still remain relevant for many modern investigations. The annual review article by \citet{magnaudet2000} is the most recent review that covers some aspects of bubble dynamics, but focuses mostly on spherical bubbles at large Reynolds number. However, this paper is now more than 20 years old. {\citet{takagi2011} wrote a comprehensive review on the effect of surfactants on single bubble motion and bubbly flows. \citet{zenit2018drinks} discuss different issues on bubbly drinks, with an emphasis on Carbon Dioxide bubbles evolving from supersaturated liquids.}  

The objective of the present review is {to} summarize  recent findings on the dynamics of gas bubbles. We pay special attention to the presentation of results and to the description of the physical mechanisms involved, making it accessible to students and junior scientists.

\section{What are bubbles and how are they generated?}

In this paper, we define a bubble as a discrete gas volume of density $\rho_b$ and viscosity $\mu_b$, surrounded by a liquid with density $\rho$ and viscosity $\mu$. Since the gas and liquid are immiscible, the strength of the interface tension is characterized by a surface tension $\sigma$ (see Fig. \ref{fig_pb_statment}). Although the gas could be the vapor of the surrounding fluid, for which mass transfer across the interface could occur, we will only consider  the case in which the gas is non-condensable in the liquid. Note that in some cases mass transfer may occur, {if the bubble gas dissolves into the surrounding liquid or if  gas is already dissolved in the liquid.}

\subsection{Conditions at a bubble surface}

We {first} turn our attention to the conditions at the bubble surface. {The surface plays a significant role in the dynamics because it acts as the boundary between the bubble and the fluid motion around it}. Conservation of mass and momentum across the surface {are used} to identify the sets of conditions that must be considered at the surface.

\subsubsection{Mass conservation}
A balance of mass across the interface can be written as
\begin{equation}\label{eq_mass1}
\rho_b (\mathbf{u_g} - \mathbf{v})\cdot\mathbf{n_b}  = \rho  (\mathbf{u}-\mathbf{v})\cdot\mathbf{n_b} 
\end{equation}
where $\mathbf{v}$ is the velocity of the interface, $\mathbf{u_g}$ is the gas velocity inside the bubble, $\mathbf{u}$ is the liquid velocity outside the bubble and $\mathbf{n_b} $ is the normal vector pointing outwards on the bubble surface.
{This general relation is derived by assuming that the interface has no mass and that there may be a phase change process (e.g., boiling or condensation) or transfer of dissolved gases at the bubble surface. Following  some manipulation, Eq. \ref{eq_mass1} can be rewritten as:
\begin{equation}\label{eq_mass2_new1}
   \mathbf{u}\cdot\mathbf{n_b}   
     = \frac{ \rho_b}{\rho} \mathbf{u_g} \cdot\mathbf{n_b} +  \left(1-  \frac{ \rho_b}{\rho}\right) \mathbf{v}\cdot\mathbf{n_b}. 
\end{equation}
This equation illustrates that a discontinuity (or jump) exists in the normal velocity across the bubble interface.}

{In the absence of mass transfer $\rho_b (\mathbf{u_g} - \mathbf{v})\cdot\mathbf{n_b}  = \rho  (\mathbf{u}-\mathbf{v})\cdot\mathbf{n_b} =0$ or when ${ \rho_b}/{\rho} \ll 1$ (a condition valid for most bubbles), the liquid velocity satisfies the condition of impermeability at the interface:
\begin{equation}\label{eq_mass2_1}
 \mathbf{u} \cdot \mathbf{n_b}  =\mathbf{v} \cdot \mathbf{n_b}
 \end{equation} 
However, the continuity with the gas velocities at the interface
\begin{equation}\label{eq_mass2}
 \mathbf{u} \cdot \mathbf{n_b}  =\mathbf{u_g} \cdot \mathbf{n_b}.
 \end{equation} 
is only satisfied in the absence of mass transfer across the interface.}
Considering also the tangential velocity continuity at the bubble surface $\mathbf{u} \cdot \mathbf{t_b}= \mathbf{u_g} \cdot \mathbf{t_b}$, {and in the absence of mass transfer, the gas and liquid velocities  are equal at the bubble surface}:
\begin{equation}\label{eq_mass3}
\mathbf{u} = \mathbf{u_g}.
\end{equation}
As discussed below, this condition combined with a low viscosity in the gas induces a motion inside the bubble driven by the external liquid flow. The domain formed by the bubble being closed, an internal toroidal recirculation develops. 

\subsubsection{Momentum Conservation}
Similarly, the conservation of  momentum across a fluid interface can be used to complete the jump conditions at the bubble surface. Neglecting mass transfer across the interface, the normal component of the momentum conservation equation leads to 
\begin{equation}
\label{eq:surface_mometum}
    P_g-\mathbf{n_b}\cdot \mathbf{\Sigma_g }\cdot\mathbf{n_b}= 2 H\sigma  +P-\mathbf{n_b}\cdot \mathbf{\Sigma}\cdot\mathbf{n_b}
\end{equation}
where $P$ and $P_g$ are the pressure in the liquid and gas phases, respectively. $\mathbf{\Sigma}$ and $\mathbf{\Sigma_g}$ are the deviatoric viscous stress tensors for the liquid and gas phases, respectively. $H={1/R_c}$ is the {local mean curvature} of the interface, {$R_c$ being the mean radius of curvature}. The surface tension of the interface is represented by $\sigma$. If the normal viscous stresses of both phases are neglected, it is possible to establish a condition for the pressure jump across the interface
\begin{equation}
    \Delta P = \frac{2 \sigma}{R_c}.
\end{equation}
{For a spherical bubble of radius $R=R_c$, the pressure jump  $\Delta P=P_g-P$ is constant across the interface}. This relation is often referred to as the Young-Laplace equation.  

As discussed below, the spherical shape of a bubble is the consequence of the surface tension which tends to minimize the surface area. However, when a bubble rises in a liquid (due to buoyancy or other forces), the forces induced by the motion can result in deformation of the bubble shape. In such a case, the resulting shape is determined by the balance between pressure, viscous and surface tension forces at the interface. 
The relative importance of  pressure and viscous forces at the interface for a bubble (neglecting viscous and inertial effects of the gas phase) compared to surface tension forces, can be assessed by values of {the Bond number $\Bo$}, {the Weber number} $\We$ and {the capillary number} $\Ca$ ({defined below } in Eqs. \ref{eqn:Bo}, \ref{eqn:We} and  \ref{eqn:Ca}, respectively).
A spherical shape is maintained as long as the surface tension {stress $\sigma/d_b$  can resist 
the dynamic pressure ($\We<1$), the hydrostatic pressure ($\Bo<1$) and the normal viscous stress ($\Ca<1$).}

Considering now the tangential component of the momentum conservation at the interface, we can write
\begin{equation}\label{Eq_moment_tagent}
\mathbf{n_b} \cdot \mathbf{\Sigma_g} \cdot  (\mathbf{I} -  \mathbf{n_b}\mathbf{n_b}) = \nabla_I \sigma + \mathbf{n_b} \cdot \mathbf{\Sigma} \cdot  (\mathbf{I} -  \mathbf{n_b}\mathbf{n_b}).
\end{equation}
When the interface is {surfactant-free} and at a uniform temperature, $\sigma$ is constant, and therefore $\nabla_I \sigma=0$. {Under such conditions, and} considering  that $\mu_g \ll \mu$, the above equation reduces to:
 \begin{equation}\label{Eq_moment_tagent2}
\mathbf{n_b} \cdot \mathbf{\Sigma} \cdot  (\mathbf{I} -  \mathbf{n_b}\mathbf{n_b}) = 0,
\end{equation}
which indicates that the liquid is subjected to a zero shear stress condition at the bubble surface. In other words, the bubble surface has a perfect slip condition. The presence of surfactants on the interface induces a
gradient of surface tension, $\nabla_I \sigma\neq 0$, which significantly modifies the  surface mobility and,  consequently, the bubble dynamics. This is the so-called Marangoni effect, {as discussed in Section \ref{section_contamination}}.

\subsection{Generation of bubbles}

\begin{figure}[h]
\includegraphics[width=6.5in]{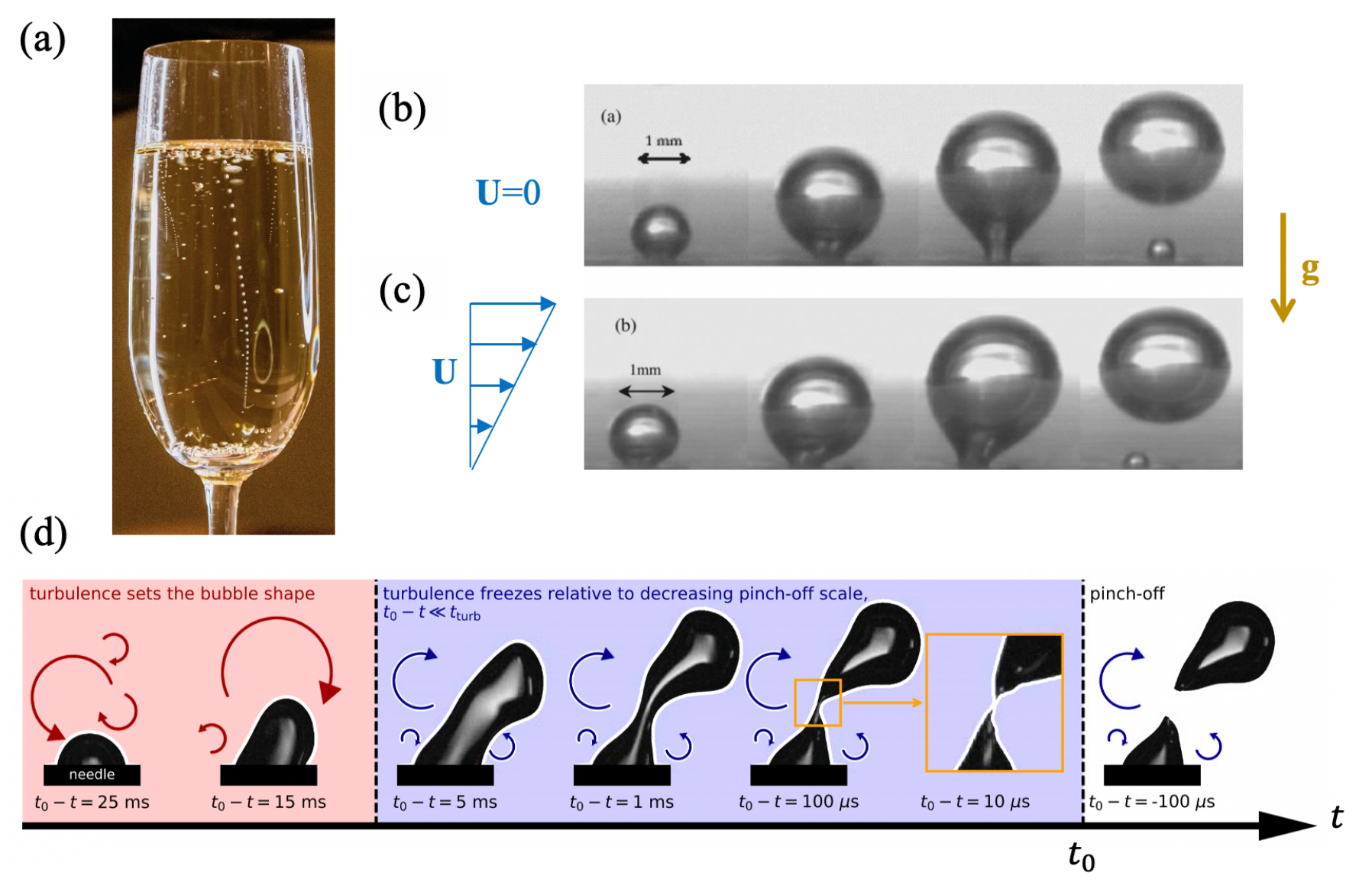}
\caption{{Bubble generation on a wall in a vertical gravity field $\mathbf{g}$: (a) in carbonated drinks, bubbles are formed at nucleation sites,  surface roughness or faults that allow the dissolved gas to grow into bubbles \cite{Atasi2023}; (b) from a capillary in a fluid at rest, or (c) in a shear flow, which induces the bubble deformation along the flow and creates a vertical lift force in favor of the bubble detachment  \cite{Duhar2006}; (d) by pinch off in a turbulent shear flow, in conditions where the turbulence shapes the bubble but has no significant impact on the pinch-off event, which occurs much faster ($<10$ms) than the turbulence time scale (here, from 15 to 70 ms)  \cite{Ruth2019}}.}
\label{fig_def}
\end{figure}

{Bubbles can be produced as a result of complex breakup events such as observed during breaking waves \cite{Ruth2022} (see Section \ref{Sec_bubble_rupture}) or under very well controlled conditions as first discussed here.} A bubble can be formed by injecting gas into a liquid or by mass diffusion from dissolved gas within the liquid (see Fig. \ref{fig_def}). If mass transfer effects are not considered, then the  bubble will grow by forcing gas into the liquid using a capillary, a hole, or a flexible membrane. Consider first the case in which the liquid is stagnant and the gas flows slowly (i.e., neglecting inertial effects in both phases). A bubble can be formed at the tip of a capillary, connected to an external gas container at a pressure $P_c$. At first, the bubble will grow if $P_c$ is larger than $P_b$. The pressure inside the bubble is 
\begin{equation}
    P_b=P_\infty+\frac{4\sigma}{D_c}
\end{equation}
where $P_\infty$ is the pressure of the liquid around the bubble and $D_c$ is the diameter of the capillary. As the bubble grows, its diameter will become larger than that of the capillary, $D_c$. At that point the pressure inside the bubble will be larger than that of equilibrium conditions. Hence, for the bubble to grow steadily, $P_c$  must change dynamically. In practice, $P_c$ is always larger than $P_b$ (overpressure condition). This is the reason why most devices that generate bubbles do so under non-equilibrium conditions, leading to jetting and the production of bubbles with different diameters. This issue was discussed in depth by {  \citet{oguz_prosperetti_1993}}.

Assuming that the steady growth condition can be met, the bubble 
will remain attached to the tube by capillary forces, represented by $C \sigma$, where $C$ is the wet perimeter of the capillary or hole. The bubble will continue to grow, attached to the capillary, until its buoyancy, $\rho g \vartheta_b$, matches the capillary force. For a capillary with an inner diameter $D_c$, the detachment bubble diameter, {often referred to as the Fritz diameter \cite{fritz1935}}, would be:
\begin{equation}
    d_b = k  \left({\frac{D_c \sigma}{\rho g}}\right)^{1/3} \label{eqn:steadybubblesize}
\end{equation}
where $k=\sqrt[3]{6}$, for the case of circular tube. The value of $k$ changes for other geometries. If the bubble is formed at a hole rather than a thin-walled capillary tube, the wet perimeter depends on the details of the geometry of the hole \cite{simmons2015formation}. Similarly, if the bubble is formed at a flexible membrane, the size of the wet perimeter will depend on the pressure $P_c$. Since there are many techniques for producing bubbles by injecting pressurized gas through holes and tubes, numerous empirical correlations exist. While these correlations are useful, they often lack a strong physical basis. A good reference for such correlations can be found in \citet{kulkarni2005bubble}. The effects of the wall inclination and the type of substrate when bubbles nucleate on a wall are discussed by \citet{Lebon2018}.

In addition to buoyancy and inertial effects, as described above, a bubble may detach from the surface of a capillary or hole if a flow is imposed onto it while it is growing, {see Fig. \ref{fig_def} (c) and (d).} The flow induces a drag force onto the bubble surface that can help dislodge the bubble at a size smaller than that predicted by Eq. \ref{eqn:steadybubblesize}.  Force balance models have been developed allowing for predictions of the detachment size as a function of the detachment size as function of the gas flow rate and the liquid flow shear rate. An important aspect of such approach is the modeling of the contact line at the bubble foot \cite{Duhar2006}.

In many practical applications, gas bubbles can form in a liquid via gas diffusion. When gas is dissolved in the liquid, it may come out of solution depending on the saturation level and the pressure, {see Fig. \ref{fig_def}(a)}. By neglecting advective transport, the bubble growth can be modelled using the Epstein-Plesset equation \cite{epstein1950stability} that balances the mass flux from the fluid to the bubble by diffusion: 
\begin{equation}
    \dot R = \frac{(C_S-C_\infty) D}{\rho}  \left(\frac{1}{R}+\frac{1}{\sqrt{\pi D t}}\right),
\end{equation}
where the overdot indicates a time derivative, $D$ is the mass diffusion coefficient and $C_S$ and $C_\infty$ are the dissolved gas concentration at the surface and far from the bubble, respectively. The bubble size is proportional to $\sqrt{t}$ for long times, assuming that $C_S-C_\infty$ remains constant. {Bubbles are formed by this mechanism when the fluid pressure is reduced. In such a case, the gas in the supersaturated fluid comes out of solution spontaneously. Bubbles are formed at nucleation sites, such as small surface imperfections, crevices on the container, or small particles in the fluid. These bubbles grow until their buoyant force is large enough to dislodge them from their nucleation sites, according to Eq. \ref{eqn:steadybubblesize}. After the bubble detaches due to buoyancy, its size can evolve but the convective mass transfer effect must be taken into account \cite{legendre2017, soto2019}.}

\section{Rising bubbles: terminal velocity, deformation and interface contamination}

\begin{figure}[h!]
\includegraphics[width=6.8in]{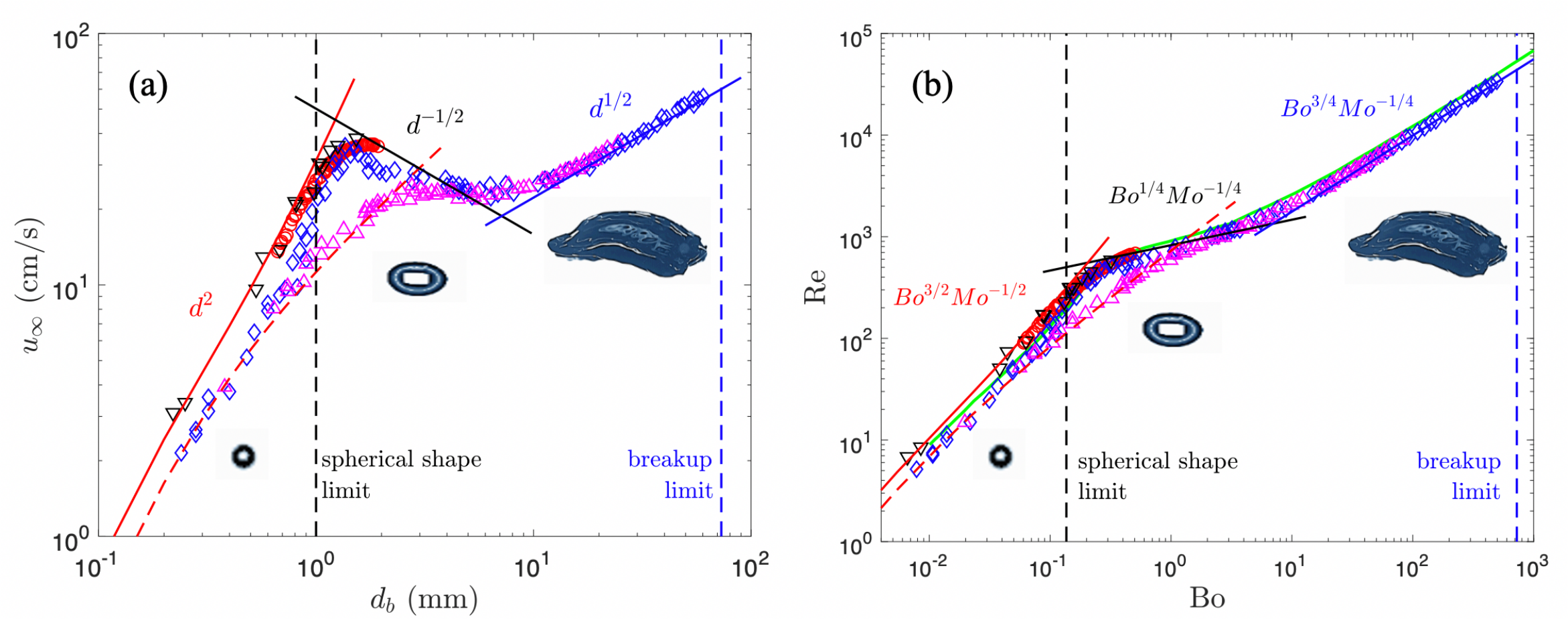}
\caption{{(a) Terminal velocity $u_\infty$ of an air bubble in water at room temperature as a function of its equivalent diameter $d_b$, defined in Eq. \ref{eqn:db}. The markers show experimental data from the literature:
(\textcolor{blue}{$\lozenge$}), filtered water at 18-21$^\circ$C \cite{haberman1956},
(\textcolor{magenta}{$\triangle$}), tap water at 21$^\circ$C \cite{haberman1956}, 
(\textcolor{black}{$\triangledown$}), de-ionized water (20$^\circ$C) with particle concentration less than 18 part/ml \cite{Huang2011}; 
(\textcolor{red}{$\circ$}),  ultra-purified water  at 19.6$^\circ$C with less than 10 p.p.b. organic particles \cite{duineveld1995}.
The solid lines  show the prediction for the terminal speed of a spherical bubble and the corresponding trends $u_\infty$ with $d_b$ are reported in the figure:
(\textcolor{red}{\rule[2pt]{18pt}{1.5pt}}) using the drag force \ref{eq:drag_mei} for a clean spherical bubble, 
(\textcolor{black}{\rule[2pt]{18pt}{1.5pt}}) relation \ref{eq_def4} for ellipsoidal bubble with $\We=3.5$, 
(\textcolor{blue}{\rule[2pt]{18pt}{1.5pt}}) relation \ref{Eq_UB_cap} for spherical cap bubble.
The red dashed line (\textcolor{red}{\rule[2pt]{5pt}{1.5pt} {\rule[2pt]{5pt}{1.5pt}} {\rule[2pt]{5pt}{1.5pt}}}) shows the prediction for the terminal speed of a solid sphere \cite{shiller1933} used to describe contaminated bubbles. 
The vertical black dashed line {at $d_b=1$ mm} (\textcolor{black}{\rule[2pt]{5pt}{1.5pt} {\rule[2pt]{5pt}{1.5pt}} {\rule[2pt]{5pt}{1.5pt}}}) shows the condition for which $\We\approx 1$, when bubble shape deviates from spherical (see Section \ref{sec_deformation}). 
The vertical blue dashed line {at $d_b=73$ mm} (\textcolor{blue}{\rule[2pt]{5pt}{1.5pt} {\rule[2pt]{5pt}{1.5pt}} {\rule[2pt]{5pt}{1.5pt}}}) shows the condition for which $\Bo\approx 730$ and  $\We\approx 370$, when bubble breakup is expected (see section \ref{Sec_bubble_rupture}).
(b) Same plot in dimensionless terms: the  ($\Bo, \Rey$) phase diagram. 
The additional line 
(\textcolor{green}{\rule[2pt]{18pt}{1.5pt}}) reports the  iso-Morton curve $\Mo=10^{-11}$ corresponding to water,  from the Clift-Grace-Weber diagram \cite{clift1978}.}
}
\label{fig_1_new}
\end{figure}

We begin our discussion by addressing the most basic question in bubble dynamics: what determines the terminal speed of an air bubble rising in water? Figure \ref{fig_1_new}(a) shows the results from numerous experimental measurements aimed at answering this question.  Once the bubble forms at the bottom of a container, its lower density will result in a buoyant force that will push the bubble upward. 
The buoyancy force, arising from Archimedes' principle, is given by 
\begin{equation}
    \mathbf{F_A} = - \rho  \vartheta_b \mathbf{g} 
    \label{eqn:buoyancy}
\end{equation}
The buoyancy force accelerates the bubble from rest and the bubble experiences a resistance from the fluid, the drag force  $\mathbf{F_D}$, in opposite direction  of its  velocity $\mathbf{u_b}$. When the drag force balances the buoyant force, the bubble no longer accelerates and reaches a constant speed, known as the terminal speed or velocity, {$u_\infty$:
\begin{equation}
   u_\infty = \mathbf{u_b}\cdot \mathbf{e_y}
    \label{eqn:uterminal}
\end{equation}
where $\mathbf{e_y}$ is the vertical coordinate along the gravity with $\mathbf{g} = - g \mathbf{e_y}$ (see Fig. \ref{fig_pb_statment}). In the case of unsteady rising motion because of a spiraling or zigzaging path (see Section \ref{sec_zigzag}), $u_\infty$ is then the mean value of the vertical rise speed. In a turbulent environment, the mean terminal rising velocity  is noted $\overline{u_\infty}$ and is compared to the corresponding terminal velocity $u_\infty$ in absence of turbulence in Section \ref{Section_bubble_turb}}. 

In general, we will neglect the effect of viscosity and density of the gas inside the bubble because $\mu_g/\mu\ll 1$ and $\rho_g/\rho \ll 1$. Note that although we consider the case of water for obvious reasons, once we express the results in dimensionless terms, the discussion is valid for any gas bubble in any viscous fluid.

The first observation from the data shown in Fig. \ref{fig_1_new}(a) is that the bubble terminal velocity changes in a non-monotonic way with the bubble diameter, $d_b$. Initially, for diameters smaller than 0.5 mm, the bubble terminal speed increases with bubble size. For bubbles in between 0.5 and 1.5 mm, the velocity continues to increase, but data shows large dispersion. For a diameter of approximately 1.5 mm, the bubble velocity appears to decrease, or remains roughly constant, with increasing bubble size until reaching approximately 10 mm. For bubbles with sizes larger than this, the speed increases monotonically with size but at a different rate than small bubbles.  As detailed below, these different behaviors are the consequence of bubble deformation and surface contamination by the presence of surfactants in the liquid. {Let us first consider the  bubble rising velocity in dimensionless terms. Then, we introduce the drag force $F_D$, which is the main force that determines the value of the terminal rise velocity.} Note that $d_b$, in general, represents the bubble equivalent diameter, which is defined as
\begin{equation}
    d_b=\left(\frac{6 \vartheta_b}{\pi}\right)^{1/3}
\label{eqn:db}
\end{equation} 
where $\vartheta_b$ is the bubble volume.

\subsection{{Dimensional Analysis}}
In most cases of practical interest, the density and viscosity of the gas are much smaller than that of the liquid ($\rho_b/\rho \ll 1$ and $\mu_b/\mu \ll 1$ ); therefore, their effect on bubble motion can be neglected. We can \emph{a priori} consider the following fundamental variables that control the bubble rising velocity $u_\infty$ {in a fluid at rest}: a characteristic size $d_b$, liquid with density and viscosity, $\rho$ and $\mu$, gravity $g$, and surface tension $\sigma$ so that
\begin{equation}
u_\infty = f(d_b, \rho, \mu, g, \sigma).
\end{equation}
According to the $\Pi$-Vaschy-Buckingham theorem, the dimensionless bubble speed depends only on two non-dimensional groups. 
In the famous Clift-Grace-Weber diagram \cite{clift1978}, the normalized bubble rising velocity is reported using the Reynolds number $\Rey$ and plotted as a function of the bubble Bond (or Eotvos) number $\Bo$ (Eo) for fixed values of the Morton number $\Mo$:
\begin{equation}
\Rey = \mathcal{F}(\Bo, \Mo),
\end{equation}
which are defined as:
\begin{align}
\Rey & = \frac{\rho d_b u_\infty}{\mu} \label{eqn:Re} \\
\Bo &=  \frac{\rho g d_b^2 }{\sigma} \label{eqn:Bo}\\
\Mo & = \frac{g\mu^4}{\rho \sigma^3} \label{eqn:Mo}.
\end{align}
The Reynolds number, \Rey,  classically compares inertial to viscous effects. 
{The Bond number (also called Eotvos number in many references) is a comparison of the bubble diameter to the capillary length scale $\ell_c=\sqrt{\sigma/\rho g}$. The Morton number has a unique value for a given liquid-gas combination; it is obtained by combining inertia, viscous, surface tension and gravity effects {to eliminate $d_b$ and $u_\infty$}.}

{The experimental data reported in Fig. \ref{fig_1_new}(a) for air bubbles in water correspond to a unique value of the Morton number, $\Mo\approx 2 \times 10^{-11}$. In Fig. \ref{fig_1_new}(b), the same data are presented in terms of  ($\Bo$, $\Rey$). We also show a comparison with the iso-Morton curve $\Mo=10^{-11}$ extracted from the Clift-Grace-Weber diagram \cite{clift1978}. For each liquid-gas combination, a relationship $\Rey = \mathcal{F}(\Bo, \Mo)$ exists.}

Clearly, there are other dimensionless numbers that can be formed using these variables. They are used in the literature depending on the regime at which bubbles move {and the physical terms they compare}. Some often-used numbers are:
\begin{align}
  \We & = \frac{d_b u_\infty^2 \rho}{\sigma} = \frac{\sqrt{\Mo} \Rey^2}{\sqrt{\Bo}} \hspace{0.8cm} \text{Weber number} \label{eqn:We}\\
  \Oh & = \frac{\mu}{\sqrt{\rho\sigma d_b}} = \frac{\sqrt{\We}}{\Rey} \hspace{0.8cm} \text{Ohnesorge number} \label{eqn:Oh}\\
  \Ca & = \frac{\mu u_\infty}{ \sigma} = \frac{\We}{\Rey} \hspace{0.8cm} \text{capillary number} \label{eqn:Ca}\\
  \Ga & = \frac{g d_b^3\rho^2}{\mu^2} =\frac{\Mo \Rey^6}{\We^3} \hspace{0.8cm} \text{Galilei number} \label{eqn:Ga}\\
  \text{Fr} & = \frac{u_\infty}{\sqrt{g d_b}} = \frac{\We^{3/2}}{\Rey^2 \Mo^{1/2}} \hspace{0.8cm} \text{Froude number} \label{eqn:Fr}
\end{align}

The Weber number, $\We$, compares the inertial to surface tension forces. In contrast, the capillary number, $\Ca$, compares the viscous to the surface tension forces.  Bubbles with small values of $\Bo$, $\We$ and $\Ca$ would be spherical, due to the dominance of surface tension forces. In many practical applications, the Ohnesorge number, \Oh, is used to identify the conditions at which a droplet or bubble can fragment, either by viscous stresses or {inertial} forces. The Galilei number $\Ga$ (often also called Archimedes number) is the square of the Reynolds number using $\sqrt{g d_b}$ as the characteristic speed, instead of $u_\infty$; this quantity is used in many studies of bubble dynamics when the bubble terminal velocity is unknown to begin with \cite{Tripathi2015}. 

{It should be noted that when the fluid is not at rest, the terminal velocity $u_\infty$ in the definitions of the dimensionless numbers listed above should be replaced by the norm of the relative velocity $U_\infty = \| \mathbf{U_\infty} \| = \| \mathbf{U} - \mathbf{u_b} \|$, as defined in Fig. \ref{fig_pb_statment}.}

\subsection{Terminal velocity of a spherical bubble}

\subsubsection{Drag force}
Considering that for bubbles in liquids $\mu_g/\mu\ll 1$, it is reasonable to assume that the external fluid is free to slip at the bubble surface as illustrated in Fig. \ref{fig_3} (a). {This slip condition resembles the one that results from the potential flow around objects}. Hence, a naive interpretation would conclude that the drag force around the bubble is zero, considering the d'Alembert paradox. However, the viscous fluid is displaced by the moving bubble, resulting in deformation and thus viscous dissipation. The drag force $\mathbf{F_D}$ experienced by the bubble can therefore be inferred from the total viscous dissipation in the liquid considering \cite{levich1962, batchelor1967}:
\begin{equation}\label{eq:dissipation}
\mathbf{F_D} \cdot \mathbf{u_b} = - \int_{\vartheta} 2 \mu \mathbf{S}:\mathbf{S} d\vartheta 
\end{equation}
where $\vartheta$ is the liquid volume and $\mathbf{S}$ is the strain-rate tensor. 
Note that Eq. \ref{eq:dissipation} is valid for any flow regime, i.e. any Reynolds number $\Rey$. 

Let first consider the solution from creeping flow $Re \ll1$ of a bubble moving at velocity $\mathbf{u_b}$ in a fluid at rest $\mathbf{U}=0$. In that case, the velocity field decays from the bubble center as $u_b R/r$ and the rate of deformation evolves as $ u_b R/r^2$, where $r$ is the radial distance from the bubble center. Therefore, $\mu \mathbf{S}:\mathbf{S} \sim \mu u_b^2 R^2/r^4$ and, from the integral in Eq. \ref{eq:dissipation}, $F_D \propto \mu R u_b$, (linearly with viscosity, bubble size and speed), which clearly has the same scaling as the Stokes drag force of a solid sphere in same condition : $\mathbf{F_D} = - 6 \pi \mu R \mathbf{u_b}$ \cite{stokes1851}. For the case of a bubble, from the exact flow field obtained by \cite{hadamard1911, rybczynski1911}, the drag force and the drag coefficient $C_D$ considering its classical definition, $C_D= 2 F_D/\pi R^2 \rho u_b^2$, are
\begin{equation}
    \mathbf{F_D} = - 4 \pi \mu R \mathbf{u_b}; \quad   C_D=\frac{16}{\Rey}
\end{equation}

As expected, the drag force on a bubble is smaller than the one of a solid sphere because of the slip boundary condition at the bubble surface, which reduces the friction on the interface. However the difference is not so large; the prefactor is $4 \pi$ instead of $6 \pi$. This may be explained by the fact that in both cases  a similar amount of liquid displacement occurs. {This motion contributes to the integral in Eq. \ref{eq:dissipation} in a similar manner, as the momentum associated with the bubble/particle motion is diffused over a domain  around the moving bubble/particle of size $d/\Rey$, which is much larger than the bubble or particle size.}

In the limit of large bubble Reynolds number ($\Rey \gg 1$), the vorticity generated at the bubble surface is finite and of order $O(u_b/R)$ (discussed below). More importantly, vorticity remains confined to a layer of thickness $d_b/Re^{1/2}$, $d_b/Re^{1/4}$, and $d_b/Re^{1/6}$ at the bubble surface, near wake and far wake, respectively \cite{moore1963}.  Consequently, these  regions of vorticity decrease in size when increasing $\Rey$   and the potential flow can then be considered a good approximation for calculating the drag force using Eq. \ref{eq:dissipation}. 
The velocity field obtained from potential flow decays as $u_b R^3/r^3$. Therefore, the deformation rate evolves as $u_b R^3/r^4$ and thus $\mu \mathbf{S}:\mathbf{S} \sim \mu u_b^2 R^6/r^8$. From Eq. \ref{eq:dissipation}, the drag force scales as  $F_D \sim \mu R u_b$, which has the same scaling as a Stokes drag force but  differs from the drag of a solid sphere {at large Reynolds number}. This point is discussed below. The exact calculation, considering the potential flow solution around a bubble, was first conducted  by \citet{levich1962}, resulting in:
\begin{equation}\label{eq:drag_levich}
 \mathbf{F_D} = -12 \pi \mu R  \mathbf{u_b} ; \quad   C_D=\frac{48}{\Rey}.
\end{equation}

\begin{figure}[h]
\includegraphics[width=5in]{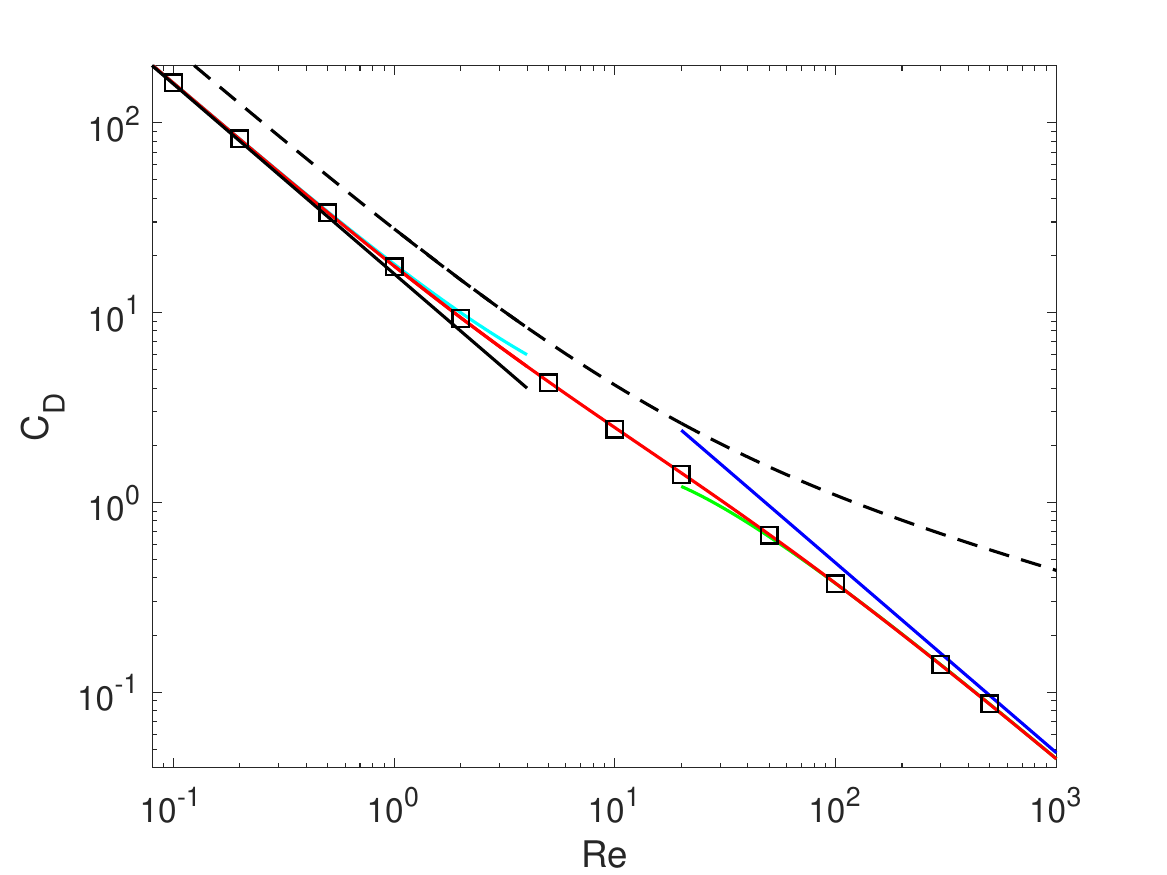}
\caption{{Drag coefficient, $C_D$, as a function of Reynolds number, $\Rey$, for a spherical bubble: ($\square$) direct numerical simulations from \cite{legendre1998}. The continuous lines show the relations for a clean bubble (see Table \ref{tab_1} for each line symbol). The red line shows the prediction from Eq. \ref{eq:drag_mei_K}).  The dashed line is the prediction for a solid sphere used to describe contaminated bubble; see also Table \ref{tab_1}.}}
\label{fig_CD}
\end{figure}

We can now discuss the evolution of the drag force, and correspondingly the drag coefficient,  from small to large $\Rey$ for a spherical bubble. Different analytical expressions have been derived to extend the Stokes solution \cite{hadamard1911,rybczynski1911} to  Reynolds number of order unity \cite{Acrivos1962}, and to extend the Levich solution \cite{levich1962} to  Reynolds number of order 50 \cite{moore1963}. Clearly, it is not possible to find an analytical expression for intermediate values of the $\Rey$ number but {an} accurate fit was obtained using {direct} numerical simulations{, i.e. by solving the entire Navier-Stokes equations,} by \citet{mei1992}. In its general form when the liquid is moving at velocity  $\mathbf{U}$, the expression is:
\begin{equation}\label{eq:drag_mei}
    \mathbf{F_D} =  4 \pi \mu R \, \mathcal{K}(\Rey) \, \left( \mathbf{U} - \mathbf{u_b}\right), \quad  C_D=\frac{16}{\Rey} \mathcal{K}(\Rey)
\end{equation}
where $\mathcal{K}(\Rey)$  connects the small $\Rey$ to the large $\Rey$ drag force evolution 
\begin{equation}\label{eq:drag_mei_K}
  \mathcal{K}(\Rey) = \frac{16+3.315 \Rey^{1/2}+ 3 \Rey}{16+3.315 \Rey^{1/2}+ \Rey}.
\end{equation}
All these expressions are shown in Table \ref{tab_1} and plotted in Fig. \ref{fig_CD}. As also indicated in Eq. \ref{eq:drag_mei}, the relevant characteristic velocity to consider is $\mathbf{U} - \mathbf{u_b}$ so that the Reynolds number is now written as:
\begin{equation}\label{def_Re_correct}
\Rey = \frac{d_b   \|  \mathbf{U} - \mathbf{u_b}  \|  \rho}{\mu}.
\end{equation}

\begin{table}[h]
\caption{Drag force, {$\mathbf{F_D}$}, and drag coefficient, {${C_D}$},  for a spherical bubble over a wide range of Reynolds numbers. 
{The expressions are given for the general configuration of a bubble moving at velocity $\mathbf{u_b}$ in a fluid of local velocity $\mathbf{U}$. The drag force is then directed along the relative velocity $ \mathbf{U_\infty} = \mathbf{U}-\mathbf{u_b} $,  as defined in Fig. \ref{fig_pb_statment}. The function $\mathcal{K}(\Rey)$ is given by Eq. \ref{eq:drag_mei_K}.}
\label{tab_1}}
\begin{center}
\begin{tabular}{|c|c|c|c|}
 \hline
Range & Drag force, {$\mathbf{F_D}$} & Drag coefficient, $C_D$ &  Method and source \\ 
 \hline
$\Rey \ll 1$ (\textcolor{black}{\rule[2pt]{18pt}{1.5pt}}) & $ 4 \pi \mu d_b  {\mathbf{U_\infty}}$ & $\frac{16}{\Rey}$ &  Analytical solution  \cite{hadamard1911,rybczynski1911}\\ \hline 
$\Rey \le 1$ (\textcolor{cyan}{\rule[2pt]{18pt}{1.5pt}}) &  $ 4 \pi \left(1+ \frac{\Rey}{8}\right)  \mu d_b {\mathbf{U_\infty}}$ & $\frac{16}{Re} +2 $ &  Analytical solution \cite{Acrivos1962}\\  \hline
all $\Rey$ (\textcolor{red}{\rule[2pt]{18pt}{1.5pt}}) & $4 \pi \mathcal{K}(\Rey)  \mu d_b {\mathbf{U_\infty}}$ & $\frac{16}{\Rey} \mathcal{K}(\Rey)$ &  Numerical data fit \cite{mei1992} \\  \hline
$\Rey \ge 50$ (\textcolor{green}{\rule[2pt]{18pt}{1.5pt}})& $ 12 \pi \left(1- \frac{2.211}{\sqrt{\Rey}}\right) \mu d_b {\mathbf{U_\infty}}$ & $\frac{48}{\Rey} \left(1- \frac{2.211}{\sqrt{\Rey}}\right)$ &  Analytical solution \cite{moore1963}  \\  \hline
$\Rey \gg 1$ (\textcolor{blue}{\rule[2pt]{18pt}{1.5pt}})& $ 12 \pi \mu d_b {\mathbf{U_\infty}}$ & $\frac{48}{Re}$ & Analytical solution \cite{levich1962}  \\  \hline
$\Rey \le 800$ ({\rule[2pt]{5pt}{1.5pt}} {\rule[2pt]{5pt}{1.5pt}} {\rule[2pt]{5pt}{1.5pt}}) &  \, $ 6 \pi\left(1+0.15 \Rey^{0.687}\right)  \mu d_b {\mathbf{U_\infty}}$ \, & \, $\frac{24}{Re} \left(1+0.15 \Rey^{0.687}\right)$ \, &  Empirical fit \cite{schiller1933} \\
\, Contaminated bubble \, & &  &  \\ \hline 
\end{tabular}
\end{center}
\end{table}

\subsubsection{Terminal velocity}
Since the drag and buoyancy forces are known from Eqs. \ref{eq:drag_mei} and \ref{eqn:buoyancy}, respectively, an explicit expression for the terminal  speed $u_\infty$ of a spherical bubble can be obtained considering a steady balance, leading to:
\begin{equation}\label{eq_u_terminal}
u_\infty= \frac{1}{12 \, \mathcal{K}(\Rey)} \frac{\rho g d_b^2}{\mu}
\end{equation}
where the prefactor {$({12 \, \mathcal{K}(\Rey)})^{-1}$} varies from $1/12$ to $1/36$ from small to large $\Rey$. The expression above shows that $u_\infty$ is proportional to $d_b^2$, which is in accordance with the trend shown by the data in Fig. \ref{fig_1_new} (a) for diameter smaller than 1 mm.

\subsubsection{Scaling of the drag force  at large Reynolds numbers\label{section_draglargeRe}}

For a spherical bubble moving at large \Rey, the drag coefficient is inversely proportional to the Reynolds number, as indicated by Eq. \ref{eq:drag_levich}.  This behavior at large Reynolds number clearly differs  from the drag force experienced by a solid sphere moving at same speed, $u_S$. If the flow  is dominated by inertial effects, we expect  $F_D \propto \rho R^2 u^2_S$ which would result in a constant drag coefficient. The drag coefficient of a solid sphere  is $C_D \approx 0.45$  for $\Rey>800$, while the drag is found to be $F_D \propto \mu R u_b$ for a spherical bubble resulting in $C_D=48/\Rey$.
This significant difference can be explained by the flow {separation and the induced} recirculation that develops in the wake of a solid sphere when $\Rey$ increases. The flow around a spherical bubble, on the other hand, resembles that predicted by potential flow without any recirculation {\citep{dandy1986,blanco1995}}, as it is discussed  in the next  section.
At large Reynolds number the drag force is mainly given by the pressure distribution on the surface and it can be estimated from the pressure evolution as  $F_{D, P} \approx \Delta P \,S$ where $\Delta P$ is the characteristic pressure variation between the top and bottom part {(respectively $\theta=0$ and $\theta=\pi$ according to Fig. \ref{fig_3}.a)} of the sphere and $S=4 \pi R^2$ is the sphere surface. 
In both cases, the pressure on the top of the surface is given by the dynamic pressure $\rho u_b^2/2$ (resp. $\rho u_S^2/2$).
In the case of a bubble,  the pressure distribution on the surface is close to being up-downstream symmetrical  as in the case of potential flow, with some deviation  resulting from  viscous effects located in a boundary layer of thickness $\sim d_b/ \Rey^{1/2}$, so that $\Delta P = O (\mu u_b/d_b)$ and $F_{D, P} \sim \mu R u_b$. In contrast, for a solid sphere the dynamic pressure is not recovered downstream   because of the {flow separation and the resulting} recirculation in the wake, so that  $\Delta P = O(\rho u_S^2)$ and $F_{D, P} \sim \rho R^2 u^2_S$, leading to a constant value of the drag coefficient. {To support this discussion, a comparison of pressure distribution between a spherical bubble and a solid sphere can be found in \citet{magnaudet1995}.}

\subsubsection{Drag and production of vorticity at the bubble surface}\label{section_dragvorticity}

\begin{figure}[h!]
\includegraphics[width=6.5in]{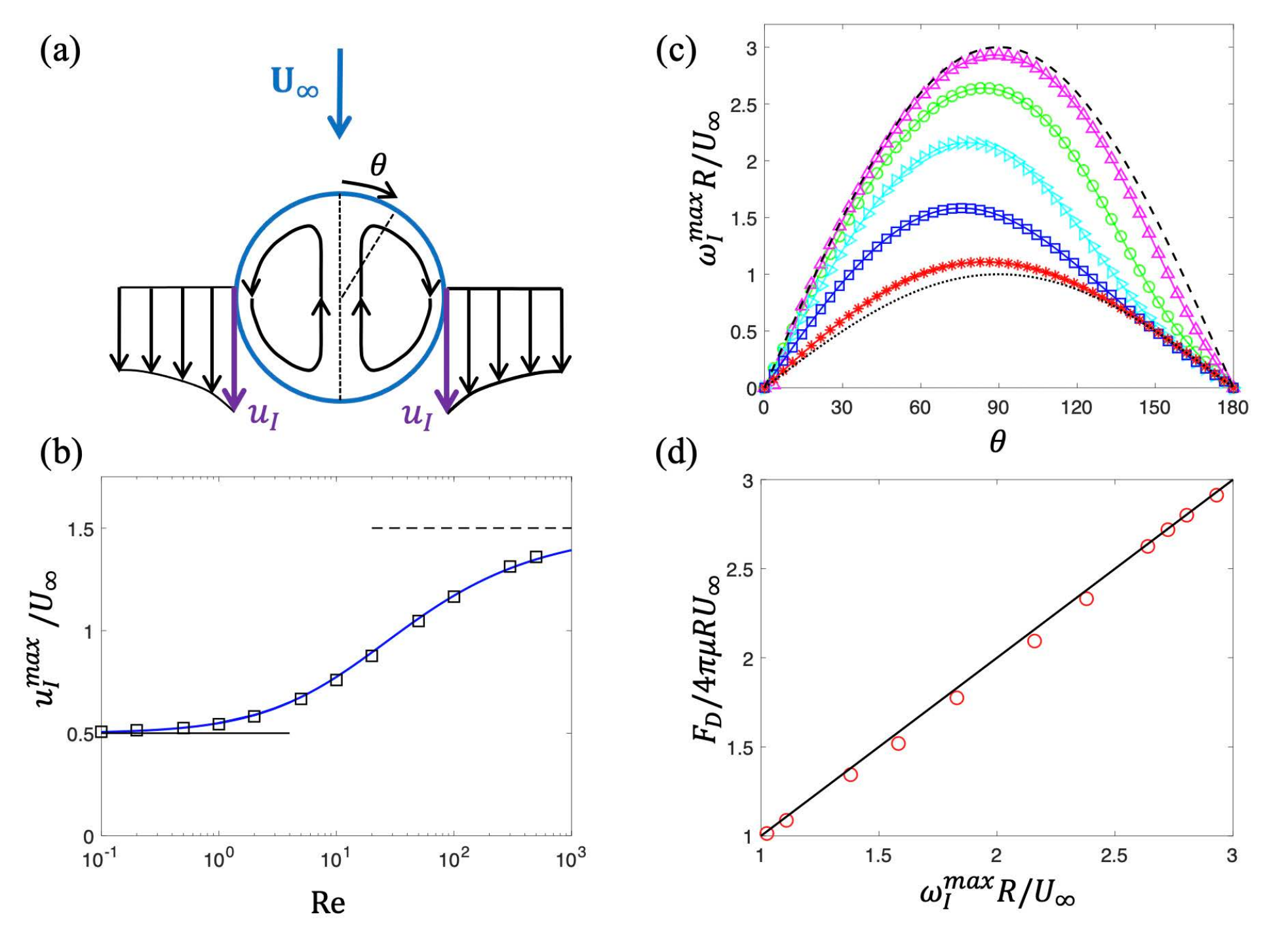}
 \caption{(a) Schematic illustration of the flow around and inside a bubble shown in the frame of reference {$\mathcal{R}_b$ (see Fig. \ref{fig_pb_statment})} moving with the bubble. 
 (b) Normalized maximum interfacial velocity $u_I^{\text{max}}$ as a function of  $\Rey$: (\textcolor{blue}{\rule[2pt]{18pt}{1.5pt}}), Equation (\ref{eq_uimax}); (\textcolor{black}{\rule[2pt]{18pt}{1.5pt}}),  $u_I^{\text{max}}=1/2 U_\infty$, Stokes solution from \citet{hadamard1911} and \citet{rybczynski1911}; ({\rule[2pt]{5pt}{1.5pt}} {\rule[2pt]{5pt}{1.5pt}} {\rule[2pt]{5pt}{1.5pt}}) $u_I^{\text{max}}=3/2 U_\infty$, potential flow solution;  ($\square$) direct numerical solution taken from \citet{legendre2007}.
 (c) Surface distribution of vorticity, {obtained from direct numerical simulation from \citet{legendre2007}:} (\textcolor{red}{*}) $\Rey=1$, (\textcolor{blue}{$\square$}) $\Rey=10$, (\textcolor{cyan}{$\triangleright$}) $\Rey=50$, (\textcolor{green}{$\circ$}) $\Rey=300$,
 (\textcolor{magenta}{$\triangle$}) $\Rey=5000$, (\makebox[8mm]{\dotfill}) Stokes flow,  ({\rule[2pt]{5pt}{1.5pt}} {\rule[2pt]{5pt}{1.5pt}} {\rule[2pt]{5pt}{1.5pt}}) potential flow.
 (d) The normalized drag force $F_D / 4\pi\mu R U_\infty$ as a function of the normalized maximum surface vorticity $\omega_I^{\text{max}} R/U_\infty$. 
 {Symbols are from the numerical results shown in (c). (\textcolor{black}{\rule[2pt]{18pt}{1.5pt}}) is the parity line.}}
\label{fig_3}
\end{figure}

According to \citet{legendre2007}, the magnitude of the drag force on a bubble can be written in terms of the maximum of the surface vorticity as:
\begin{equation}\label{eq_FDomegamax}
F_D= 4\pi\mu R^2 \omega_I^{\text{max}}.
\end{equation}
For an axisymmetric flow for a spherical bubble, the surface vorticity is directed along the azimuthal direction and can be expressed as a function of the surface velocity $u_I$ as
\begin{equation}\label{eq_wI}
\omega_I = \frac{1}{R}\frac{\partial u_I}{\partial \theta}+ \frac{u_I}{R}
\end{equation}
$u_I$ is the tangential fluid velocity at the bubble surface in the reference frame moving with the bubble where the fluid velocity far from the bubble is then $\mathbf{U_\infty} = \mathbf{U} - \mathbf{u_b}$. Since the gas viscosity inside the bubble has a negligible effect {and considering a clean surface},  the liquid experiences a zero shear stress at the bubble surface  $\frac{1}{R}\frac{\partial U_I}{\partial \theta} - \frac{u_I}{R}=0$. Therefore, the interfacial vorticity is then
\begin{equation}\label{eq_omegamax}
    \omega_I =2 \frac{u_I}{R}.
\end{equation}
Figure \ref{fig_3}(c) shows the surface vorticity distribution for different Reynolds numbers varying from 0.1 to 5000, calculated from direct numerical simulations \cite{legendre2007}. These results show that $\omega_I$ varies continuously from the Stokes solution ($U_\infty \sin (\theta) /R$ from \citet{hadamard1911} and \citet {rybczynski1911}) to the potential flow solution ($3 U_\infty \sin (\theta) / R$) which was obtained from the potential flow on a sphere surface $u_I= 1.5 U_\infty \sin (\theta)$ in Eq. \ref{eq_omegamax}. This plot shows that, due to the absence of any recirculation in the  wake, for $\Rey=1000$ the solution is close to the one imposed by the potential flow. This remarkable result differs drastically  from the classical picture of the flow around a solid sphere where a recirculation develops at large $\Rey$ increasing the drag force.
From Fig. \ref{fig_3}(c) the maximum vorticity $\omega_I^{\text{max}}$ is measured and  the evolution of the normalized drag force $F_D / 4\pi\mu R U_\infty$ is reported in Fig. \ref{fig_3}(d) as a function of the normalized maximum interfacial vorticity  $\omega_I^{\text{max}} R/U_\infty$ to demonstrate relation (\ref{eq_FDomegamax}). This reveals that the drag force is directly proportional  to how much vorticity is produced at the bubble surface. As discussed below, the surface vorticity is very relevant to understand many aspects  of the bubble dynamics  considered in this review.

Finally, comparing Eqs. \ref{eq_FDomegamax} and  \ref{eq:drag_mei}, the  maximum vorticity at the bubble surface varies as: 
\begin{equation}\label{eq_uimax}
\omega_I^{\text{max}} = \frac{U_\infty}{R}  \mathcal{K}(\Rey) 
\end{equation}
with $\mathcal{K}(\Rey)$ given in Eq. \ref{eq:drag_mei}.
Figure \ref{fig_3}(b) shows how the maximum value of $u_I^{\text{max}} = \omega_I^{\text{max}} R/2 $ evolves with \Rey, calculated from direct numerical simulations \cite{legendre2007}. This evolution is well fitted with Eq.\ref{eq_uimax}.  {Note that the relation between maximum surface vorticity and drag force is also observed for ellipsoidal bubbles with an axisymmetric wake, as discussed by  \citet{legendre2007}}.

\subsection{Bubble deformation {and rupture}\label{sec_deformation}}
\begin{figure}[h]
\includegraphics[width=6.5in]{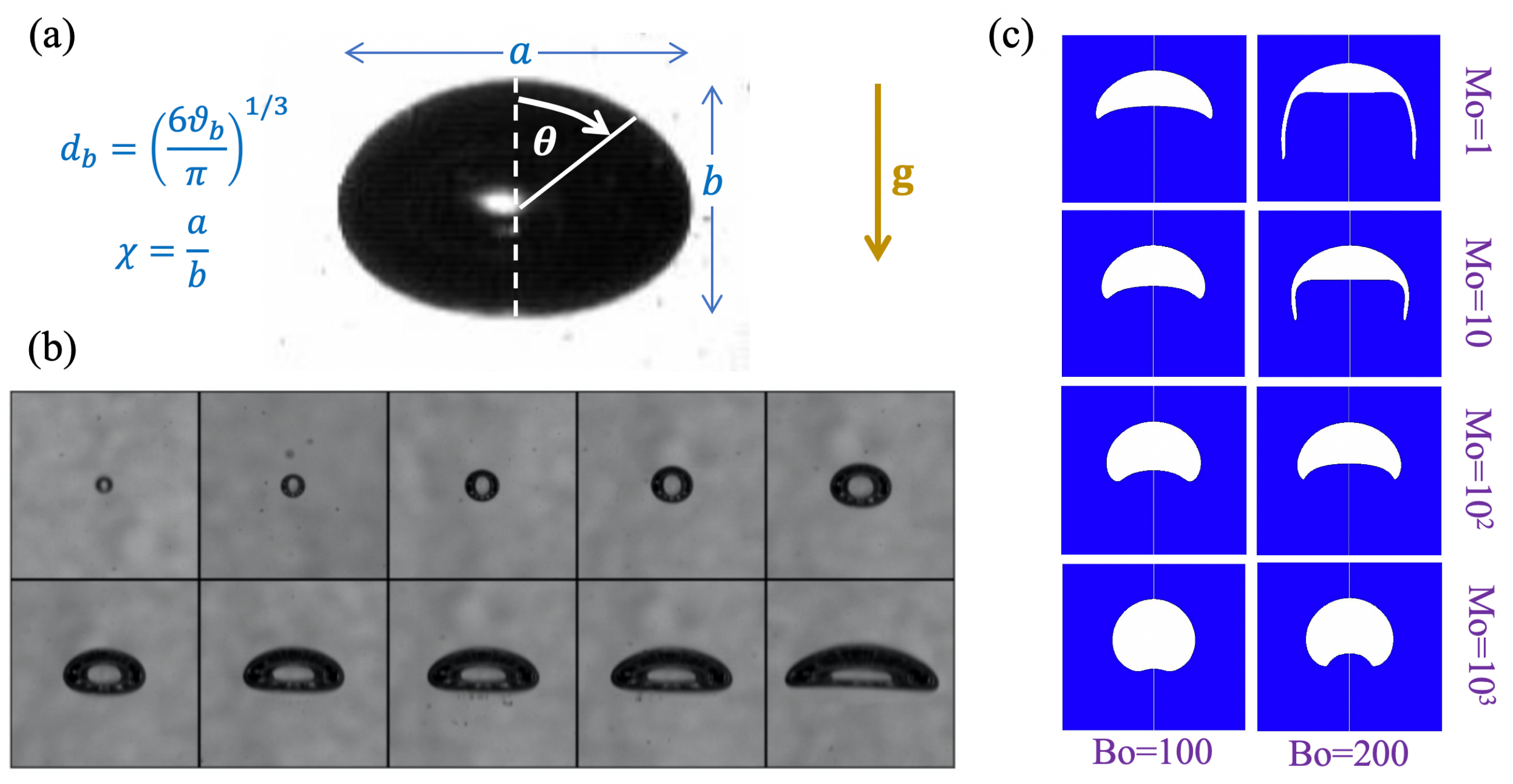}
\caption{{(a) Definition of bubble equivalent diameter $d_b$ and aspect ratio {$\chi=a/b$}. (b) Evolution of the bubble shape with increasing equivalent bubble diameter $d_b$ in a 90\% glycerin-water solution.  $d_b$ ranges from 2.5 to 10 mm. Images taken from \citet{ravisankar2021}. 
(c) Bubble shapes for  $\Bo \ge 100$ {and Morton number ranging from $\Mo =1$ to $\Mo =10^3$} showing spherical cap, dimpled and skirt bubbles. Images {from direct numerical simulation taken from \citet{legendre2022}}. In all cases, the images show the side view of bubbles}.}
\label{fig_deformation}
\end{figure}

The shape of a gas bubble is determined by a balance between surface tension forces, hydrostatic pressure,  viscous, and inertial forces which grow in magnitude depending on the flow condition, flow regime, and bubble size. Figure \ref{fig_deformation} shows some examples of different shapes that bubbles can adopt. Bubble deformation has been extensively described in the literature \cite{clift1978, maxworthy1996} and we report here the main findings considering bubbles rising at their terminal velocity $u_\infty$ in a liquid at rest.

The spherical shape is a consequence of the dominance of surface tension, which acts to minimize the surface area. When a bubble rises in a liquid its shape then results from the normal stress balance at its interface. As discussed above, the effect of the inertia and normal viscous stress from the gas phase can be neglected in the momentum balance (Eq. \ref{eq:surface_mometum}) resulting in
\begin{equation}\label{eq_def1}
    P_g= 2 H\sigma  +P-\mathbf{n_b}\cdot \mathbf{\Sigma}\cdot\mathbf{n_b}
\end{equation}
where $P_g$ is the pressure inside the bubble. {Because both gas inertia and viscosity effects are negligible inside the bubble, $P_g$} is considered constant and uniform in the following discussion.

From this equation we can argue that the bubble will depart from a spherical shape  when the magnitude of either the pressure or viscous stress becomes of order $\sigma/d_b$.  The contributions to bubble deformation (second and third terms in the right-hand-side of Eq. \ref{eq_def1}) include pressure changes due to a hydrostatic head, $\rho g d_b$, dynamic pressure, $\rho u_b^2$, and viscous stresses, $\mu u_b/d_b$. Comparing these effects to  $\sigma/d_b$, we can form three dimensionless groups: $\Bo, \We$ and $\Ca$, defined by Eqs. \ref{eqn:Bo}, \ref{eqn:We} and \ref{eqn:Ca}, respectively. Hence, we can expect a deviation from a spherical bubble shape when either of these conditions is met:
\begin{equation}\label{eq_def_bubble}
\Bo > 1, \quad \We > 1, \quad \Ca >1.
\end{equation}
Note that when inertial effects are negligible ($\Rey\rightarrow 0$), the shape of a rising bubble is spherical regardless of the values of $\Bo$, $\We$, or $\Ca$. In this case, the viscous and gravitational contributions are balanced across the bubble surface, as shown by \citet{Taylor1964}.

When inertia is present ($Re>1$), the conditions in Eq. \ref{eq_def_bubble} can be used to determine the critical diameter at which a bubble is expected to deform while rising in a liquid.  The condition \Bo $>1$ implies that the bubble diameter is larger than the capillary length, $\ell_c=\sqrt{\sigma/\rho g}$. For an air bubble in water, this implies that the bubble should have a diameter $d_b$ (resp. volume $\vartheta_b$) larger than 2.7 mm (resp. 1.4 mm$^3$).  The condition $\Ca>1$, for air-water, implies that the bubble velocity would have to be larger than $\sigma/\mu=72$ m/s, which is well beyond what is observed experimentally under standard gravitational conditions (see Fig. \ref{fig_1_new}). Therefore, viscous induced deformation is expected to be negligible for water but may be relevant to more viscous fluids. For $\We>1$, we can consider the terminal velocity $u_\infty=\rho g d_b^2/36 \mu$ inferred from Eq. \ref{eq_u_terminal}, in the $\Rey\gg1$ regime since viscous effects are small. Therefore,  for a bubble to deform due to inertial effects, $d_b \ge \left( 36^2 \sigma \nu^2 / g^2 \rho \right)^{1/5}$. For air-water properties, $u_\infty=27$ cm/s and $d_b>1$ mm ($\vartheta_b > $ 0.52 mm$^3$). This limit is shown in Fig.\ref{fig_1_new}, which coincides  with the transition when the bubble velocity departs from the $d_b^2$ dependence. This indicates that when the bubble shape is no longer spherical, its terminal speed is affected by the deformation.

When bubbles are deformed their characteristic size can be defined based on their volume $\vartheta_b$, which is independent of the deformation. We define the  equivalent bubble diameter as:
\begin{equation}\label{eq_def_deq}
d_b = \left(\frac{6 \vartheta_b}{\pi} \right)^{1/3}.
\end{equation}
Bubble deformation is usually characterized by the bubble aspect ratio, $\chi$, which is defined as:
\begin{equation}
    \chi=\frac{a}{b}
\end{equation}
where $a$ and $b$ are the larger and smaller dimensions of the bubble, respectively, {as depicted in Fig. \ref{fig_deformation}.} Note that in some studies, the bubble deformation is described using the inverse of $\chi$, namely, $E=b/a$, resulting in values smaller than 1 \cite{Aoyama2017}.

\subsubsection{Ellipsoidal shape}\label{section_ellpsoidal}
Based on the above discussion we expect that, {for a gas bubble in a liquid}, the effect of dynamic pressure will first induce a change of the bubble shape.  If the $\Rey$ number is large, we can assume that the potential flow prediction models the flow around the bubble correctly. From such a solution, the pressure distribution can be used to write the normal stress balance, Eq. \ref{eq_def1}, as:
\begin{equation}\label{eq_def2}
    P_g= 2 H\sigma  + P_o- \frac{9}{8}\rho u_\infty^2 \sin\theta
\end{equation}
where  $\theta$ is the azimuthal angle measured to be zero at the bubble top. For $\theta=0$, we have $P_g = P_o + 2 H_{top} \sigma$, which fixes the reference pressure. For other values of $\theta$, moving along the bubble surface, we have
\begin{equation}
    H=H_{top} + \frac{9}{16}\frac{\rho u_\infty^2}{\sigma} \sin\theta.
\end{equation}
Since  the surface tension remains the same, the radii of curvature of the bubble surface is observed to increase. Therefore, at $\theta=0$ or $\pi$ the bubble curvature  is minimum $H_{top}$, but at $\theta=\pm\pi/2$, at the bubble equator, $H_{equator}-H_{top} = \frac{9}{16}\frac{\rho u_\infty^2}{\sigma}$. When normalized by the bubble diameter, we see that the relative change in shape is controlled by the Weber number $\We$. As a result, the bubble becomes flattened along the rising direction, taking on a shape similar to an ellipsoid with the small axis parallel to the direction of its ascent. This change in shape increases the bubble's drag force. The bubble velocity then decreases, as shown in the velocity evolution in Fig. \ref{fig_1_new}, for bubble diameter ranging from 1 mm to 1 cm.

The bubble deformation and the induced change in drag force were first studied by \citet{Taylor1964} for small Reynolds numbers and by \citet{moore1965} for large Reynolds numbers. For small bubble deformations, the bubble aspect ratio and drag coefficient include $\We$ corrections in both large and small $\Rey$ limits. Hence, the Weber number is the relevant non-dimensional number for deformation effects when inertia is {present}. It is important to note that in some studies, the bubble deformation is characterized using the Bond number,  $\Bo$, \cite{Aoyama2017}. When considering a bubble velocity scale of $u_\infty \propto \sqrt{g d_b}$, $\We$ and $\Bo$ become equivalent.

For $\Rey \ll 1$, \citet{Taylor1964} report that 
\begin{equation}
      \chi=1+ \frac{5}{32} {\We}; \quad \, C_D=\frac{16}{\Rey} \left[ 1+ \frac{\Rey}{8} + \frac{\We}{12}  \right].
\end{equation}

In the limit $\We \rightarrow 0$ ($\chi \rightarrow 1$), the analytical expression by Taylor \& Acrivos for spherical bubbles (see Table \ref{tab_1}) is recovered. For the opposite limit of a large Reynolds number ($\Rey \gg 1$), \citet{moore1965} obtained
\begin{equation}
      \chi=1+\frac{9}{64} {\We}; \quad  \, C_D=\frac{48 G(\chi)}{\Rey} \left[ 1+ \frac{H(\chi)}{\Rey^{1/2}}  \right]
\end{equation}
where $G(\chi)$ and $H(\chi)$ are implicit functions of $\chi$, that can be found in \citet{moore1965}. In the limit $\We \rightarrow 0$ ($\chi \rightarrow 1$), the functions tend to $G(\chi) \rightarrow 1$ and $H(\chi) \rightarrow -2.21$. In such a case the Moore analytical expression for spherical bubbles (see also Table \ref{tab_1}) is recovered.

Based on these findings, many studies to determine the bubble shape have been performed  for different fluids and  bubble sizes. \citet{Legendre2012} used these studies to develop a correlation describing bubble deformation over a wide range of Morton numbers, applicable from water ($Mo = O(10^{-11})$) to water-glycerin solutions ($Mo = O(1)$):
\begin{equation}
    \chi=\frac{1}{1-\frac{9}{64} {\We}{\left( 1 + 0.2 Mo^{1/10} \right)^{-1}} }.
\end{equation}

For small to moderate deformation, the surface tension and dynamic pressure effects balance at the surface, resulting in a shape for which $\We$ is approximately constant \cite{maxworthy1996}.
A constant Weber number indicates that the terminal velocity is now decreasing with the diameter, as described by
\begin{equation}\label{eq_def4}
u_\infty \propto \left(\frac{\sigma}{\rho d_b} \right)^{1/2}.
\end{equation}
This trend is shown in Fig. \ref{fig_1_new}, which appears to agree with experimental observations for  bubble diameters between 1 and 10 mm. In Fig. \ref{fig_1_new}, the Eq. \ref{eq_def4} is plotted with the prefactor 1.87 corresponding to a Weber number $\We \approx 3.5$. For diameters larger than 10 mm, the bubble velocity increases again, which corresponds to the next behavior regime described below (spherical cap regime).

\subsubsection{{Zigzagging bubbles}\label{sec_zigzag}}

{An remarkable behavior appears during the rise of such ellipsoidal bubbles. After a critical diameter, $d_b \approx 1.8$mm ($\vartheta_b \approx $ 0.94 mm$^3$) for an air bubble in water, the trajectory ceases to be rectilinear, and instead follows zigzagging or spiraling paths \cite{hartunian1957, saffman1956,aybers1969,lunde1997,DeVries2002}. Leonardo da Vinci observed and documented this behaviour in the first scientific reference of the phenomenon \cite{Leonardo}. These remarkable images, shown in Fig. \ref{fig_examples}(a), were discovered by Professor Andrea Prosperetti \cite{prosperetti2004}, who coined the term `Leonardo's Paradox'.  The review by Magnaudet and Eames \cite{magnaudet2000} gives a clear summary of the many experimental studies and attempts to understand the nature of the instability, which was not fully understood when the review was written.}

\begin{figure}[h]
\includegraphics[width=5.in]{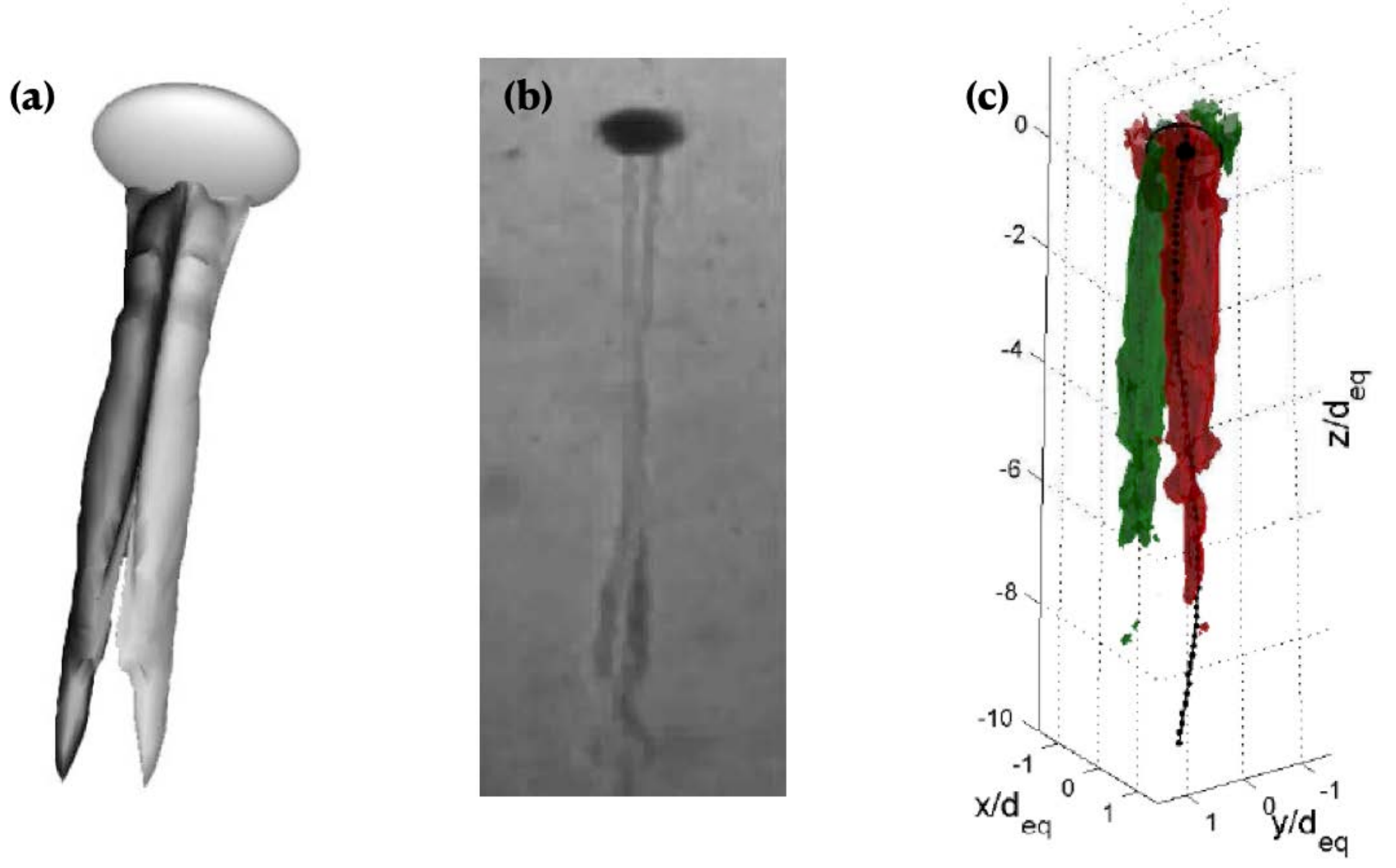}
\caption{{The wake behind zigzagging bubbles: (a) Numerical result for a fixed-shaped ellipsoidal bubble from \citet{mougin2002}; (b) Schlieren images of the wake behind a 2 mm diameter air bubble in zigzagging motion in purified water, from \citet{DeVries2002}; (c) Reconstruction of the wake behind a zigzagging bubble rising in silicone oil, from \citet{zenit2009b}.}
} 
\label{fig_zigzag}
\end{figure}

{It is now clear that the path instability originates from the instability of the wake behind the bubble. The conditions for the appearance of the wake instability remained unclear because most of the experimental evidence was for air bubbles in water, which is subject to surface contamination. The publication of experimental measurements for ultra-purified water \cite{duineveld1995}, along with the development of numerical techniques capable of reproducing the structure of the wake behind bubbles \cite{blanco1995}, reignited interest in the subject. When the bubble shape exceeds a critical aspect ratio, its wake transitions from being axi-symmetric to displaying a double threaded with counter-rotating stream-wise vorticity, as shown in Fig. \ref{fig_zigzag} \cite{lunde1997,DeVries2002,zenit2009b}. In a series of papers \cite{mougin2002,mougin2006,mougin2007,tchoufag2013,tchoufag2014,cano2016,cano2016b}, Magnaudet and collaborators, using direct numerical simulations and linear stability analysis, developed a detailed picture of the nature of this instability. In their most recent paper, \citet{Bonnefis2023} demonstrate that the path instability always arises from the destabilization of a non-axisymmetric mode. As the bubble size increases, this primary instability, tied to a fixed bubble shape, remains unchanged for Morton numbers larger than $10^{-7}$, which are associated with weak, time-dependent changes in bubble shape. They also found that time-dependent bubble deformation does not play a causal role in the first stages of the path instability for low-Mo liquids.
}

\subsubsection{Spherical cap shape}
When the bubble volume increases, the shape evolves into a spherical cap shape { (see Fig. \ref{fig_deformation}) and the bubble trajectory regains a rectilinear rising path}. The shape is now governed by the balance between the fluid inertia and the hydrostatic pressure balance at the interface, $\rho u_\infty^2 \propto \rho g d_b$,  resulting in 
\begin{equation}\label{eq_cap1}
 u_\infty \propto \sqrt{ g d_b}.
\end{equation}
This dependence of $u_\infty$ with $d_b$ is shown in Fig. \ref{fig_1_new} for $d_b > 10$ mm, again in good agreement with experiments. {By balancing the buoyancy force, $\rho g \frac{\pi}{6} d_b^3$, with the drag force, $C_D \frac{1}{8}\pi d_b^2 \rho u_\infty^2$, we can write}
\begin{equation}\label{eq_ub_cd}
\Fr^2 =\frac{u_\infty^2}{{gd_b}} = \frac{4}{3} \frac{1}{{C_D}}.
\end{equation}
Combining Eqs. \ref{eq_cap1} and \ref{eq_ub_cd}, results in a constant Froude number and a constant drag coefficient. Note that for a solid object, a regime of constant drag coefficient, sometimes called the Newton regime, is also observed but its origin is completely different {as detailed in section \ref{section_draglargeRe}}. 

Considering the potential flow approximation at the front of a spherical cap bubble, the pressure distribution, $\frac{9}{8} \rho u_b^2 \sin^2{\theta}$, {is balanced} by the variation in the liquid's hydrostatic pressure, $\rho g a (1 - \cos{\theta})$, where $a$ is the radius of the spherical cap, and $\theta$ is the angle measured from the stagnation point. The  limit of $\theta \rightarrow 0$ gives an expression for the terminal velocity obtained by \citet{Davies1950} and \citet{Joseph2003}:
\begin{equation}\label{eq_cap2}
u_\infty=\frac{2}{3}\sqrt{g a}.
\end{equation}
This relation is  expected to hold for $\Rey \gg 1$, $\We\gg1$ and $\Bo\gg1$ because the flow is assumed  to be potential at the bubble front but the surface tension effect is considered to be negligible compared to both inertia and gravity. For a spherical cap bubble, the front curvature radius, $a$, can be related to the bubble diameter as 
$ a=c(Re) d_b$ according to \citet{clift1978}. In fact, $c(Re)$ depends {on the angle $\theta_c$ made by the cap} that itself depends on the Reynolds number. In the limit of large Reynolds number, the cap angle reaches a constant value of $\theta_c \approx 50^o$. Correspondingly,   $c(Re)=1.14$. Therefore, the velocity can be simply related to the bubble diameter as
\begin{equation}\label{Eq_UB_cap}
u_\infty=0.707 \sqrt{g d_b}.
\end{equation}
This expression indicates that the bubble rising velocity is again increasing with the bubble size and scales with the characteristic velocity $\sqrt{g d_b}$, in agreement with the trend observed in Fig. \ref{fig_1_new} for $d_b>10$ mm ($\vartheta_b >  524$ mm$^3$). As a consequence of the balance between inertia and gravity, the spherical cap bubble motion is characterized by a constant Froude number $\Fr=0.707$ and a constant drag coefficient of $C_D=8/3$.

At large Bond numbers, $\Bo,$ and for sufficiently viscous fluid a peculiar bubble shape is observed:  a thin film of gas, commonly referred to as a `skirt', forms and extends downwards from the rim of the spherical cap, as shown in Fig. \ref{fig_deformation}. Experiments \cite{Guthrie1969} indicate that the  thickness of the skirt film is around 50$\mu$m. Direct numerical simulations have been useful to investigate the flow structure, the skirt thickness and length and the corresponding terminal velocity for such bubbles \cite{legendre2022}. The rising velocity of spherical, skirt and spherical cap bubbles can be fitted by:
\begin{equation}\label{Eq_UB_skirt}
u_\infty= \sqrt{g d_b} \left[ \frac{1}{(0.707)^2} +  \frac{12}{\Rey} \right]^{-1/2}
\end{equation}
as long as the Bond number is larger than 100. In the limit of small Reynolds number, the Stokes terminal velocity is recovered ($u_\infty=\rho g d_b^2/12 \mu$), while in the limit of large Reynolds number, the expression for the terminal velocity of a spherical cap is recovered (given by Eq. \ref{Eq_UB_cap}). This relation, Eq.\ref{Eq_UB_skirt}, corresponds to a drag coefficient $C_D= 8/3 + 16/Re$, where the two limits, spherical bubble and spherical cap bubble, are readily identified.

\subsubsection{Bubble rupture\label{Sec_bubble_rupture}} 

\begin{figure}[h]
\includegraphics[width=6.5in]{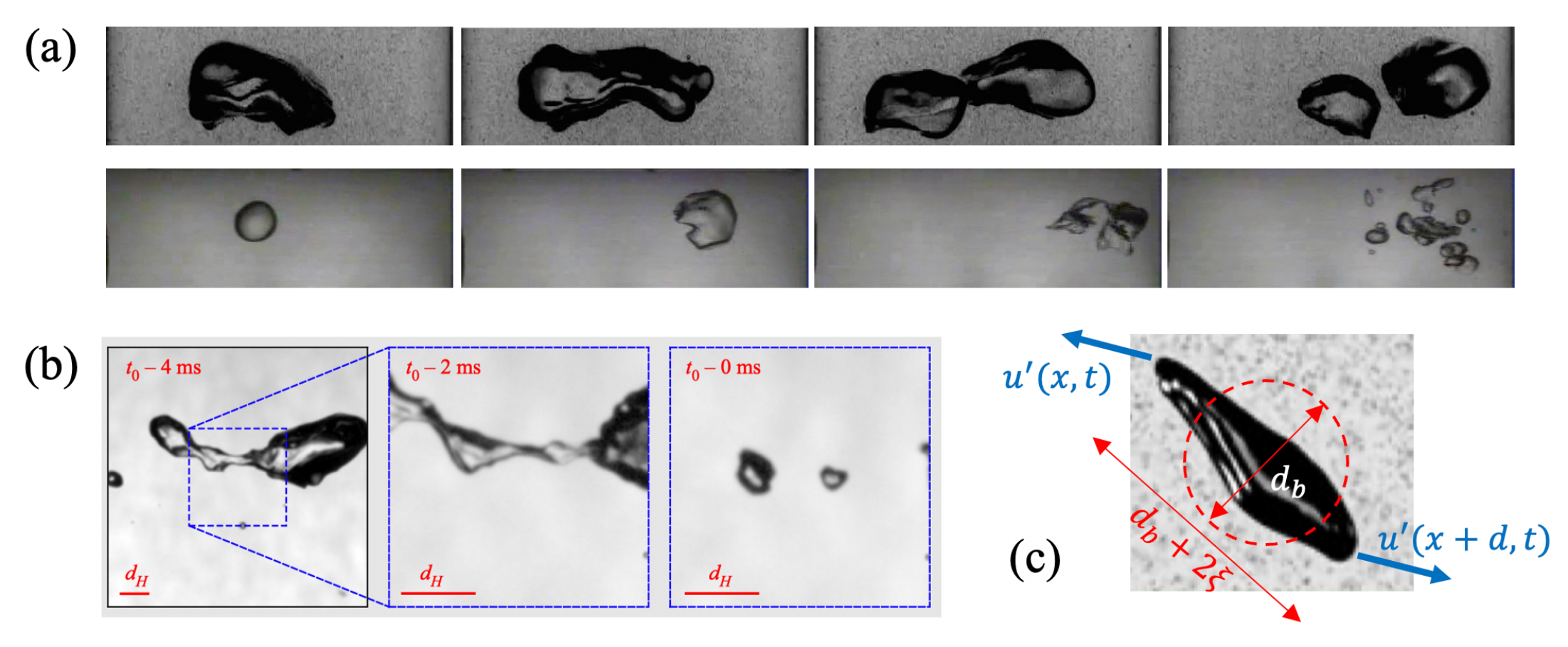}
\caption{Evolution of bubble shape up to rupture in a turbulent flow. (a) Shape evolution supporting the two approaches discussed in the text for bubble rupture scenario \cite{Risso1998}. Top row: the bubble is continuously deformed up to break up into two daughter bubbles; second row: the bubble is subject to turbulent fluctuations and finally breaks into multiple daughter bubbles. {(b) Formation of a gas ligament prior to rupture resulting in the formation of very small bubbles, {$d_H$ being the Hinze scale} (image from \citet{Ruth2022}). (c) Schematic of the scenario of bubble deformation in a turbulent field, adapted from \citet{Lalanne2019}, bubble image taken from \citet{Salibindla2020}.}}
\label{fig_rupture}
\end{figure}

When exposed to a strong shear or to an intense turbulent environment bubbles can {deform under conditions favorable for} break-up, as depicted in Fig. \ref{fig_rupture}. Different approaches have been proposed to describe and model bubble breakup and a significant body of literature has addressed this topic over the past decades \cite{Hinze1955, Risso1998, Ravelet2011, Vejrazka2018, Ruth2022}. {For detailed reviews on the subject, we refer the reader to \citet{Risso2000} and  \citet{Ni2024}}. 

{To understand the conditions for bubble rupture, we can first consider the non-dimensional numbers introduced before to describe bubble deformation, Eq. \ref{eq_def_bubble}, namely the Bond, the capillary and the Weber numbers. When a bubble is continuously submitted to an external forcing that increases its deformation, rupture may occur when it reaches a maximum deformation. A rupture criteria can then be proposed based on limit values for $\Bo$, $\Ca$ and $\We$. 
In particular, considering rising bubbles in stagnant liquids the critical Weber number $We_C$ is related to the critical Bond number $Bo_C$  by \citep{Risso2000}:
\begin{equation}\label{Eq_We_Bo_C}
\We_C=0.51 \, \Bo_C,\quad  \Bo_C=730 \, (1+\Mo^{1/4})^{1.66}
\end{equation}
where $\Mo$ is the Morton number. For an air bubble in water, $\Bo_C\approx 730$ and $\We_C\approx 370$ corresponding to a maximum bubble equivalent diameter $d_b\approx 7.3$cm ($\vartheta_b \approx $ 203 cm$^3$), shown in Fig. \ref{fig_1_new}(a) by the vertical blue dashed line. In other words, a bubble larger than this critical size will fragment as it rises. The stability of the upper surface of a large gas bubble rising steadily through liquid under gravity is analyzed by \citet{Batchelor1987}.
}

This list of relevant dimensionless parameters for deformation and then breakup can be extended for turbulent flows.  
{According to the Hinze-Kolmogorov theory \citep{Hinze1955}, the size of turbulent eddies that will induce large bubble deformation are those comparable to bubble size. The  bubble deformation is thus induced by the dynamic pressure difference between two points separated by a bubble diameter, which can be expressed as the difference of the velocity fluctuations $u^\prime$  as $\delta u^2(d_b,x,t)=\| u^\prime(x+d_b,t)- u^\prime(x,t)\|^2$, as depicted in Fig. \ref{fig_rupture}(c). From a Lagrangian point of view we can consider $\rho\overline{\delta u^2(d_b)}$ as the average turbulent pressure along the bubble motion. We can therefore define a turbulent Weber number $We_t$, as
\begin{equation}\label{eq_Wet}
\We_t= \frac{\rho \overline{\delta u^2(d_b)} d_b}{\sigma}
\end{equation}
Assuming that turbulence is isotropic at the bubble scale and that bubble diameter is within the inertial turbulent subrange, $\overline{\delta u^2(d_b)}$ can be expressed in terms of the dissipation rate $\epsilon$. In this case, $\overline{(\delta u)^2} \approx 2 \left( \epsilon d\right)^{2/3}$, allowing us to rewrite the turbulent Weber number as}
\begin{equation}\label{eq_Wet_hinze}
\We_t= \frac{2 \rho \epsilon^{2/3} d_b^{5/3}}{\sigma}
\end{equation}
{The bubble is thus expected to break if $\We_t$ exceeds a critical Weber number $\We_{tC}$ corresponding to a bubble size larger than the Hinze scale:
\begin{equation}\label{eq_dH}
d_H= \We_{tC}^{3/5} \, \left( \frac{ \sigma}{2 \rho \epsilon^{2/3}} \right)^{3/5}.
\end{equation}
Experimental and numerical investigations indicate that $\We_{tC} = O(1)$ \cite{Vejrazka2018, Ruth2022}.
For bubbles with initial sizes much larger than $d_H$, the size distribution of bubbles smaller than $d_H$ after breakup follows $N(d_b) \propto d_b^{-3/2}$ with the formation of very tiny bubbles, as shown in  Fig. \ref{fig_rupture}(b), resulting from the  capillary instability of ligaments formed when the bubble elongates \cite{Ruth2022}}. 

{Additionally, to complete our assessment of rupture, the bubble can be considered  an oscillator. This concept originated by considering a resonance mechanism between the bubble deformation and the turbulent fluctuations \cite{Sevik1973}. Balancing the second mode of oscillation $\omega_2 =2 \pi f_2=\sqrt{96 \sigma /\rho d_b^3}$ with the turbulence frequency at the bubble scale $\approx 2 \sqrt{\overline{\delta u^2(d_b)}}/d$,
the turbulent Weber number $We_t$ defined in Eq. \ref{eq_Wet} is recovered. The criterion for breakup is $We_t\approx 6/\pi^2= O(1)$, which is consistent with  experimental observations.
This concept can be extended by considering a linear oscillator}
(see also section \ref{section_bouncing} where such a dynamical system is considered for modelling bubble rebound) where the mass of the oscillator is the added mass of the liquid, the stiffness is the surface tension, the dissipation arises from the liquid viscosity {and the external forcing is imposed by the instantaneous local turbulent fluctuations \cite{Lalanne2019, Ni2024}.}  {A bubble oscillator equation can be then written to describe the amplitude of bubble oscillation $\xi(t)$, as depicted in Fig. \ref{fig_rupture}(c), as
\begin{equation}\label{eq_osc_turb}
 \frac{d^2 \xi}{dt^2} + 2 D_2 \frac{d \xi}{dt} + \omega_2^2 \xi = K \frac{\delta u^2(d_b,t)}{d}
\end{equation}
where $\omega_2$ is close to the eigenfrequency $\sqrt{96 \sigma/\rho d^3}$ and $D_2$ is related to the damping rate $80\nu/d_b^2$ of the 
mode 2 of deformation of the bubble \cite{Risso1998, Riviere2024b}.
The oscillation period and the characteristic time of damping are then $2\pi / \omega_2$  and $1/D_2$, respectively. 
The equation for the normalized deformation $\xi^*=\xi/d$ with the normalized time $t^*=t\omega_2$ indicates that the turbulent forcing is controlled by the turbulent Weber number (Eq. \ref{eq_Wet}):
\begin{equation}\label{eq_osc_turb1}
 \frac{d^2 \xi^*}{{dt^*}^2} + 2 \frac{D_2}{\omega_2}  \frac{d \xi^*}{dt^*} +  \xi^* = K \frac{\We_t}{\We_2}
\end{equation}
where $\We_2=\rho \omega_2^2 d_b^3/\sigma$ is a characteristic Weber number of the free bubble mode 2 oscillation. A critical deformation $\xi^*_C=2.5 K /We_2$ is used for the breakup criteria \cite{Lalanne2019}
 while \citet{Riviere2024} consider $\xi^*_C= 0.37$.}

The threshold at which breakup occurs depends on both the Reynolds number and the initial bubble shape. Since, in real configurations, bubble dynamics are rarely quasistatic, these results have important practical consequences: history matters. The critical Weber number at which bubbles break should always be considered together with a set of initial conditions or at least understood in a statistical sense.

\subsection{Bubble surface contamination} \label{section_contamination}

\begin{figure}[h]
\includegraphics[width=6.5in]{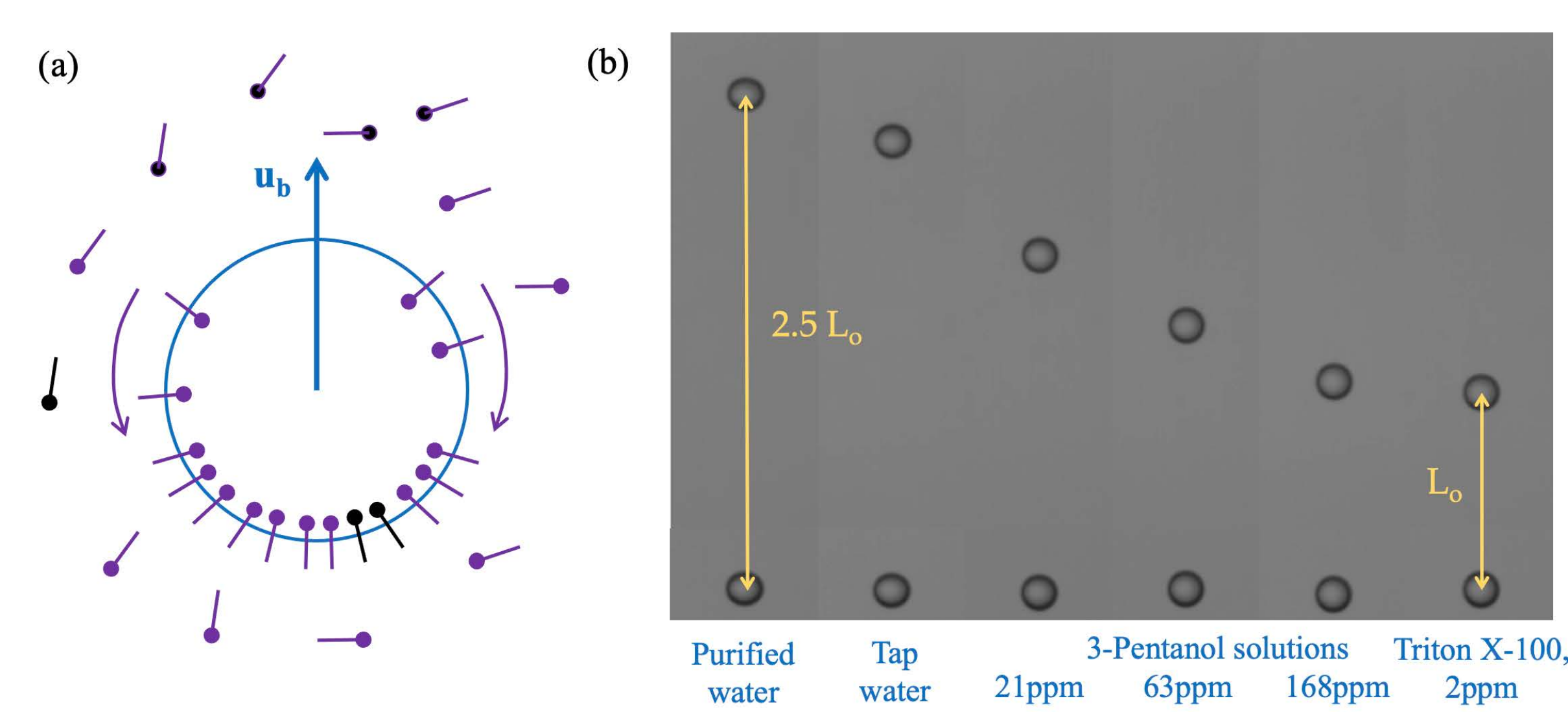}
\caption{(a) Schematic illustration of bubble interface contaminated by surfactants that accumulate in the rear part of the bubble. (b) An air bubble, with $d_b=0.9$ mm, rising in water with different surfactants. The images are shown at the same time after the bubble appears in the observation window. From left to right: ultra-purified water, tap water, 3 solutions of Pentanol, Triton X-100. {Images adapted from} \citet{takagi2009}.}
\label{fig_cont1}
\end{figure}

The effect of surfactants on the rise of a spherical bubble is clearly illustrated by the experiments of a rising bubble in water conducted with different types and amounts of surfactant \cite{takagi2008} shown in Fig. \ref{fig_cont1}(b) for ultra-purified water, tap water, water with three concentrations of Pentanol (21, 63, and 168 ppm) and with 2 ppm of Triton X-100. The bubbles are shown at the same time after entering the window of observation. As shown in the figure, the distance traveled during a given time varies significantly depending on the type of surfactant and also on their concentration. The strongest effect is observed for Triton X-100, which has a long molecule chain. The distance traveled is 2.5 times less than the distance traveled by the bubble in a purified system. The corresponding velocity is thus 2.5 smaller revealing a larger drag force for a bubble in a highly contaminated situation.

{
Considering a liquid with surfactant, as the bubble rises these molecules  accumulate  at the bubble surface. The evolution of the surfactant concentration at the bubble interface, $\Gamma_I$, is modeled by the surface advection-diffusion equation \cite{levich1962, stone1990,cuenot1997}:
\begin{equation}\label{Eq_cont1}
\frac{\partial \Gamma_I}{\partial t} =-  \nabla_I . \left( u_I \Gamma_I \right) + D_\Gamma \nabla_I^2 \Gamma_I+ S_\Gamma
\end{equation}
where $\nabla_I$ is the surface gradient operator, $D_\Gamma$ is the surface diffusion coefficient, $u_I$ is the surface velocity and $S_\Gamma=k_a C_I (\Gamma_\infty - \Gamma_I) - k_d \Gamma_I$ is the flux of surfactants from the liquid phase to the interface with  $C_I$ the surfactant concentration in the liquid in contact with the interface, $\Gamma_\infty$  the saturation value of $\Gamma$, $k_a$ and $k_d$  the coefficients of the adsorption and desorption kinetics, respectively.}
The two first terms on the right-hand side {of} Eq. \ref{Eq_cont1}, namely the interface advection and diffusion, control the surfactant distribution at the interface. The relative importance of convective to diffusive transport of surfactants is characterized by  an interfacial P\'eclet number $Pe_I= d_b u_b/D_\Gamma$. Usually, $Pe_I \gg 1$ so that surfactants accumulate at the rear of the bubble, transported by the interfacial velocity  to form a cap, as illustrated in Fig. \ref{fig_cont1}(a).

{In fact, four typical situations can be expected depending on the  values of the different parameters governing the problem \cite{cuenot1997}: (i) the unretarded velocity profile (no effect of surfactant on the  slip velocity at the bubble surface), (ii) the uniformly retarded profile (the slip velocity at the bubble surface is uniformly reduced), (iii) the stagnant-cap model (the front part of the bubble is not impacted  by surfactant while the rear part of the bubble is immobilized because of surfactant accumulation), and (iv) the completely stagnant model (the interface is entirely saturated in surfactant resulting in non-slip condition).}

{The stagnant cap case occurs for cases involving a large interfacial P\'eclet number $\text{Pe}_I$ and small values of  $\alpha_\Gamma = k_a  C_\infty d/U_b$ and $\alpha_\Gamma/\text{La}=k_d   d/U_b$ where {$\text{La}= k_a  C_\infty/k_d$ is the Langmuir number and $C_\infty$ is the bulk concentration of surfactant}. Under these conditions and at equilibrium, 
Eq. \ref{Eq_cont1} simplifies to  $\nabla_I \cdot  \left( u_I \Gamma_I \right) \approx 0$. Combined with a zero flux  boundary condition at the stagnation points, we have that $u_I \Gamma_I \approx 0$. 
On the "clean part" of the surface this condition is satisfied because $ \Gamma_I \approx 0$ thus the external fluid can slip on the surface; conversely, on the contaminated part, the condition $u_I \Gamma_I \approx 0$ can only be satisfied if $u_I=0$, which results in a no-slip condition.}

The motion of the liquid at the bubble interface {is also related  to} the viscous shear condition given by Eq.\ref{Eq_moment_tagent} where the effect of the gas has been neglected
\begin{equation}\label{Eq_cont3}
\mathbf{n_b} \cdot \mathbf{\Sigma} \cdot  (\mathbf{I} -  \mathbf{n_b}\mathbf{n_b}) = \nabla_I \sigma
\end{equation}
where  the surface tension $\sigma$  depends on the local surfactant concentration $\Gamma_I$. {$\sigma(\Gamma_I)$} is usually determined by considering the Langmuir adsorption isotherm \cite{levich1962}
\begin{equation}\label{Eq_cont4}
\sigma(\Gamma_I) = \textrm{max}\left[ \sigma_\infty, \sigma_0 \left( 1+ \frac{\Re T \Gamma_\infty}{\sigma_0} \ln \left( 1- \frac{\Gamma_I}{\Gamma_\infty} \right) \right) \right]
\end{equation}
where $\Re$ is the ideal gas constant, $T$ is the absolute temperature, and $\sigma_\infty$ is the surface tension value that corresponds to the maximum concentration of surfactants {$\Gamma_\infty$} at the interface \cite{Olgac2003}.
The dynamic surface tension measurement of various air-liquid interfaces with nonionic surfactants (Triton X-100 or $C_{12}E_{6}$) has values ranging from $\sigma_\infty/\sigma_0 \approx 0.2$ to $0.6$, for different surfactant concentrations \cite{Giribabu2007, Li2019}.

From Eqs. \ref{eq_wI}, \ref{Eq_cont3} and \ref{Eq_cont4}, we can show that the bubble surface vorticity has an additional source when the interface is contaminated \cite{Atasi2023}
\begin{equation}\label{Eq_cont7}
\omega_I    =  2 \frac{u_I}{R_c} + \frac{\mathcal{R} T \Gamma_\infty}{\mu \left(\Gamma_\infty- \Gamma \right)} \nabla_I \Gamma_I 
\end{equation}
where $R_c$ is the local bubble {radius of} curvature. The presence of a concentration gradient $\nabla_I \Gamma_I $ at the interface increases the interface vorticity. As discussed in Section \ref{section_dragvorticity}, increasing vorticity at the interface  increases the drag force and thus the terminal velocity is reduced.

This reduction in bubble speed is {observed} in Fig. \ref{fig_1_new}, {where the experimental terminal velocity of a spherical  bubble} is smaller than the clean bubble prediction given by Eq.  \ref{eq_u_terminal}. When the concentration of surfactant in the liquid is sufficiently high, the bubble surface mobility is reduced, resulting in a slower rising velocity shown in Fig. \ref{fig_cont1}. The lower curve (red dashed line) in Fig. \ref{fig_1_new} was obtained by solving 
\begin{equation}\label{Eq_cont2}
6 \pi \mu R  \left(1+0.15 \Rey^{0.687} \right) \mathbf{u_b} = \frac{4 \pi R^3}{3} \rho \mathbf{g}, 
\end{equation}
where the term on the left side is the drag force of a solid sphere \cite{schiller1933}. This reveals that the lower part of the data reported in Fig. \ref{fig_1_new} shows a  highly contaminated bubble whose interface is immobilized by the presence of surfactants. In such a situation, the bubble behaves like a sphere with a no-slip surface {and the function $\mathcal{K}(Re)$ to consider in the drag force expression (\ref{eq:drag_mei}) is then 
\begin{equation}\label{eq:drag_contaminated_K}
\mathcal{K}(Re)= \frac{3}{2} \left(1+0.15 \Rey^{0.687}\right) 
\end{equation}
instead of relation (\ref{eq:drag_mei_K}) used for a clean bubble}.

{Note that the presence of surfactant on the bubble surface also affects the bubble path instability discussed in Section \ref{sec_zigzag} \cite{Zhang2001, Tagawa2014, Pesci2018}.}

\section{Bubble interaction with non-uniform and unsteady flows: added mass and lift effect}

\begin{figure}[h]
\includegraphics[width=6.5in]{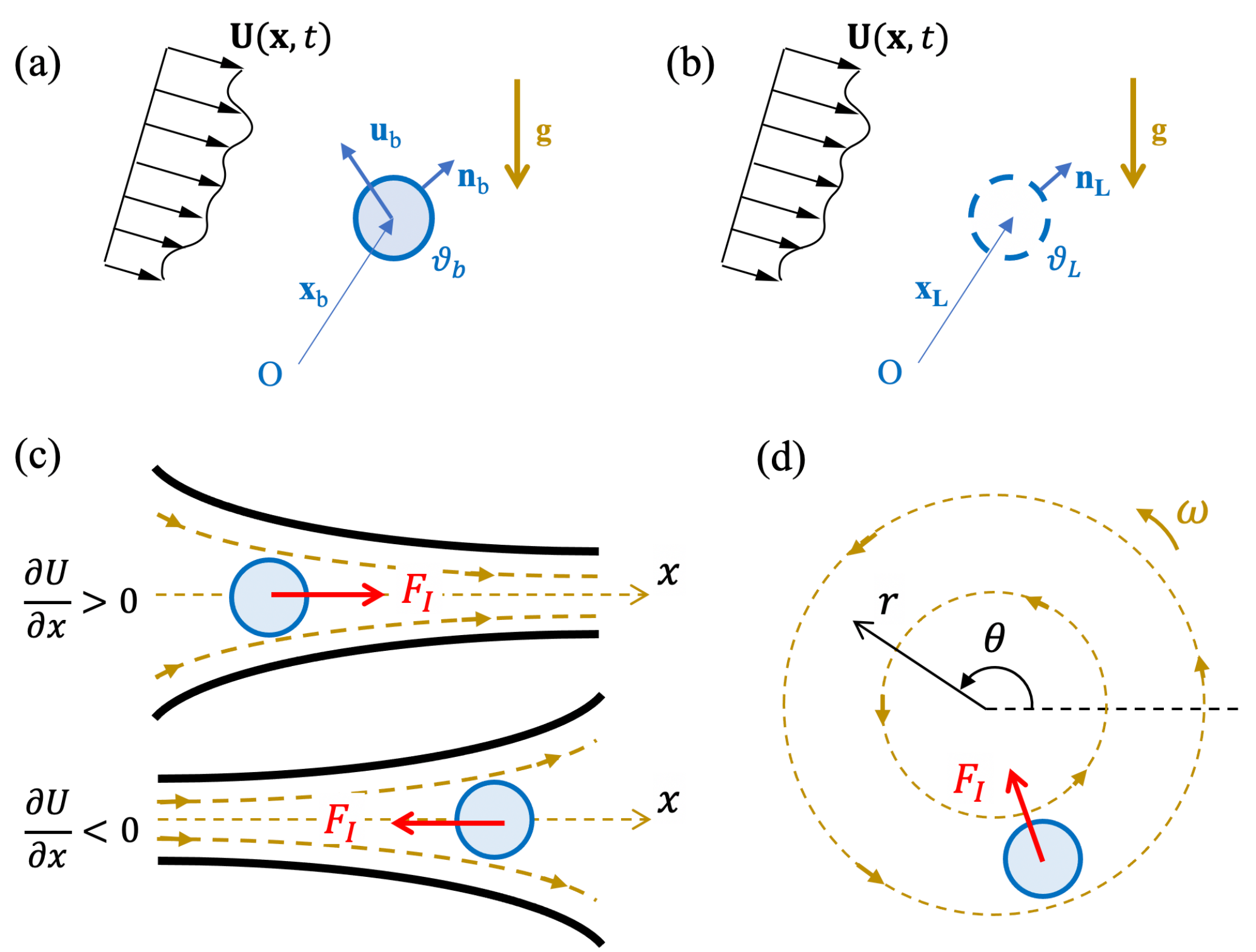}
\caption{(a) Bubble {of volume $\vartheta_b$ moving} in an unsteady non-uniform flow denoted $\mathbf{U}(\mathbf{x},t)$. {(b) Definition of a volume of fluid having the same volume as the bubble $\vartheta_L=\vartheta_b$ located at the position $\mathbf{x_L}=\mathbf{x_b}$ in the same flow field. (c) Inertial force $\mathbf{F_I}$ applied to a bubble in a convergent channel $\frac{\partial U}{\partial x}>0$ and in a divergent channel $\frac{\partial U}{\partial x}<0$ flow along the $x$-direction. (d) Inertial force $\mathbf{F_I}$ applied to a bubble in a solid rotation flow $\mathbf{U}=\omega r \mathbf{e_{\theta}}$.}}
\label{fig_nonunif_1}
\end{figure}

We now consider a gas bubble moving in a more general situation consisting of an unsteady non-uniform flow noted $\mathbf{U}(\mathbf{x},t)$, instead of a stagnant fluid,  as shown schematically in Fig. \ref{fig_nonunif_1}. 

\subsection{The generalized Archimedes force {and added mass effects} \label{sec_added_mass}}

We first consider a control volume $\vartheta_L$ in the liquid  with surface $S_L$ {and normal vector $\mathbf{n_L}$ directed out of $\vartheta_L$}, see  Fig. \ref{fig_nonunif_1}(b), to calculate the force $\mathbf{F}_{L\,\text{to}\,L}$ {applied by the external liquid on this volume of fluid $\vartheta_L$} corresponding to the last term in the trajectory equation (\ref{eq_motion}):
\begin{equation}\label{eq_nonunif_1}
\mathbf{F}_{L\,\text{to}\,L}=\int_{S_L}{ \mathbf{T}\cdot\mathbf{n_L} dS} =\int_{\vartheta_L}{ \text{div} \left( \mathbf{T} \right) d\vartheta}
\end{equation}
Since velocity and pressure fields of the fluid are defined inside {$\vartheta_L$}, the integral in Eq. \ref{eq_nonunif_1} can be transformed into a volume integral. The velocity field $\mathbf{U}(\mathbf{x},t) $  satisfies the Navier-Stokes equations such that $\text{div} \left(\mathbf{T} \right)=\rho \left(- \mathbf{g} + {\partial \mathbf{U}}/{\partial t} + \mathbf{U} \cdot  \nabla \mathbf{U}  \right)$, so the force can be written as:
\begin{equation}\label{eq_nonunif_2}
\mathbf{F}_{L\,\text{to}\,L} =\int_{\vartheta_L}{ \rho \left(- \mathbf{g} + \frac{\partial \mathbf{U}}{\partial t} + \mathbf{U} \cdot  \nabla \mathbf{U}  \right) d\vartheta}
\end{equation}
Using a series Taylor expansion of the first two terms with respect to the position of center of the control volume $\mathbf{x_L}$ and by retaining only the first term leads to:
\begin{equation}\label{eq_nonunif_3}
\mathbf{F}_{L\,\text{to}\,L} = - \rho \vartheta_L  \mathbf{g} + \rho  \vartheta_L \left(\frac{\partial \mathbf{U}}{\partial t} + \mathbf{U}\cdot \nabla \mathbf{U} \right)_{\mathbf{x_L}} 
\end{equation}
The first term in Eq. \ref{eq_nonunif_3} is the Archimedes force which results from the hydrostatic pressure in the fluid. The two other terms in the equation give an extended interpretation of this force for the case when the flow  is unsteady and non-uniform, as discussed by {\citet{batchelor1967}}.
The second term results from the time acceleration of the fluid, while the last one is induced by the convective fluid acceleration. 
Typically, this contribution can be observed in a convergent (resp. divergent) flow  induced, for example, by the reduction (resp. increase) of the container cross-section, {see  Fig. \ref{fig_nonunif_1}(c)}. For example, considering that the velocity component $U_x$  increases (resp. decreases)  in the flow direction denoted $e_x$, a term of the form $U_x \partial U_x/\partial x$ expresses the  acceleration (resp. deceleration) of the fluid particles. 
In summary, three types of acceleration can act on the {liquid} volume {$\vartheta_L$: gravity, the temporal,  and  convective accelerations of the flow}.

Consider now an object, such as a bubble of constant volume $\vartheta_b{=\vartheta_L}$, in the same location ($\mathbf{x_b}{=\mathbf{x_L}}$) as ${\vartheta_L}$ {and moving at velocity $\mathbf{u_b}$} (see Fig. \ref{fig_nonunif_1}a). This object  experiences the same {forces}, but a correction is needed because the object surface is now impermeable, which modifies the {fluid displacement} around the bubble. Remarkably, according to {\citet{auton1988} (see also \citet{magnaudet2000} for a detailed discussion)}, the  correction is simply the multiplication of the last two  terms in Eq. \ref{eq_nonunif_3} by $(1+C_M)$ where $C_M$ is the added mass coefficient, which is $C_M=1/2$ for a spherical shape. When the bubble deforms or interacts with other bubbles and/or a wall, $C_M$ increases \cite{wijngaarden1976}. {Subtracting the buoyancy force, $\rho \mathbf{g} \vartheta_b$, the resulting force then writes as}
\begin{equation}\label{eq_nonunif_4}
\mathbf{F_{I}} =\rho  \vartheta_b (1+C_M)  \left(\frac{\partial \mathbf{U}}{\partial t}  + \mathbf{U} \cdot \nabla \mathbf{U} \right)_{\mathbf{x_b}} -C_M \rho \vartheta_b \frac{d  \mathbf{u_b}}{dt}
\end{equation}
 As discussed above, the first term of Eq. \ref{eq_nonunif_4} represents the effect of the time acceleration of the surrounding liquid motion on the bubble; the second term represents spatial changes of the velocity field, which can be relevant to convergent and divergent flows, as well as in rotating flows.
{Considering first the convergent (resp. divergent) flow  shown in  Fig. \ref{fig_nonunif_1}(c)}, the velocity component $U_x$  increases (resp. decreases)  in the flow direction denoted $\mathbf{e_x}$ and a term of the form $U_x \partial U_x/\partial x$ expresses the  acceleration (resp. deceleration) of the fluid particles.
 {It results in an inertial force transmitted to the bubble along the $\mathbf{e_x}$ direction of the form $\mathbf{F_{I}} \cdot \mathbf{e_x} =\rho  \vartheta_b (1+C_M) U_x \partial U_x/\partial x$.
 Another example where the spatial contribution $\mathbf{U} \cdot \nabla \mathbf{U}$  in $\mathbf{F_{I}}$ is of great importance is the vortex bubble capture mechanism. Consider the image in Fig. \ref{fig_nonunif_1}(d) where the solid body rotation 
 $\mathbf{U}=\omega r \mathbf{e_{\theta}}$ is expressed in cylindrical coordinates $(r, \theta, z)$. 
 The contribution $\mathbf{U} \cdot \nabla \mathbf{U} = - r \omega^2 \mathbf{e_r}$ induces a centripetal force $\mathbf{F_{I}} =- r \rho  \vartheta_b (1+C_M) \omega^2 \mathbf{e_r}$ resulting in bubble trapping by vortex structures \cite{vanNierop2007, rastello2011}.} 
 
{When the flow cannot be considered uniform at the bubble scale, Eq. \ref{eq_nonunif_3} needs to be corrected by the Fax\'en correction 
 that accounts for the curvature of the surrounding flow $(\nabla^2 U)_{\mathbf{x_b}}$  at the bubble location \cite{gatignol1983, Homann2010}.}

{The additional term involving the bubble acceleration in Eq.\ref{eq_nonunif_4} (last term on the right hand side) has an  important  impact on  the bubble motion.} To clarify its contribution, let us come back to the bubble trajectory given by Eq. \ref{eq_motion}, where the term $\int_{S_b}{ \mathbf{T}\cdot\mathbf{n_b} dS}$ has been split into two parts in order to combine the last term of Eq.   \ref{eq_nonunif_4}  with the bubble acceleration 
\begin{equation}\label{eq_nonunif_6}
\left(\rho_b  + C_M  \rho \right) \vartheta_b \frac{d \mathbf{u_b}}{dt} = \rho_b \vartheta_b \mathbf{g} + \sum{\textrm{Other forces from the fluid} }.
\end{equation}
This expression shows that the bubble acceleration involves its own mass but also has an additional mass corresponding to the liquid volume ($C_M \vartheta_b $). The corresponding contribution is called the added mass force, associated with the added mass coefficient $C_M$. For bubbles, this term is much larger than the  bubble's own mass ($\rho_b  \ll C_M  \rho$). It is, therefore, {crucial} to be considered when describing the bubble dynamics. Similarly, the bubble weight in Eq. \ref{eq_nonunif_6} is very small when it is compared to the Archimedes force since $\rho_b  \ll   \rho$.

\subsection{Shear induced lift force}\label{section_lift}

\begin{figure}[h!]
\includegraphics[width=6in]{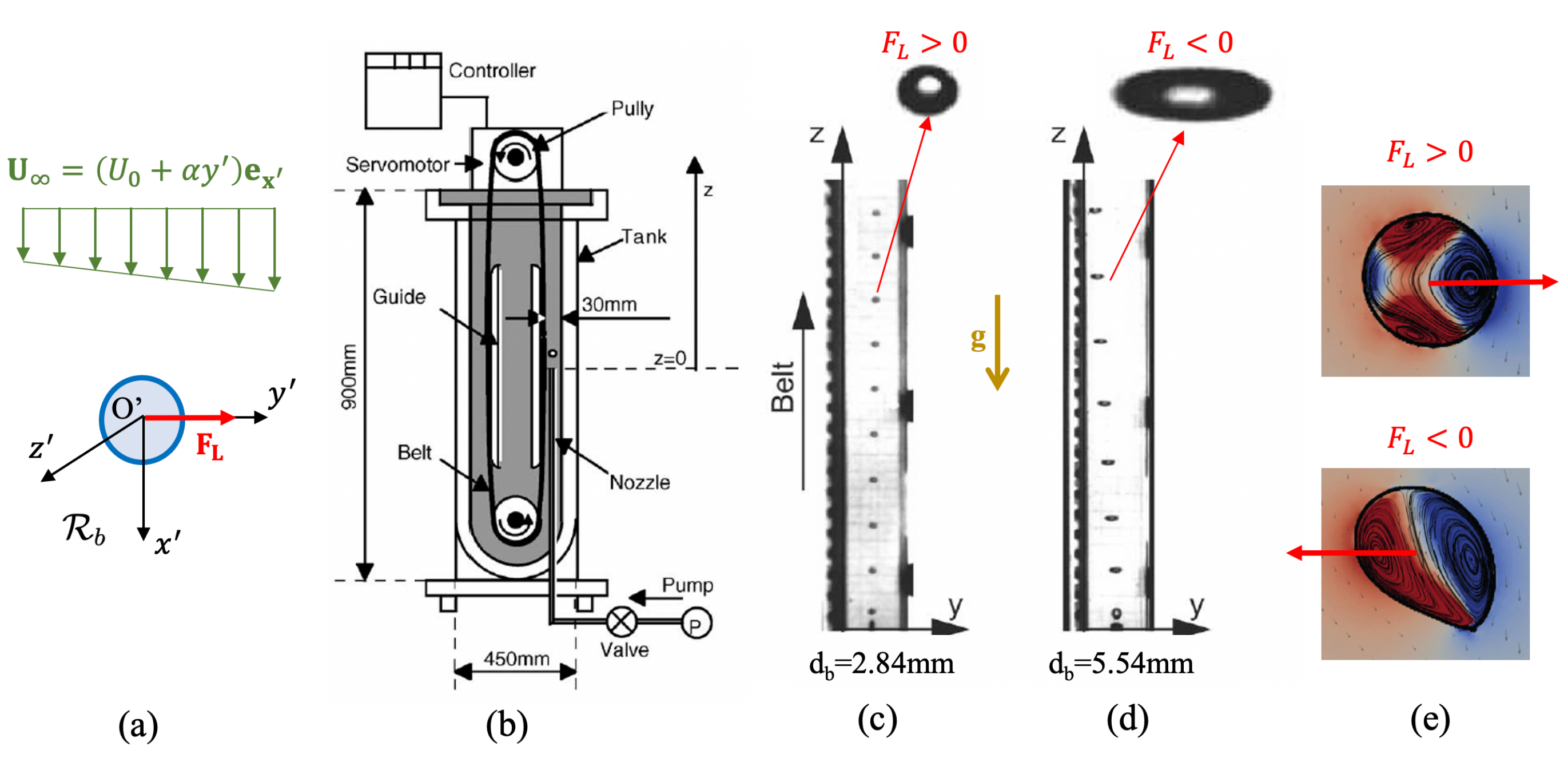}
\includegraphics[width=6in]{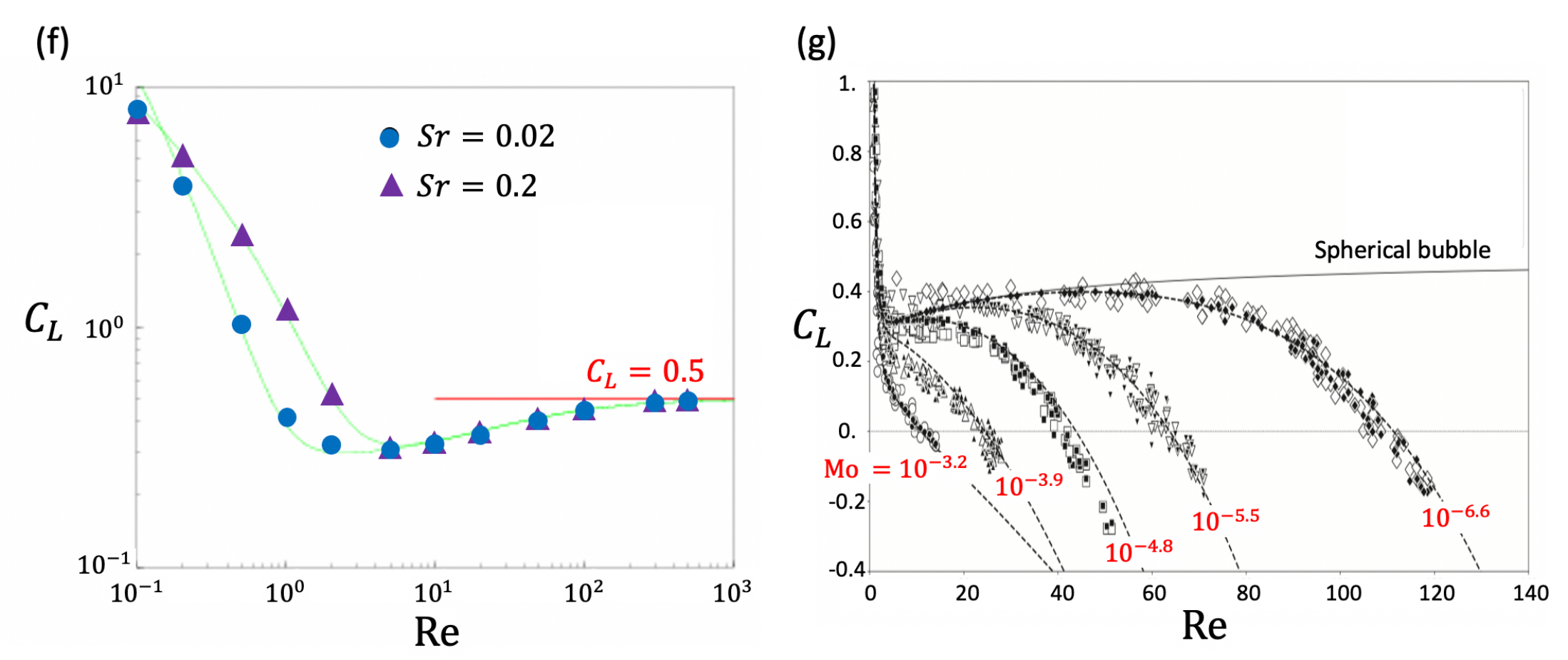}
\caption{Rising bubble in a vertical linear shear flow showing horizontal migration. (a) Problem statement in the frame of reference  {$\mathcal{R}_b$} moving with the bubble, corresponding to the experimental setup shown in (b). (b) Experimental apparatus using a belt to generate the {vertical} linear shear flow in a Glycerol–water solution with $\Mo = 10^{-5.3}$ \cite{Tomiyama_2002}; (c) bubble experiencing a positive lift force ($d_b=2.84$ mm); (d) bubble experiencing a negative lift force ($d_b=5.54$ mm). (e) Numerical simulation at low Reynolds number ($\Ga=18$) for a small deformation (top $\Bo=0.4$) and significant deformation (bottom $\Bo=20$). (f) Evolution of the lift coefficient of a fixed spherical bubble in a linear shear flow obtained from numerical simulation \cite{legendre1998}. (g) Lift coefficient of a deformable bubble rising in a linear shear flow in a setup similar to that shown in (b) \cite{hayashi_lift_2020}.
}
\label{fig_lift_3}
\end{figure} 

Here, the ``lift effect" refers to the force experienced by a bubble in a direction perpendicular to  the relative motion of the fluid with respect to the moving bubble $\mathbf{U} - \mathbf{u_b}$.
Lift effects can be observed resulting from different causes. For airplanes, the lift force that supports the airplane weight  is generated by the flow circulation induced by the wing shape. The spinning lift, or Magnus effect, is induced by the  rotation of the object, which is very relevant in sports. Bubbles can also experience lift effects when rising in a quiescent fluid due to their wake structure when their shape is ellipsoidal, as shown in Fig. \ref{fig_deformation}. {Bubbles can also experience a lift effect, even when spherical,} if they traverse a non-uniform flow. In such a situation, the lift force results from a non-axisymmetric distribution of both the pressure and normal viscous stress on the bubble interface. This force is responsible for the migration of bubbles in vertical pipes and determines the radial distribution of the gas volume fraction \cite{colin2012}. As it will be outlined here, for bubbles this lift effect depends on the  Reynolds number and the bubble shape. {For the discussion,} we only consider  unbounded linear shear flows. Details for wall-bounded linear shear flow effects can be found in \citet{Shi2020} and the references therein.

The lift force has been mostly characterized by considering a steady linear shear flow, as shown in Fig.\ref{fig_lift_3}(a).
The experiments from \citet{Tomiyama_2002} provide a clear idea of the lift force effect. Bubbles of different sizes were injected in a linear shear flow  generated by a belt moving at a constant speed as illustrated in  Fig. \ref{fig_lift_3}(b). The   flow velocity field is $\mathbf{U}=(U_o + \alpha y^\prime) \mathbf{e_{x^\prime}} $ in the frame of reference  moving with the bubble. Two bubble sizes are shown here, $d_b$=2.84 mm and $d_b$=5.54 mm, corresponding to a nearly spherical {and an ellipsoidal bubble, respectively}.  As shown in Fig.\ref{fig_lift_3}(c), both bubbles   deviate in the horizontal direction when rising, clearly revealing a lift effect. However, the two bubbles migrate in opposite directions. 

The induced lift force is usually expressed in the form \cite{Zun1980, auton1987}
\begin{equation}\label{eq_lift1}
\mathbf{F_L}= C_L \rho \vartheta_b  (\mathbf{U} - \mathbf{u_b}) \times {\mathbf{\Omega}}
 \end{equation}
where $C_L$ is the lift coefficient {and $\mathbf\Omega=\nabla\times\mathbf{U}$ is the vorticity field at the bubble location}. Based on this equation, $C_L$ is positive (resp. negative) for {a spherical} (resp. ellipsoidal) bubble shown in Fig. \ref{fig_lift_3}(c) {(resp. Fig. \ref{fig_lift_3}(d))}.  
The lift coefficient for a spherical {and clean} bubble is reported as a function of the bubble Reynolds number for different  shear rates $Sr = \alpha d_b /  U_o $ in Fig. \ref{fig_lift_3}(f).
The lift coefficient is positive for all Reynolds numbers and shear rates, so spherical bubbles are expected to always migrate following the case reported in Fig. \ref{fig_lift_3}(c). A constant value
$$C_L=0.5$$ 
was derived by  \citet{auton1987} considering a weak inviscid linear shear flow, applicable for $Re \gg 1$ and $Sr\ll 1$. The trend {shown in Fig. \ref{fig_lift_3}(f)} for $Re > 100$, where {$C_L \approx 0.5$} confirms that the term $\rho \vartheta_b  (\mathbf{U} - \mathbf{u_b}) \times {\mathbf\Omega}$ is relevant to describe lift effects at large Reynolds numbers. For the case depicted in Fig. \ref{fig_lift_3}(a),  $\mathbf\Omega=-\alpha \mathbf{e_{z^\prime}}$. Under such conditions, the  lift force is $\mathbf{F_L}= \frac{\pi}{12} \rho d_b^3 U_o \alpha \mathbf{e_{y^\prime}}$.

For $Re \ll 1$, the lift force experienced by a bubble  was calculated by  \citet{legendre1997} extending the  works of \citet{saffman1965} and then \citet{mclaughlin1991} both for a solid sphere. The lift force clearly follows a different behavior from the large $\Rey$ case showing
$\mathbf{F_L}= \frac{1}{\pi} \sqrt{\rho \mu \alpha} d^2 U_o J(\epsilon) \, \mathbf{e_{y^\prime}}$
corresponding to a lift coefficient 
\begin{equation}\label{eq_CL_saffman}
    C_L= \frac{6}{\pi^2}\sqrt{Re \, Sr} J(\epsilon)
\end{equation}
where $\epsilon = \sqrt{Sr/Re}$, and  $J(\epsilon)$ is a function numerically calculated by \citet{mclaughlin1991} that can  accurately be approximated by  $J(\epsilon) = \frac{2.255}{(1 + 0.2/\epsilon^2)^{3/2}}$ \cite{legendre1998}. This expression is valid for $Re \ll 1$ and  $\sqrt{Re\, Sr} \ll 1 $.

{Expressions for the lift coefficient to be used in the description of bubble motion in non-uniform flow can be found in \citet{legendre1998} for spherical clean bubbles and in \citet{hayashi_lift_2020} and \citet{hayashi_lift_2021} for deformed bubbles.}
{Numerical simulations considering a stagnant cap model for the contamination of spherical bubbles report a monotonous decrease of the bubble lift coefficient with the increase of  the contamination  \cite{takagi2008, takagi2011}.
 {Experiments of ellipsoidal bubbles rising in water in a shear flow (generated at the side of a bubble column) with different amounts of 1-Pentanol and Triton X-100 solutions confirm a strong dependency of the lift coefficient with surface contamination \cite{Hessenkemper2021}. However, the experiments also reveal that the lift coefficient can be increased in some particular conditions of surfactant concentration and bubble deformation. In the same set-up but with saline aqueous solutions of NaCl, the lift coefficient dependency on the bubble Reynolds number is found to only be weakly affected by the salt concentration \cite{Hessenkemper2020}.  This indicates that the combined effect of surfactants, electrolytes and deformation requires further careful analysis for the particular case of the lift force.}}

{The induced vorticity mechanisms play a key role in the production of lift; they are discussed below considering the vorticity field, $\mathbf{\omega}=\nabla\times\mathbf{u}$}.

\begin{figure}[h]
\includegraphics[width=6.5in]{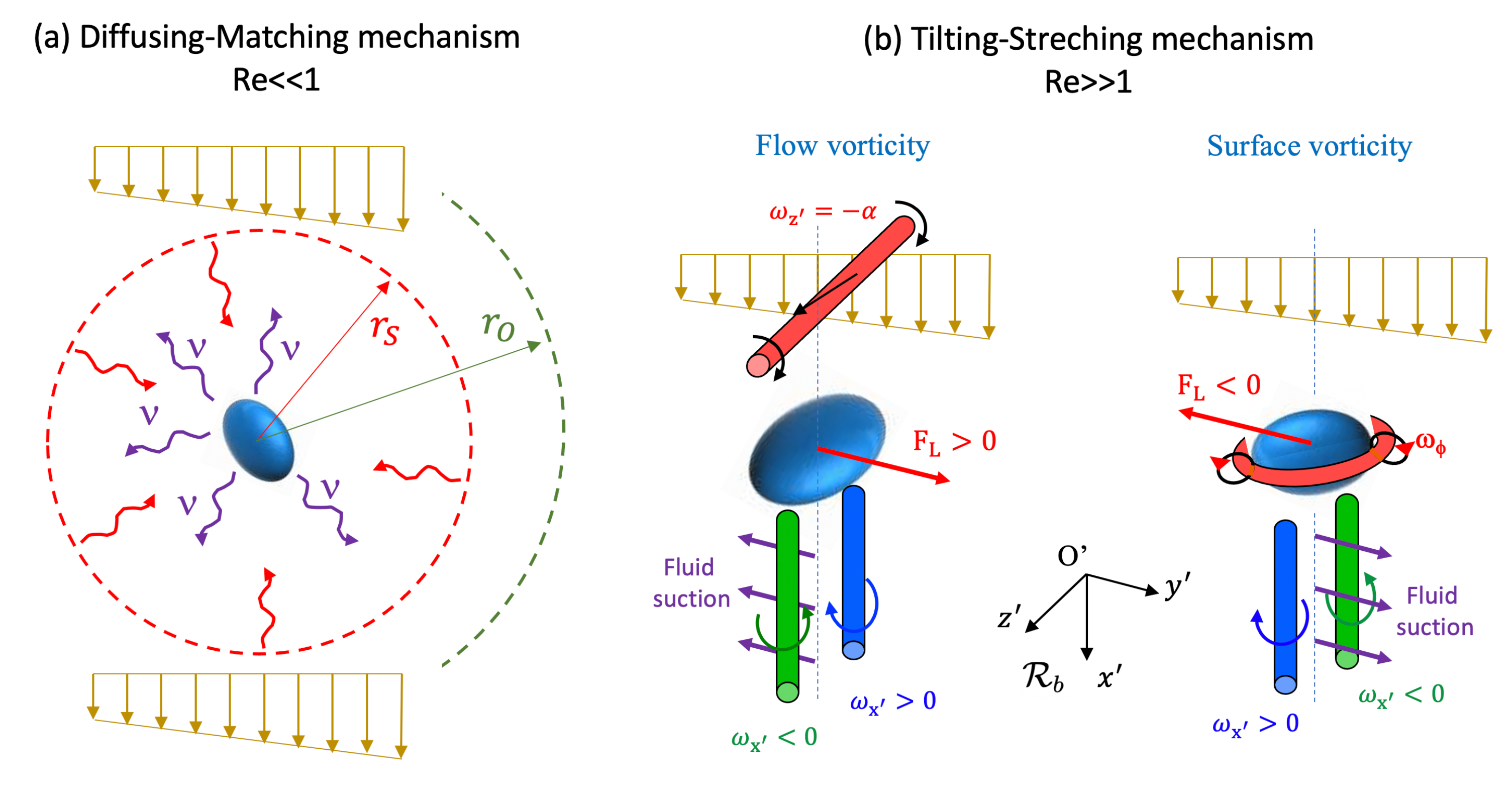}
\caption{Illustration of lift force mechanisms resulting in positive or negative lift effect depending on the interaction of the interfacial vorticity with the shear flow. (a) Diffusing-Matching Mechanism (or Saffman Mechanism) {showing the Oseen radius $r_O=O(d_b/Re)$ and the Saffman radius $r_S=d_b/\sqrt{Re Sr}$ for $\epsilon = \sqrt{Sr/Re} >1$}. (b) The Tilting-Stretching mechanism  \cite{adoua2009} {applied  to the vorticity field  of the flow {$\omega_{z^\prime} = - \alpha$}  (Lightill L-mechanism)  resulting in a positive lift contribution (left), and  applied to the vorticity {$\omega_{\phi}$} generated at the bubble surface (Surface S-mechanism)  {resulting in a negative lift contribution} (right).}}
\label{fig_lift_2}
\end{figure}

\subsubsection{Lift effect at large Reynolds: The tilting-stretching mechanism}
The process of tilting and stretching of the {incident} vorticity  around a sphere has been described by \citet{lighthill1956}, and  \citet{auton1987} calculated the corresponding lift force in an inviscid weak linear shear flow. These calculations have been  confirmed by numerical simulation \cite{legendre1998, adoua2009, Hidman2022}. {For the case depicted in Fig. \ref{fig_lift_3}(a), the incident flow vorticity component $\omega_{z^\prime} = - \alpha$} is  tilted and stretched when passing the bubble resulting in a streamwise vorticity field $\omega_{x^\prime}$ characterized by two intense vorticity tubes rotating in opposite directions  in the bubble wake. This induces  suction of the fluid between the two tubes in the negative $y^\prime$-direction, inducing a reaction on the bubble in the positive $y^\prime$-direction as  illustrated in Fig. \ref{fig_lift_2}(b). 

{However,} the production of vorticity at the bubble surface at large Reynolds numbers reduces the effect of the inviscid tilting-stretching mechanism described above. {As first explained in \citet{adoua2009}, the vorticity produced at the bubble surface is mainly azimuthal, $\omega_\phi$, and  is also tilted in the wake by the velocity gradient as depicted in Fig.\ref{fig_lift_2}(b). This process results in the production of two tubes of streamwise vorticity field $\omega_{x^\prime}$, which induce suction (in between the two tubes) in the positive $y^\prime$-direction and a reaction on the bubble in the negative $y^\prime$-direction}. 

For spherical {and clean} bubbles, the surface vorticity contribution is not strong enough to counterbalance the tilting-stretching mechanism {of the  incident flow vorticity component $\omega_{z^\prime} = - \alpha$}; therefore,  the lift coefficient remains positive for all Reynolds {numbers}.
Thanks to numerical simulations \cite{legendre1998}, the lift force {has been} obtained for a viscous case with relatively large Reynolds numbers (typically $Re>50$). In this case, {the correction to the inviscid value is}  $C_L = 0.5 - 6.5 Re^{-1}$ {confirming} a decrease of the lift effect {as discussed above. The lift decrease is} proportional to $Re^{-1}$, which {follows} the viscous drag evolution at such Reynolds numbers, $C_D \propto Re^{-1}$, related {to} the vorticity produced at the bubble surface, as discussed in Section \ref{section_dragvorticity}.

However,  a  lift reversal  is observed in Fig. \ref{fig_lift_3}(d) for large bubbles \cite{Tomiyama_2002} as well as   for surfactant-covered bubbles \cite{fukuta2008}. 
Indeed, as discussed in Section  \ref{section_contamination} with Eq. \ref{Eq_cont7}, 
the interfacial vorticity, $\omega_I$, is {increased} when the bubble surface is  contaminated {and/or} deformed, and its contribution can become dominant causing the  induced lift to change direction. {For a rising bubble with a  rear surfactant cap, the gradient of surfactant concentration  $\nabla_I \Gamma_I$ is maximum and positive at the cap angle, generating a significant increase in $\omega_I$. 
As the bubble size increases,  the bubble deforms, and} the minimum for the local bubble {radius of} curvature $R_c$ is located at the bubble equator, generating a maximum for the production of $\omega_I$.
Such a change in the lift direction, {as bubbles become more deformed}, has been confirmed  experimentally for  air bubbles in different fluids, including water \cite{hayashi_lift_2020, hayashi_lift_2021}, as shown in Fig. \ref{fig_lift_3}(g), as well as using direct numerical simulations \cite{adoua2009, Hidman2022}.

\subsubsection{Lift effect at small Reynolds: The diffusing-matching mechanism}
We can  now consider  the lift mechanism at small Reynolds numbers, {still considering the linear shear flow $\mathbf{U}=(U_o + \alpha y^\prime) \mathbf{e_{x^\prime}}$}.
In this regime, the mechanisms involved in generating the lift force are clearly different.  First, the Stokes solution around a solid sphere or a spherical bubble is not valid {far from the object (as $r \rightarrow \infty$, where $r$ is the distance from the bubble center)} and needs to be matched with an outer solution where inertial effects have the same order of magnitude as the viscous terms. In the outer region,  the moving object is seen as a source term of momentum (a point force) whose magnitude is described at first order by the drag force. Second, the Stokes solution for the linear shear flow around  a solid sphere, a spherical droplet or a spherical bubble produces zero lift force \cite{legendre1997}. It follows that the lift force originates from small but non-zero inertial effects and results from the matching of the inner solution managed by the Stokes equations with the outer solution, where inertial effects have to be considered \cite{saffman1965}. The inertial  term of the Navier-Stokes equation from the base flow $U_o$ becomes comparable to the viscous term at the Oseen distance, $r=r_O=O(d_b/Re)$,  while the shear flow term balances the viscous term  at the Saffman distance $r=r_S=d_b/\sqrt{Re Sr}$, {see Fig. \ref{fig_lift_2}(a)}. The ratio of the two gives the parameter $\epsilon$ introduced above. The solution for the lift force was derived considering that the shear term  must first be balanced, resulting in the condition 
$\sqrt{Re Sr} \ll 1 $ for the validity of Eq. (\ref{eq_CL_saffman}).

The matching process of the inner solution with the outer solution requires correcting the inner solution with a velocity field that satisfies the Stokes equation. Thus, the lift force results from how the {bubble point force is impacting the flow  outside the Saffman distance}. For a spherical bubble, the vorticity distribution at the bubble surface is at first order axisymetric and the {point force representing the bubble in the outer region is aligned with the flow direction}. However, its matching with the external flow differs because the inertial term is not axisymmetric, and the resulting feedback to the inner solution generates a positive lift force as shown in Fig. \ref{fig_lift_3}. \citet{Hidman2022} recently reported numerical simulations of lift reversal at low bubble Reynolds numbers. {In this case}, the bubble is  deformed in a non-axisymmetric manner (see Fig. \ref{fig_lift_3}), which leads to significant differences in vorticity production on the bubble surface and disrupts the fore-aft symmetry of the flow. This results in a {bubble point force} not aligned with the flow direction, impacting the matching of the inner solution  with the  outer solution, {and the feedback to the bubble}. Consequently, depending on the bubble's {deformation and} orientation in the shear flow, a lift reversal can be observed.

\section{Bubble-bubble and bubble-wall interactions}

Different types of bubble interactions can be considered. The objective here is to focus  on the mechanisms that govern bubble-bubble interaction in two situations: "long-range" interaction where the interaction results from hydrodynamic mechanisms and short-range  interaction where the interaction is controlled by interface deformation and lubrication. 

Bubble pair interactions have been extensively investigated by  careful experiments \cite{duineveld1995, sanada2005, Sanada2009, kong2019hydrodynamic, Kusuno2021} as well as by the use of  direct numerical simulation with fixed bubbles \cite{legendre2003, hallez2011interaction} and free-to-move bubbles \cite{zhang2021b, zhang2022}.  These studies have revealed different  scenarios for interaction and the stability of  bubble pairs. The pair interaction problem has many interesting aspects: in-line interaction can be stable or unstable, side-by-side bubble pairs can attract or repel, and free-to-move pairs exhibit the so-called  drafting-kissing-tumbling interaction, among others. {The  pair interaction mechanisms are frequently used to discuss when clustering is expected to occur in bubbly flows \cite{zenit2001,figueroa2018, ma2023fate}.}

\subsection{Side-by-side interaction}

\begin{figure}[h]
\includegraphics[width=6in]{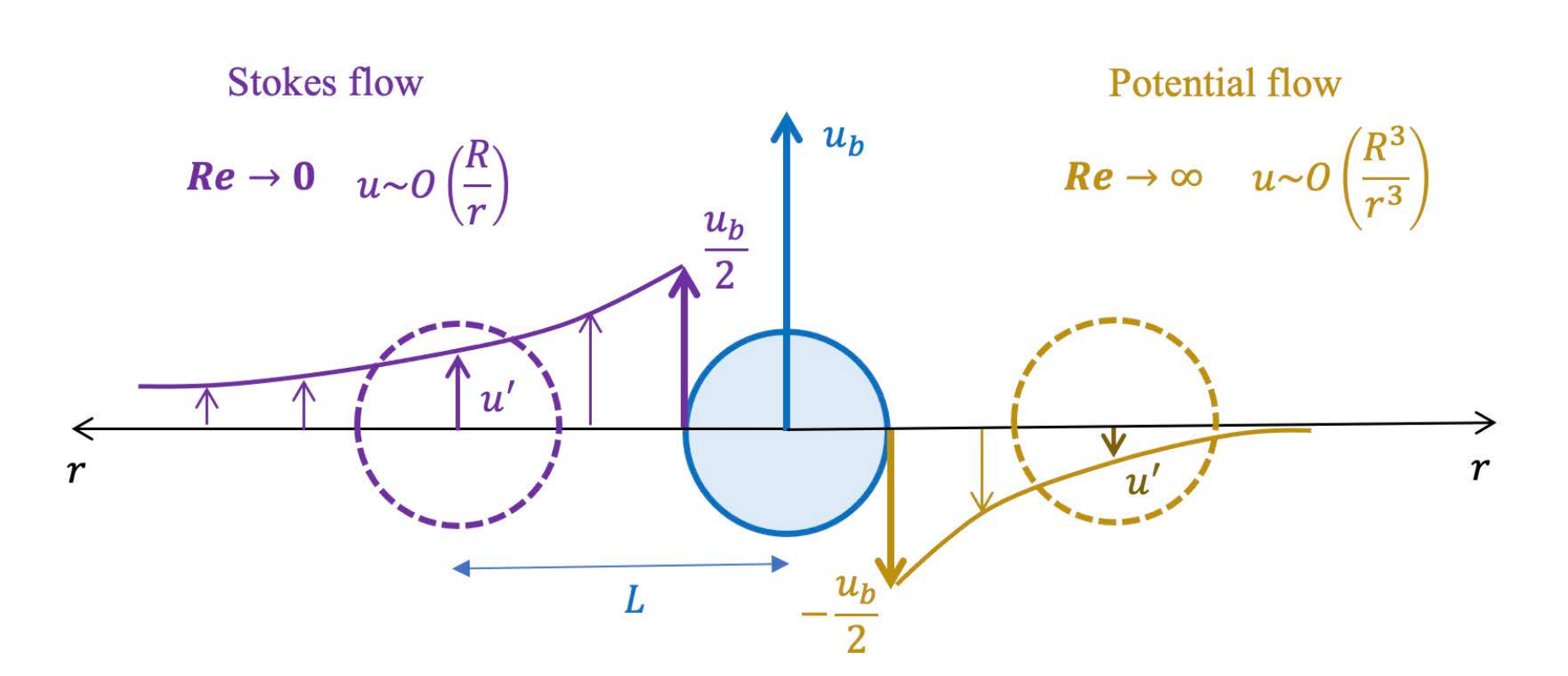}
\caption{Fluid displaced by a rising spherical bubble and the corresponding disturbance decay with distance $r$ from the bubble: (left) Stokes flow and (right) potential flow {for} small and large Reynolds numbers, respectively. The flow disturbance, $u'$, is felt by the second bubble (in dashed line) located at the distance $r=L$.}
\label{fig_displaced_fluid}
\end{figure}

We can start by considering the side-by-side interaction case, which corresponds to a bubble rising in the vicinity of another bubble of the same size and  at the same speed. The separation distance between the two bubbles is $L$. This situation can also be used to understand bubble interaction with a wall by  considering $L/2$ as the distance between the bubble and the wall, with the bubble interacting with its mirror image located at the distance $L$.  As shown below, the long-range interaction  results in a change in the rising velocity (drag force), in a sideways attractive or repulsive effect, as well as a possible bubble deformation (not discussed here). The interaction is, of course, different for small and large bubble Reynolds numbers. This can  be anticipated by considering how a bubble displaces the fluid around it, as shown schematically in Fig. \ref{fig_displaced_fluid}.

We first consider the interaction  at  high bubble Reynolds number. As discussed earlier, the potential flow is a good approximation for the flow around a bubble at  Reynolds numbers larger than $\Rey>100$.
As depicted in Fig.\ref{fig_displaced_fluid}, the flow displacement far from the bubble is directed in the downward direction with a magnitude $-u_b R^3/(2 r^3)$. At a distance $r=L$ the fluid perturbation seen by the second bubble would be $u^{'}= -u_b R^3/(2 L^3)$. Therefore, the liquid  velocity in the reference frame of the second bubble is increased to 
$u_b \left[1+R^3/(2 L^3) \right]$ resulting in an increase in the drag force to $- 12 \pi \mu R u_b \left[1+R^3/(2 L^3) \right]$. Therefore the additional drag force correction decays as $L^{-3}$. In the radial direction the potential flow prediction induces a velocity gradient along the $r$-direction which results in $- 3 u_b  R^3/(2 r^4)$. Considering the definition of the induced shear lift (Eq. \ref{eq_lift1}),  the second bubble experiences an attractive force of the form $F_a \sim \rho u_b^2 R^6/L^4$.
The potential flow analytical solution of \citet{wijngaarden1976}  and direct numerical simulations from \citet{legendre2003} both confirm the previous discussion summarized as:
\begin{eqnarray}
    F_D \approx - 12 \pi \mu R u_b \left( 1 - \frac{2.211}{\sqrt{\Rey}}  \right) \left[1+\frac{R^3}{L^3} \right], \quad
    F_a \approx 3 \pi \rho  u_b^2 \frac{R^6}{L^4} \left[ 1+\frac{R^3}{L^3}  \right]. 
\end{eqnarray}

We can now consider the limit of small Reynolds number where the development of the flow around the bubble is extended further away from the bubble, due to the diffusion process. The Stokes approximation remains valid up to the Oseen distance, $r_O$. Beyond this region, non-inertial effects become comparable to viscous ones, and corrections must be considered. Depending on the value of $L/r_O$  different types of interaction will be observed, as the second bubble will be either  in the inner (Stokes) or outer (Oseen) region of the disturbance produced by the first bubble. 

When two bubbles are within the Stokes region, considering the Stokes flow evolution as depicted in Fig. \ref{fig_displaced_fluid}, the velocity disturbance induced by a bubble  is now $u^{'}= u_b R/(2 L)$. The second bubble then experiences a reduction in its relative velocity with the surrounding fluid as $u_b \left[1-R/(2 L) \right]$ and its drag force is reduced to $-4 \pi \mu u_b \left[1-R/(2 L) \right]$. The correction shows  decay in $L^{-1}$ which is {more} pronounced than that observed at high Reynolds numbers.
The Stokes flow also induces a velocity gradient along the separation direction between bubbles like $ u_b R/(2 r^2)$ but in opposite sign compared to the  high Reynolds case. Hence, in this case, the bubbles experience  a repulsive interaction. The analytical  solution for $Re L/r \ll 1 $ \cite{legendre2003} confirms this scaling discussion with a  corrected drag force and a repulsive force $F_r$ of the form :
\begin{eqnarray}
    F_D  \approx - 4 \pi \mu R u_b \left[1 - \frac{R}{2 L} \right], \quad F_r \approx \frac{1}{2} \pi \rho R^2 u_b^2 
\end{eqnarray}

The wall interaction shows similar trends for both  small and large Reynolds numbers. 
{In the case of a wall, the flow displacement caused by the bubble motion illustrated in Fig.\ref{fig_displaced_fluid} needs to be corrected by the non-slip condition  at the wall.} In close proximity to the wall, some significant modifications  in the force evolution are observed,  as well as in  bubble deformation \cite{magnaudet2003, Figueroa2008, Takemura2009}. {The interaction of bubbles in close contact with a wall is discussed in section \ref{section_contactwall}}.

\subsection{In-line interaction} 

\begin{figure}[h]
\includegraphics[width=6.5in]{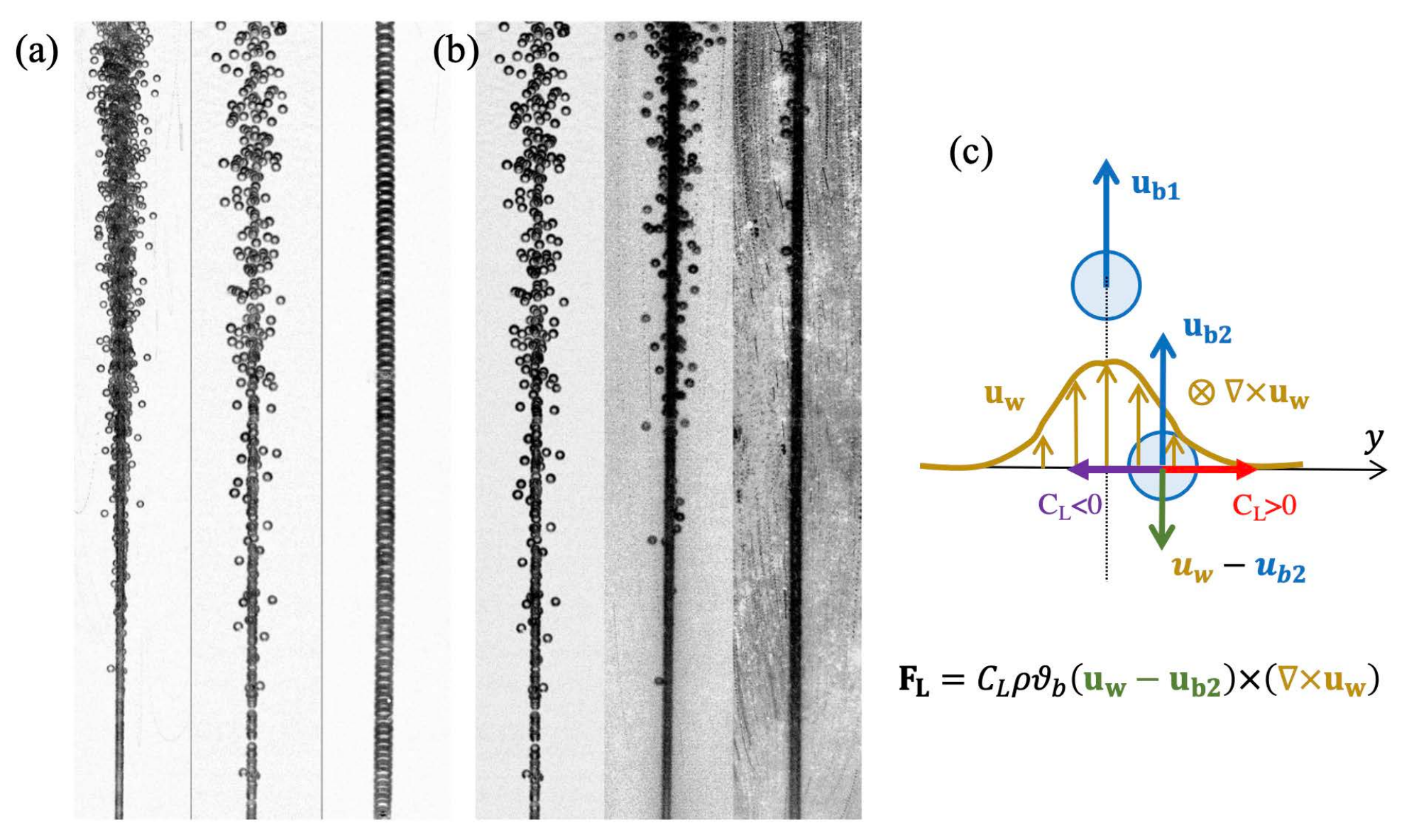}
\caption{(a and b) Experiments showing the effects of bubble size and surface contamination on the stability of bubble chains, taken from \citet{Atasi2023}. (a) The bubble
size increases (from left to right), for a liquid without surfactants, La = 0; {\Ga=(1200, 1920, 8400)}
and \Bo=(0.18, 0.25, 0.66) for (left, center, right). (b) The bubble sizes are approximately the same,
but the amount of surfactant increases (from left to right), La=(0, 0.1, 0.5), {\Ga=(1920, 1200, 1800)}
and \Bo=(0.24, 0.21, 0.32) for (left, center, right). (c) Illustration of the wake-induced lift effect. {In the wake of the first bubble, the second bubble experiences a lift force, as described in section \ref{section_lift}. Depending on the sign of the lift coefficient, the inline motion is stable ($C_L>0$ for a spherical clean bubble) or unstable ($C_L<0$ for ellipsoidal and/or contaminated bubble).} 
}
\label{fig_inline}
\end{figure}

We now consider the configuration of a bubble preceding a second bubble. 
According to the potential solution, the inline configuration is unstable \cite{harper1970, hallez2011interaction}. However, the study of in-line bubble configuration reveals that either very stable bubble chains or  very dispersed rising bubbles can be observed, depending on several factors, as shown in Fig. \ref{fig_inline}. The explanation for these two opposing   behaviors lies in a subtle coupling between  bubble deformation and surface contamination, both of which affect the wake-induced lift effect \cite{Atasi2023}. 

A bubble rising in the wake of another bubble experiences a reduction of its relative velocity, and therefore a reduction of its drag. It is  also subject to a lift effect induced by the velocity gradient inside the wake \cite{hallez2011interaction}, as depicted in Fig. \ref{fig_inline}(c). If $\mathbf{u_w}$ is the velocity profile in the wake of the leading bubble, the  velocity  experienced by the second bubble (i.e. the liquid velocity in its moving frame of reference) is then $\mathbf{u_w-u_b}$. As a first approximation (see \citet{batchelor1967} and the appendix of  \citet{hallez2011interaction}, where the wake of a single bubble is detailed), the bubble wake magnitude decays as $u_w \sim u_b R/r$, such that the vorticity field experienced by the second bubble at $r=L$ is $\omega_w=\nabla \times u_w \sim u_w/R \sim u_b/L$. This effect is thus  decreasing as $L^{-1}$, where $L$ is the distance between the two bubbles centers.

From Eq. (\ref{eq_lift1}), the lift force experienced by the second bubble in the direction perpendicular to its ascend {is then}
\begin{eqnarray}\label{eq_lift_wake}
    \mathbf{F_L}  \approx C_L \rho  \frac{4 \pi}{3} \frac{ R^3 u_b^2}{L} \left(1- \frac{R}{L} \right) \mathbf{e_y} 
\end{eqnarray}
A negative value for the lift coefficient $C_L<0$ (i.e., for deformed  and/or contaminated bubbles) as discussed in section \ref{section_lift}) results {in a force that retains the second bubble in the wake of the first one and} 
in the formation of a stable bubble line, as reported in Fig. \ref{fig_inline}{(a)}.
A positive value for the lift coefficient $C_L>0$ (i.e., for moderate bubble deformation and reduced interface contamination) induces  bubble migration from the wake resulting in the formation of a cone of bubbles, as observed in Figure \ref{fig_inline}{(b)}.
If $w_b$ is the  bubble's transverse migration velocity, the balance between the lift force \ref{eq_lift_wake} and the viscous drag $\sim \mu R w_b$ reveals that the observed cone angle $\alpha \approx {w_b}/{u_b}$ varies linearly with  the bubble detachment frequency $f=u_b/L$ {as well as with the lift coefficient $C_L$}:
\begin{eqnarray}\label{eq_angle_inline}
    \alpha  \sim f {C_L} \frac{ R^2 \mu}{\rho}.
\end{eqnarray}
{When $C_L$ is reduced, by increasing bubble contamination or by increasing bubble deformation, the dispersion decreases (i.e., $\alpha$ is decreased) as shown in Fig. \ref{fig_inline}(a and b)}. The validity of Eq.  \ref{eq_angle_inline} has been confirmed by experiments involving different bubble sizes and levels of surface contamination \cite{Atasi2023}.

\subsection{{Bubble dynamics in contact with a wall}\label{section_contactwall}}
We now consider the interaction of a bubble rising at its terminal velocity as it collides  with an inclined wall, as illustrated in Fig. \ref{fig_wall_bouncing}. {This problem has received recent attention due do its importance in understanding the bubble-induced reduction of wall friction \cite{TANAKA2021109909} and for surface cleaning \cite{Hooshanginejad2023}.}
Depending on the wall inclination, from $\alpha=0^{o}$ (horizontal wall) to $\alpha=90^{o}$ (vertical wall)  different behaviors are observed,  ranging from sliding to bouncing motion  \cite{barbosa2016conditions}. 

\subsubsection{Criteria transition from sliding to bouncing regimes}

\begin{figure}[h]
\includegraphics[width=6.5in]{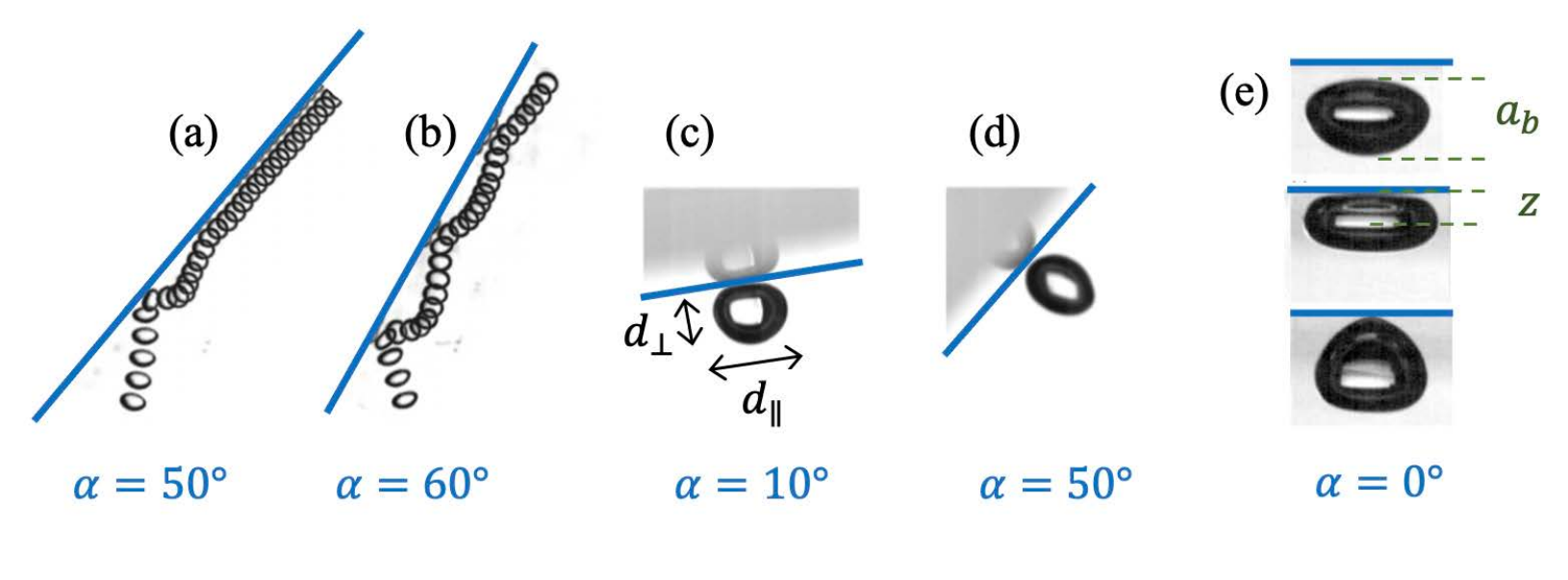}
\caption{Bubble interacting with a wall. A $d_b=1.6$ mm bubble in a water-glycerin mixture: (a) sliding  after collision, (b) bouncing repeatedly against  the wall.  Shape of a $d_b=2.2$ mm air bubble in silicon oil while sliding under an inclined wall, {forming a lubrication film between the bubble and the wall:} (c) $\alpha=10^{\circ}$ and (d) $\alpha=50^{\circ}$. (e) A $d_b=2.62$ mm bubble rising in a water-glycerin mixture bouncing under an horizontal wall. Three instants are shown from top to bottom: shape at terminal velocity (before sensing the presence of the wall), maximum deformation during impact, and maximum rebound velocity.}
\label{fig_wall_bouncing}
\end{figure}

The criterion between bouncing and sliding is obtained by considering the force balance for the motion normal to the wall and along the wall. Let us first consider the force balance in the normal direction.
The sliding motion, shown in Fig.  \ref{fig_wall_bouncing}(a), is  observed when  buoyancy is sufficient  to keep  the bubble in contact with the wall. The bouncing motion, depicted in Fig.  \ref{fig_wall_bouncing}(b), occurs when the bubble can depart from the wall due to an inertial lift-type force that results from the interaction of the bubble wake with the wall (see the wake structure in Fig. \ref{fig_zigzag}). The induced force normal to the wall, denoted as $F_{w\perp}$, depends on the strength of the circulation of the vortex filaments in the wake and can be scaled as $F_{w\perp} \sim \rho d_b^2 u_{bw}^2$ \cite{DeVries2002}, where $u_{bw}$ is the bubble velocity along the wall. The force can be compared to the buoyancy acting normal to the wall, $\rho d^3 g \cos \alpha$, to establish a condition for the bubble's departure from the wall 
\begin{equation}\label{eq_normal}
 \rho d_b^2 u_{bw}^2 \ge \rho d_b^3 g \cos \alpha.
 \end{equation}
Now, let us consider the motion parallel to the wall. For spherical bubbles, as shown in Fig. \ref{fig_wall_bouncing}(c), the balance parallel to the wall is obtained by the component of buoyancy along the wall $\rho d_b^3 g \sin \alpha$ and the bubble viscous drag $\sim \mu d_b u_{bw}$:
\begin{equation}\label{eq_tangent}
\mu d_b u_{bw} \sim \rho d_b^3 g \sin \alpha
\end{equation}
Taking the ratio of these two conditions, Eqs. (\ref{eq_normal}) and (\ref{eq_tangent}), yields a simple criterion based on the bubble sliding Reynolds number $Re_w= \rho d_b u_w/\mu$ and the wall inclination as 
$Re_w \ge \cot \alpha$.
Comparison with data from the literature \cite{barbosa2016conditions} reveals that the  effective transition can be fitted to 
\begin{equation}\label{eq_criteria}
Re_w= Re_0 + 310 \cot \alpha 
\end{equation}
where $Re_0 \approx 80$ is the transition observed for vertical wall, $\alpha = 90^{o}$ \cite{DeVries2002, takemura2003transverse}.

When bubbles are sufficiently deformed, as depicted in Fig. \ref{fig_wall_bouncing}(d), the viscous drag force is replaced by an inertial drag force of the form $\rho d_b^2 u_{bw}^2$, for which the motion parallel to the wall now satisfies
\begin{equation}\label{eq_tangent2}
\rho d_b^3 g \sin \alpha \sim \rho d_b^2 u_{bw}^2
\end{equation}
When combined with relation (\ref{eq_normal}), the condition becomes $\cot \alpha \sim \text{constant}$, resulting  in a fixed angle for the transition  that is independent of the bubble sliding Reynolds number. Considering the experiments of \citet{barbosa2016conditions}, the criteria is given by 
$\cot \alpha \approx 1$.
The transition between viscous and inertial motion is observed at a threshold sliding Weber number given by $We_w=\rho d u_w^2/\sigma \approx 1.2$, which is based on the  bubble wall velocity. The transition between the sliding and bouncing regimes for a bubble rising under a wall can be summarized as follows
\begin{equation}
 We_w \le 1.2 \quad   Re_w= Re_0 + 310 \, \cot \alpha, \qquad We_w \ge 1.2 \quad  \cot \alpha  \approx 1
    \label{eqn:transition}
\end{equation}

\subsubsection{Bubble deformation when sliding under a plane wall}
When sliding under a wall, the bubble deformation changes depending on the wall inclination as observed in Figs. \ref{fig_wall_bouncing}(c and d). For small  inclination angles, as shown in Fig. \ref{fig_wall_bouncing}(c), buoyancy 
pushes the bubble against the wall and controls the bubble deformation. In such condition, the bubble deformation is determined by balancing the potential energy used for deformation with the gain in surface energy \cite{madavan1985, Barbosa2019}, resulting in deformation controlled by  the Bond number $\Bo$. For large inclination angle, as shown in  Fig. \ref{fig_wall_bouncing}(d), the bubble is flattened along the sliding direction due to inertial effects generated by its motion, resulting in a deformation controlled by the bubble sliding Weber number $We_w$, similar to what is observed for rising bubbles \cite{moore1965} (see section \ref{section_ellpsoidal}). Both of these two modes of deformation contribute to the overall bubble deformation, and can be combined to express the {bubble-wall} aspect ratio $\chi_w = d_{\parallel}/d_{\perp}$ where $d_{\parallel}$ and  $d_{\perp}$, {as depicted in  Fig. \ref{fig_wall_bouncing}(c),} represent the bubble's length parallel and normal to the wall, respectively \cite{Barbosa2019}:
\begin{equation}
 \chi_w = \frac{1 - \beta_1 We_w }{\left(1 + \beta_1 We_w / 2 \right) \left( 1 - \beta_2 \Bo \cos \alpha \right) }
    \label{eqn:deformation_wall}
\end{equation}
with $\beta_1=3/32$ and $\beta_2=0.1$.

\subsubsection{Bubble bouncing under a horizontal wall}\label{section_bouncing}

The process of bubble deformation during bouncing under an horizontal wall is shown in Fig. \ref{fig_wall_bouncing}(e).  The  kinetic energy of a bubble moving at its terminal velocity $u_b$  (contained in its fluid added mass) is transformed into surface energy when the bubble is compressed against the wall. Then this surface energy is restored into kinetic energy in the direction opposite to the wall. A restitution coefficient can be defined as 
\begin{equation}
\epsilon = \frac{u_r}{u_b}
\end{equation}
where $u_r$ is the maximum velocity at restitution. The evolution of the bubble shape, during the bouncing process, can be  described using a mass-spring model {(see also section \ref{Sec_bubble_rupture} which considers a similar dynamical system for modelling bubble deformation and breakup in a turbulent environment)}. It is derived considering the motion of the bubble center $z=a_b/2- \zeta$ where $a_b$ is the vertical axis of the bubble before the impact (its radius when spherical) and $\zeta(t)$ is the deformation during the rebound. The  inertia {(added mass)} involved in the  motion of the bubble's center of mass $\sim \rho d^3 \ddot{z}$ varies due to two main effects: (1) the damping force induced by the liquid drained between the bubble and the wall   $-\mu d \,\dot{z}$,  and (2) the restoring force to sphericity   $ \sim \sigma \zeta$ due to bubble deformation. The bubble deformation $\zeta$ can be described by the following second-order equation \cite{Zenit2009}: 
\begin{equation}
 \frac{d^2 \zeta}{dt^2} + K_1 \frac{\mu}{\rho d^2} \frac{d \zeta}{dt} + K_2 \frac{\sigma}{\rho d^3} \zeta =0
    \label{eqn:mass_spring}
\end{equation}
where $K_1$ mostly depends on the bubble surface mobility and $K_2$ on the initial bubble shape. This equation reveals that this problem has two characteristic times: the viscous relaxation time $\tau=\rho d^2/\mu$ {that controls} the damping effect and the period of the oscillation $T=\sqrt{\rho d^3/\sigma}$ that controls the rebound duration. The velocity at the end of the semi-period $t=T/2$, which corresponds to the departure of the bubble from the wall, can  be expressed as a function of the velocity at impact. This allows the restitution coefficient to be determined \cite{Zenit2009} 
\begin{equation}
 \epsilon = \frac{u_r}{u_b} \approx \exp{(- K \Oh)}
    \label{eqn:restitution}
\end{equation}
where $K\sim K_1/\sqrt{K_2}$ is a constant that can be determined from experiments and $\Oh$ is the Ohnesorge number defined in Eq. \ref{eqn:Oh}. Interestingly, the restitution coefficient is independent of the bubble impact velocity due to the linear dependence of the forces on the velocity, {that appear during} the rebound. Experiments of \citet{Zenit2009}  indicates that $K\approx 30$.
The restitution coefficient behaves in a different way to that of solid spheres where the restitution coefficient is described using the Stokes number $\St=(\rho+ C_M \rho_p) u_s d / \mu$, thus being dependent on the impact velocity {$u_s$} \cite{joseph2001, legendre2006a}. Note that the Ohnesorge number can be expressed as function of the Stokes number as $\Oh=\sqrt{\Ca/\St}$. The main difference between the bubble and the solid sphere results from the significant deformability of the bubble surface during the interaction with the wall {that, in turn, increases the contact time with the wall \cite{legendre2006a}}.

\section{Bubbles in complex fluids \label{bubblesnonNewtonian}}

When the fluid that surrounds a bubble is not Newtonian, all of its dynamic properties can change. A fluid is considered {Newtonian} if the relationship between the stress tensor, $\mathbf{\Sigma}$ and the strain rate tensor $\mathbf{S}$, is {linear}:
\begin{equation}
    \mathbf{\Sigma} = 2 \mu \mathbf{S}
    \label{eqn:Newtonian}
\end{equation}
where $\mu$ is the shear viscosity and it has a constant value. In any other case, the fluid is considered non-Newtonian. Non-Newtonian fluids can generally be divided into two broad categories: {fluids with variable viscosity and fluids with an elastic stress component.} {In this section, we first describe  models for classical non-Newtonian rheologies and then discuss their impact on bubble dynamics.}

\subsection{{A brief introduction to non-Newtonian liquids}}
The viscosity of a fluid can vary under different conditions. Neglecting the effect of temperature, the fluid viscosity can be  shear-dependent. Fluids whose viscosity decreases as the shear rate increases are called shear-thinning (or pseudo-plastic), while fluid that become more viscous with increasing shear rate are called shear-thickening (or dilatant). A simple model that captures this behavior is the so-called power-law fluid (also called Ostwald-de Waele model). Using Eq.(\ref{eqn:Newtonian}) as a reference, the effective viscosity can be defined as:
\begin{equation}\label{eq_power_law}
    \mu_{\mbox{eff}}= \kappa \left(\sqrt{\mathbf{S}:\mathbf{S}}\right)^{n-1}
\end{equation}
where $n$ and $\kappa$ are the power index and the consistency. For $n<1$  the effective fluid viscosity decreases as $\mathbf{S}$ increases. When $n=1$, the Newtonian rheology is recovered. Thus, the shear-dependent behaviour is characterized by the values of $n$ and {$\kappa$}, which can be determined experimentally.

Shear-thinning fluids are the most commonly found in both industrial and natural environments. While a precise physical model to explain the reason behind a shear-thinning (or shear thickening) behavior does not exist, it can be argued that the reduction of viscosity results from micro-structural changes of the fluid that facilitate flow as shear is increased. For instance, in polymers and polymer solutions, the shear-thinning behaviour results from the disentanglement of polymeric strands during flow \citep{cross1979}.

Note that the behavior predicted by the power-law model (\ref{eq_power_law}) is often valid only at an intermediate range of shear rates. For both low and high shear rates, the viscosity often recovers a constant value. To capture this non-monotonic change of viscosity with shear rate, models with several more parameters are often used. See, for instance, the Carreau fluid model \cite{carreau1972rheological}.

The viscosity of fluids can also change over time. Fluids that exhibit a decrease in viscosity with time, at constant temperature, are called thixotropic. For anti-thixotropic behavior (also called rheopectic), an increase of viscosity with time can also be observed.

\subsubsection{{Bubble shape for} shear-dependent viscosity fluids}

\begin{figure}[h]
\includegraphics[width=6in]{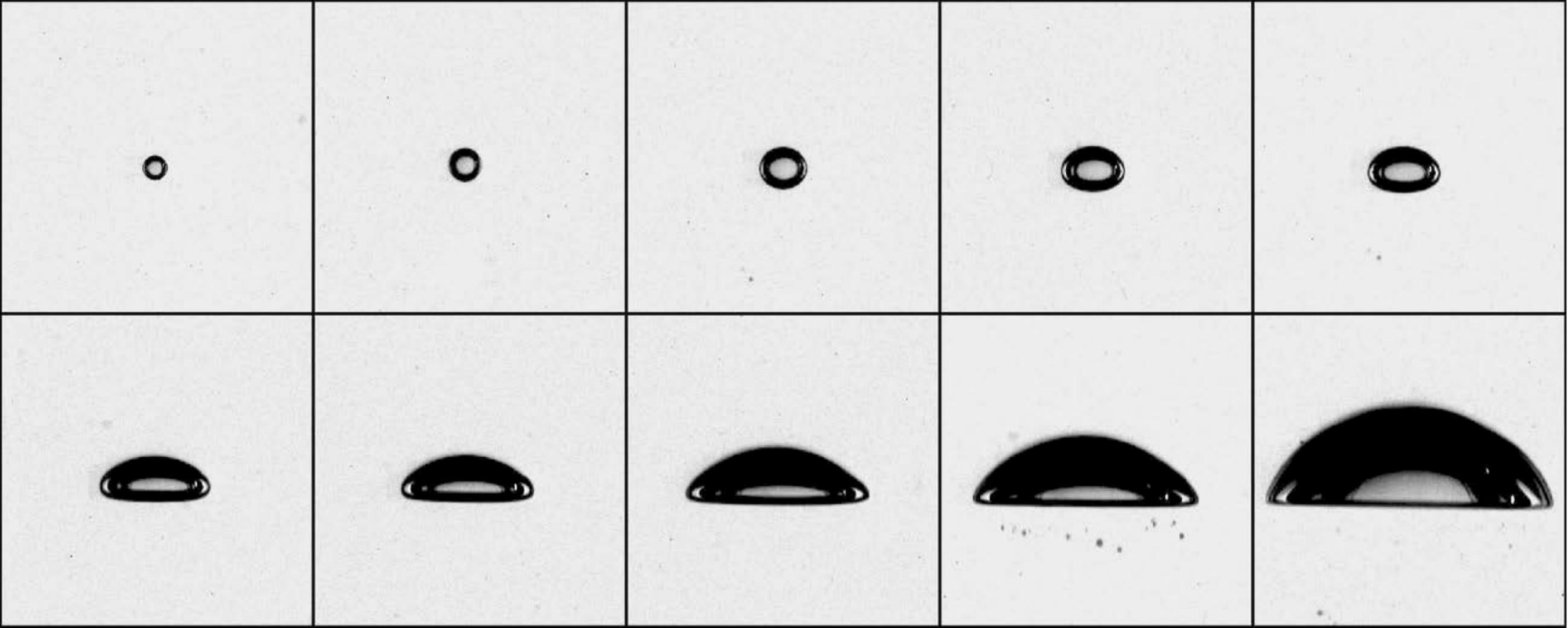}
\caption{Shape of bubbles in a shear thinning fluid. The volume ranges from 2.65 to
22 mm$^3$, from the smallest to the largest bubble shown, corresponding to bubble diameter  ranging from 1.7 to 3.5 mm. The fluid is a mixture of ethylene glycol and Carbopol for which {$\kappa$}=0.19 Pa s$^{0.8}$, $n$=0.8. \Rey $\sim$ O(1) and {\Mo $\sim$ O(10$^{-2}$)}. Original data. }
\label{fig_deformation_ST}
\end{figure}

The bubble shape, as discussed above, is the result of the balance of surface tension forces and other fluid forces. In the case of bubbles moving in complex fluids, we can expect these properties to affect the bubble shape. There have been numerous studies that have documented the bubble shape in non-Newtonian liquids \cite{chhabra2006bubbles}, but the results are sparse and, in general, lack a fundamental understanding of the underlying physical mechanisms. In most cases, the experiments are conducted for fluids that exhibit both shear thinning and viscoelastic effects simultaneously, which makes the physical interpretation more difficult.  Figure \ref{fig_deformation_ST} shows how the bubble shape evolves as the diameter increases for a fluid that is inelastic but shear-thinning \cite{Zenit2018}.  The bubble shape evolves from spherical to ellipsoidal to spherical cap, which resembles what is observed in viscous Newtonian fluid, as shown in Fig. \ref{fig_deformation}. {Note that a proper comparison would need to consider the value of the Morton number as discussed in Section \ref{sec_deformation} above. However, since the fluid is shear-thinning, as bubble size increases the effective viscosity decreases. Therefore, even for the same liquid, the Morton number is no longer constant making direct comparisons challenging.
Nevertheless, the images indicate that for this moderate value of the power index, $n$, the bubble deformation is qualitatively similar to that observed in a Newtonian liquid, in accordance with the numerical results of \citet{zhang2010bubbles_st}. A general map that shows the shapes of bubbles in shear-dependent liquids, similar to the classical one from  \citet{clift1978}, does not exist.}

In contrast, even if the bubble shape is not significantly altered by shear-thinning viscosity, the bubble terminal velocity does change with values of the power index and consistency, $n$ and  {$\kappa$}, respectively.  \citet{velez2011} showed that for moderate values of the $\Rey$ number, the drag coefficient of bubbles decreases with the power index, $n$. Hence, bubbles in such shear-thinning fluids would have a larger terminal speed, $u_\infty$, in accordance with several other studies \cite{chhabra2006bubbles}. This increase is expected since the bubble motion creates a low-viscosity region around it. Hence, {the viscous dissipation around the bubble is smaller. }

\subsubsection{{Bubble shape for} viscoelastic fluids}
The typical case for which the stress {in the fluid is not purely viscous} is an intermediate state between an elastic solid and a viscous fluid. In such a case, the stress in the fluid results from both deformation and deformation rate. The simplest model of viscoelasticity is the linear viscoelastic model, for which these two effects are linearly superposed:
\begin{equation}
     \mathbf{\Sigma} +\lambda_1 \frac{\partial \mathbf{\Sigma}}{\partial t}= \mu \left(\mathbf{S}+\lambda_2 \frac{\partial \mathbf{S}}{\partial t}\right)
     \label{eqn:jeffrey}
\end{equation}
where $\lambda_1$ and $\lambda_2$ are the elastic relaxation and retardation times, respectively. This model is the so-called  Jeffrey model. Although it captures some features of viscoelasticity, its applicability is limited. The relative importance of elastic effects in a flow is often given by the dimensionless relaxation time:
\begin{equation}
    \Wi = \frac{\lambda_1 u_b}{d_b}.
\end{equation}
The Weissenberg number $\Wi$ measures whether the elastic relaxation time is larger that the flow time. Hence, a Newtonian fluid will have $\Wi=0$, while a purely elastic solid would have $\Wi \rightarrow \infty$. {Linear viscoelastic models are often valid only when \Wi$\rightarrow 0$}.

In particular, the main issue with these linear viscoelastic models is that they are not independent of the frame of reference. This lack of frame-independence arises from the non-objectivity of total derivatives for tensorial quantitites. To address this issue, total derivatives that are objectively defined in the frame of reference of fluid particles can be proposed. These convective derivatives are used to ensure the objectivity of non-linear viscoelastic models.  By replacing the time derivatives in Jeffrey's model (Eq. \ref{eqn:jeffrey}) with the upper-convective derivative \cite{oldroyd1950}, and we arrive at:
\begin{equation}
     \mathbf{\Sigma} +\lambda_1 \overset{\kern0.25em\triangledown}{\mathbf{\Sigma}}= \mu \left(\mathbf{S}+\lambda_2 \overset{\kern0.25em\triangledown}{\mathbf{S}}\right)
     \label{eqn:jeffrey_1}
\end{equation}
where the upper-convected time derivative is defined as
\begin{equation}   \overset{\kern0.25em\triangledown}{\mathbf{A}} = \frac{\partial \mathbf{A}}{\partial t} + (\mathbf{u}\cdot \nabla) \mathbf{A} - (\mathbf{G})^T\cdot \mathbf{A} - \mathbf{A}\cdot \mathbf{G}
\end{equation}
where $\mathbf{A}$ is a tensor, $\mathbf{u}$ is the velocity field and $\mathbf{G}=\nabla \mathbf{u}$ is the velocity gradient tensor. There are other objective derivative operators (such as the lower-convected and the co-rotational derivatives) that are valid and can also be used \cite{bird1987}. 

One of the consequences of this type of rheological constitutive equations is that the normal stress differences in the fluid can be different from zero: 
\begin{eqnarray}
    N_1 & = \sigma_{xx}-\sigma_{yy} \\
    N_2 & = \sigma_{yy}-\sigma_{zz}. 
\end{eqnarray}
Both $N_1$ and $N_2$ are generally a function of the amount of shear in the flow but, most importantly, are equal to zero in Newtonian fluids \citep{bird1987}. 
The normal stress differences are responsible for many of the unusual experimental observations in viscoelastic flows such as the Weissenberg effect \cite{lodge1988weissenberg}, the die swell effect \cite{tanner1970}, the vortex rotation reversal \cite{palacios2015}, and the bubble velocity discontinuity \cite{Zenit2018}, discussed below.

\begin{figure}[h]
\includegraphics[width=6in]{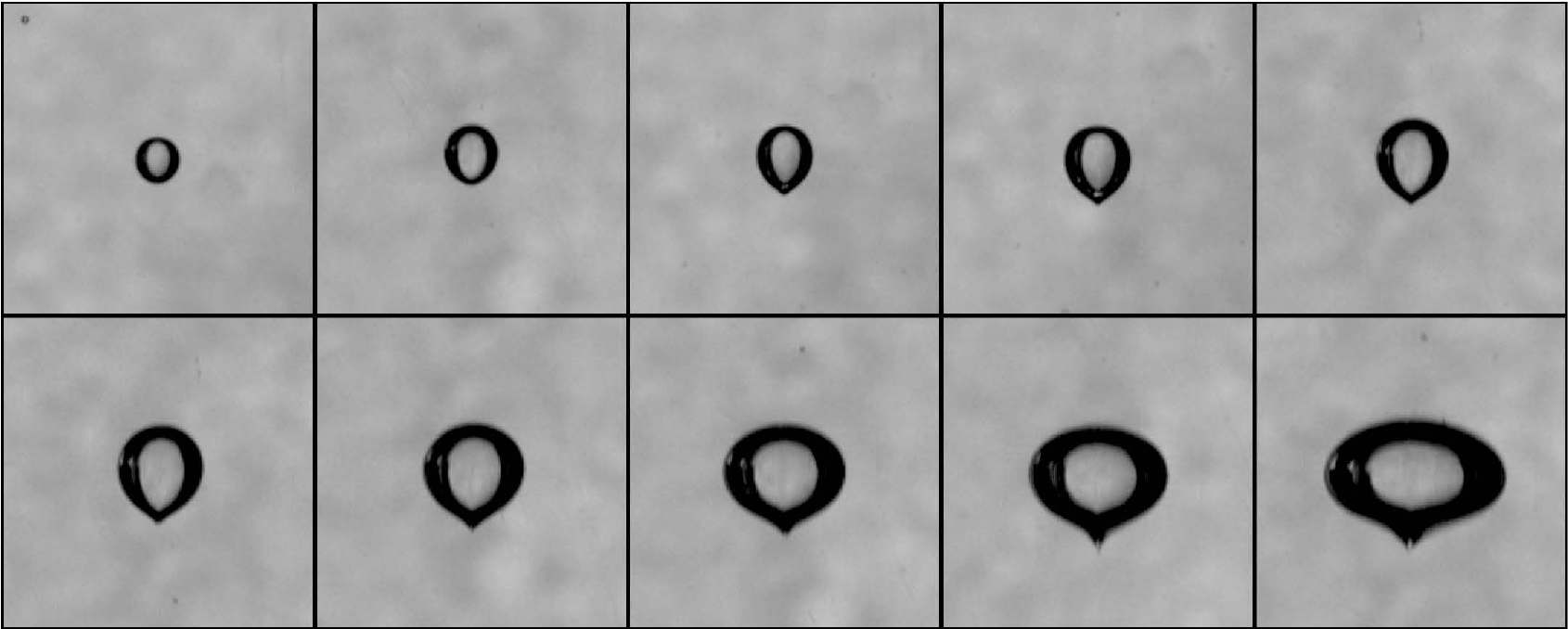}
\caption{Shape of bubbles in a viscoelastic fluid. 
The volume ranges from 12 to 440 mm$^3$, {corresponding to bubble diameter ranging from 2.8 to 9.5 mm}, from the smallest to the largest bubble shown. {The critical bubble size in this sequence, {corresponding to the velocity discontinuity as shown in Fig. \ref{fig_terminalvelocity_VE}},  is in between the third and fourth images on the top row.} The fluid is a mixture of water and polyacrilamide for which {$\kappa$}=1.92 Pa s$^{0.12}$, $n$=0.12, $\lambda_1$=12 s and  $\Rey \sim $O(1), \Mo$\sim$ O(10) and $\Wi\sim$O(10). Adapted with permission from \citet{ravisankar2021}.
}
\label{fig_deformation_VE}
\end{figure}

The effect of these elastic normal stress differences can be readily observed in flows dominated by shear. In the flow around bubbles, there are parts of the flow, such as the wake,  that are dominated by shear. Consequently,  we can expect to observe the manifestation of elastic effects in the flow around bubbles in such liquids. Figure \ref{fig_deformation_VE} shows the shape of bubbles in a viscoelastic liquid. Note that the shape of bubbles in such liquids deviates significantly from what is observed in a Newtonian one (see Fig. \ref{fig_deformation}). In general, in these fluids the bubble shape is more elongated in the flow direction, due to the appearance of normal stresses due to high shear in the bubble equatorial region. More interestingly, as bubbles reach a critical size, the rear side develops a tear-drop shape {which, as explained below, is related to the bubble velocity discontinuity phenomenon.} The tip of the bubble can become very sharp and elongated \cite{Soto2006}. As the bubble size increases, inertial effects become important which make the bubble adopt an ellipsoidal {shape that retains the pointy tip.} Again, many studies have addressed the effect of liquid rheology on the bubble shape \cite{chhabra2006bubbles}, but a fundamental understanding of the process is still lacking. {A bubble shape map for viscoelastic fluids, akin to Clift's map for Newtonian liquids \cite{clift1978}, does not yet exist. Such a map would have to include a third axis to account for changes of shape resulting from different values of the Weissenberg number, $\Wi$.}

\subsection{Bubble velocity discontinuity and negative wake}
As in the case of bubbles moving in Newtonian liquids, as the bubble size increases, the terminal speed also increases as long as the deformation is not significant.

\begin{figure}[h]
\includegraphics[width=5in]{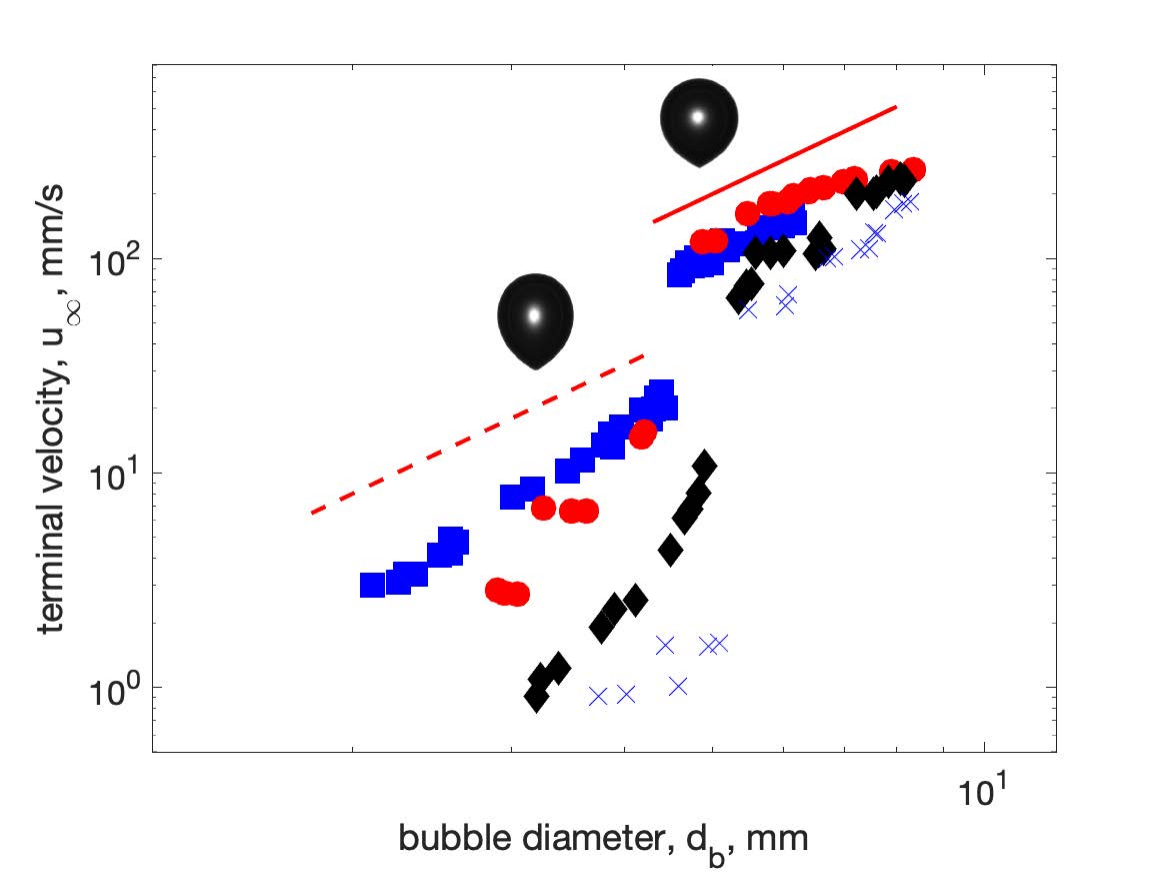}
\caption{Bubble terminal velocity in different viscoelastic fluids as a function of the bubble diameter. Experimental data  from \citet{pilz2007}: (\B{$\blacksquare$}), $\Rey\sim 10, \Wi\sim 10$, $d_b^{crit}=4.49$ mm; bubble images:  below (left)
$d_b=4.43$ mm and above (right), $d_b=4.61$ mm. {The fluid is an aqueous solution of 0.8 wt.\% polyacrylamide Praestol 2500, which is viscoelastic ($\lambda_1=0.21$ s) and shear thinning ($n\approx 0.5$).}
Experimental data from \citet{ravisankar2022}: ({$\bullet$}), $\Rey\sim 1, \Wi\sim 20$, $d_b^{crit}=4.54$ mm; ($\blacklozenge$) $\Rey\sim 0.5, \Wi\sim 10$, $d_b^{crit}=5.10$ mm; and (\B{$\times$}) $\Rey\sim 0.2, \Wi\sim 10$, $d_b^{crit}=5.29$ mm. {These fluids are aqueous solutions of polyacrylamide Separan AP30, with concentrations of 0.15, 0.20 and 0.25 wt.\%, with relaxation times ranging from $5<\lambda_1<39$ s and power indices $0.08<n<0.21$ (viscoelastic and shear thinning.)} The lines show $u_\infty \sim d_b^2$.}
\label{fig_terminalvelocity_VE}
\end{figure}

Figure \ref{fig_terminalvelocity_VE} shows the terminal velocity for an air bubble rising in different viscoelastic shear-thinning fluids. For bubble sizes smaller than a {certain} size, the bubble velocity increases {monotonically} with bubble diameter but at a rate {larger} than $d_b^2$, as observed in Newtonian fluids, depending on the fluid. This different trend is an indication of the non-linear nature of the drag force. Recall that the $d_b^2$ dependency of bubble speed was determined from the balance between a linear drag and buoyancy, hence, a deviation from such a trend indicates that $F_D$ is no longer linear with $d_b$ and/or $u_b$. {Note also that, as discussed above, the bubble shape becomes more elongated in the motion direction. \citet{soto2008} found that the drag coefficient can be reduced by up 40 \% only from the change of shape due to elastic effects. The different trend of $u_\infty$ with $d_b$ for small bubble sizes gives already an indication of the different behavior observed in viscoelastic fluids, but a detailed physical explanation is still lacking.} 

{However, the most outstanding feature in the data  shown in Fig. \ref{fig_terminalvelocity_VE} is the appearance of discontinuity of the bubble velocity. At a certain critical bubble diameter, $d_b^{crit}$, the bubble terminal velocity  increases sharply. For instance, for the data shown in Fig. \ref{fig_terminalvelocity_VE}, for the blue squares, the critical bubble diameter is $d_b^{crit}\approx 2.94$ mm. The terminal velocity jumps from 23.8 mm/s to 83.7 mm/s with a slight increase in diameter.} This phenomena, discovered by  \citet{astarita1965}, remained unexplained for years until recent studies offered insights \citet{fraggedakis2016} and \citet{bothe2022}.  The increase in bubble size leads to a change in bubble shape that has less viscous drag, which causing a slight increase in velocity. This triggers the shear-thinning nature of the fluid, which further boosts the bubble velocity which in turn induces further changes in the shape. These two effects, combined, with the reduction of the accumulation of surfactants for faster bubbles, leads to the significant increase in velocity \cite{Zenit2018}.
The data in Fig. \ref{fig_terminalvelocity_VE} shows that this phenomena is observed for moderate values of the $\Rey$ number and for $\Wi$ of order 1. {Interestingly, for bubbles beyond the critical size, the bubble velocity shows a dependence closer to $d_b^2$, suggesting that at larger diameters and higher speeds, viscoelastic effects become less significant. Despite these observations, a clear physical explanation of how elasticity influences bubble terminal velocity remains elusive.}

\begin{figure}[h]
\includegraphics[width=4in]{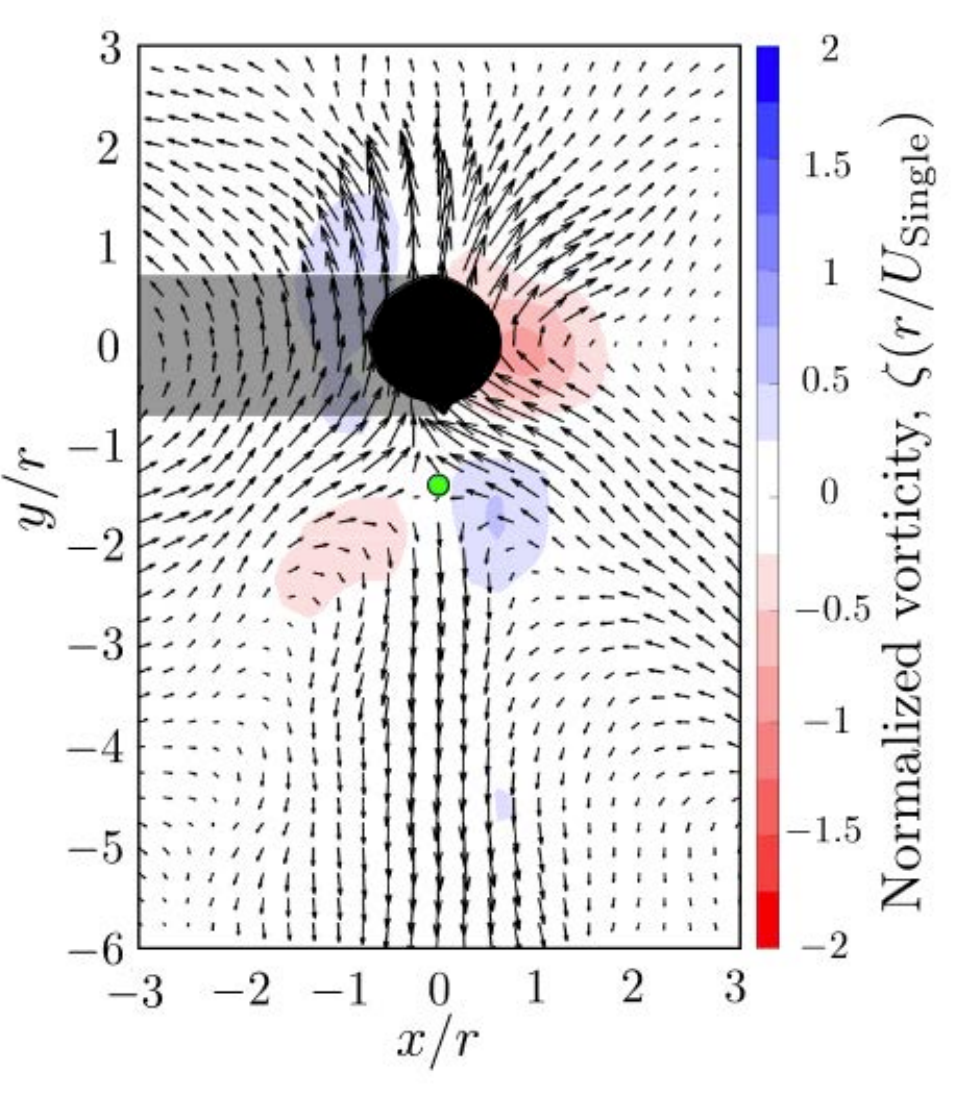}
\caption{Velocity field  behind a bubble rising in a viscoelastic, shear thinning fluid at $\Wi \sim 5$, {in the laboratory reference frame $\mathcal{R}$}. Colors represent vorticity, and the green dot shows the location of the stagnation point {between the forward-moving fluid and the reversed flow in the negative wake}. Taken with permission from \citet{ravisankar2022}.}
\label{fig_negative}
\end{figure}

An important feature of the motion of bubbles in viscoelastic media is the appearance of a negative wake. In viscoelastic fluids, stress accumulates and relaxes over time, rather than instantly as in Newtonian fluids. When a bubble traverses the liquid, it causes flow and continuous deformation which induce fluid stress. Unlike Newtonian fluids, where the fluid relaxes immediately once the bubble passes, the memory effect of the viscoelastic fluid causes the stress to remain present for a certain time, proportional to the elastic relaxation time, $\lambda_1$.  This phenomenon is illustrated in Fig. \ref{fig_negative}, showing the velocity field around a bubble rising in a viscoelastic fluid. In contrast to a Newtonian fluid, where the fluid follows the bubble, viscoelastic fluids exhibit flow reversal at a certain distance behind the bubble.  In other words, the fluid returns to the configuration that it had, before the passage of the bubble. This so-called negative wake was first reported by \citet{hassager1979} but has been studied by many since then \cite{Zenit2018}. This flow reversal can be observed in the wake behind particles, drops or bubbles in viscolastic liquids. However, due to the small bubble mass,  changes in the fluid stress have a significant impact on the bubble terminal velocity $u_\infty$.  \citet{herrera2003} argued that the negative wake contributed to the appearance of the bubble velocity discontinuity, because they appeared simultaneously.

\subsection{Hydrodynamic interactions among bubbles in complex fluids}
Given the complexity introduced by the non-Newtonian properties of the liquid on the motion and shape of bubbles, it is not surprising to find that the interactions among bubbles are significantly affected. Although in many practical applications the motion of bubbles in non-Newtonian liquids is prevalent, a general understanding of bubble clustering and induced mixing is not currently available.  {Some of the issues in understanding bubble-bubble pair hydrodynamic interactions in non-Newtonian liquids have been addressed by Zenit and co-workers \cite{velez2011,velez2011b,velez2012,ravisankar2022}, but a mature level of understanding has not been reached.} 
\begin{figure}[h]
\includegraphics[width=6.5in]{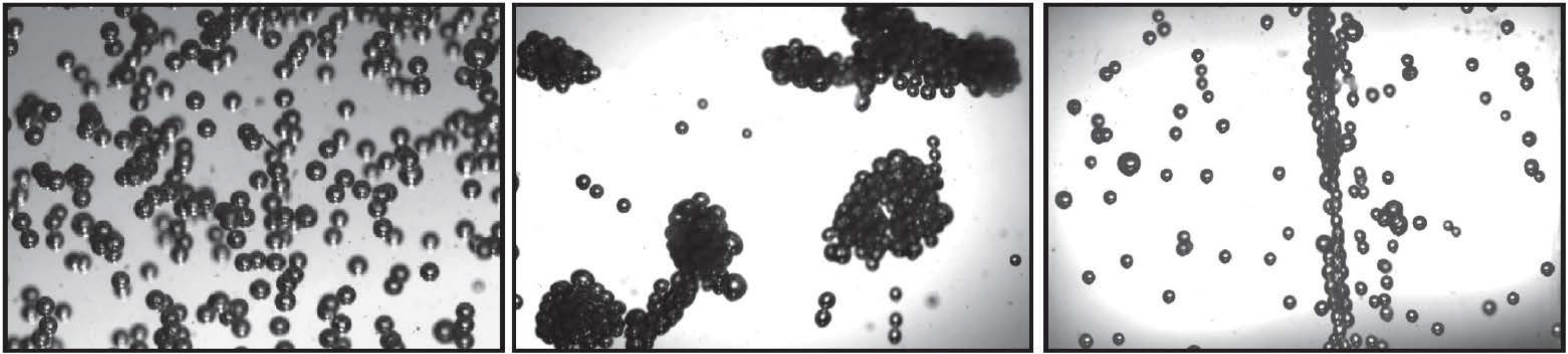}
\caption{Clusters in non-Newtonian bubbly flows: (left) Newtonian liquid; (center), shear-thinning liquid; (right) viscoelastic liquid. The three images are for the same gas volume fraction, $\alpha_G=0.05$. Adapted with permission from \cite{Zenit2018}.
}
\label{fig_bubblyflow_NN}
\end{figure}

In general, shear-thinning and viscoelastic effects result in interaction forces that promote the clustering and agglomeration of bubbles, as shown in Fig. \ref{fig_bubblyflow_NN}. The low-viscosity region around a bubble in shear-thinning fluids can serve to capture neighboring bubbles. The elastic stresses in the equatorial region of bubbles, where shear dominates, can attract bubbles to form clusters. Clustering leads to coalescence, which, in turn, leads to the formation of large bubbles that produce regime changes \cite{Zenit2018} and result in velocities and mixing of different magnitudes than those expected for dispersed bubbly liquids \cite{risso2018agitation}. 

\section{Equation of bubble motion, viscous relaxation time, Stokes number, and  history force}

\subsection{{Equation of bubble} motion\label{sec_traj}}

{Following the work of \citet{maxey1983, gatignol1983, Thomas1983, auton1988}, we can now rewrite Eq. \ref{eq_motion}, as complete equation of motion for a} spherical bubble moving at the velocity $\mathbf{u_b}=d \mathbf{x_b}/dt $ in a Newtonian liquid with a velocity $\mathbf{U}(\mathbf{x},t)$:
\begin{eqnarray}\label{eq_traj_2_1}
 C_M \rho \vartheta_b \frac{d \mathbf{u_b}}{dt} &= & -\rho \vartheta_b \mathbf{g} \, \quad \quad \quad \quad \quad \quad  \quad\quad \quad \quad \quad \quad \text{Buoyancy} \\\label{eq_traj_2_2}
&+& 2 \pi \mu d_b \, \mathcal{K}(Re) \, (\mathbf{U} - \mathbf{u_b})\quad \quad \quad \quad  \quad\text{Drag}  \\\label{eq_traj_3}
&+& \rho  \vartheta_b (1+C_M)  \left(\frac{\partial \mathbf{U}}{\partial t}  + \mathbf{U}.\nabla \mathbf{U} \right)_{\mathbf{x_b}} \,  \,  \,  \quad\text{Inertial \& Added mass}  \\\label{eq_traj_2_4}
&+& C_L \rho \vartheta_b  (\mathbf{U} - \mathbf{u_b}) \times \Omega. \quad \,  \, \quad \quad \quad \quad  \quad \text{Lift} 
\end{eqnarray}
{This force decomposition extends the expression derived by \citet{auton1988} for the case of a rigid sphere in an unsteady inviscid rotational flow. Its validity has been confirmed by numerical studies as discussed in  \citet{magnaudet2000}.}

We can analyze each of the contributions to  this equation. {The bubble acceleration of the added mass (on the left-hand side) is balanced by the forces (on the right-hand side, from top to bottom)}: the {buoyancy force $\mathbf{F_B}$ reduced to the} Archimedes force, the drag force {$\mathbf{F_D}$}, the generalized inertia force due to the time and space acceleration of the fluid and, finally,  the lift force {$\mathbf{F_L}$}. Not included in this equation, because negligible in most cases, are the bubble's own acceleration, its weight and the history force {$\mathbf{F_H}$} (discussed below).
{This equation is usually rewritten by dividing all the terms by  $C_M \rho \vartheta_b$ to express the bubble acceleration $ \mathbf{a_b}$ as
\begin{equation}\label{eq_traj_tau}
\mathbf{a_b} = \frac{d \mathbf{u_b}}{dt} =  \frac{1}{\tau_b} (\mathbf{U} - \mathbf{u_b})
+ \gamma   \left(\frac{\partial \mathbf{U}}{\partial t}  + \mathbf{U}.\nabla \mathbf{U} \right)_{\mathbf{x_b}}
+ \gamma^\prime (\mathbf{U} - \mathbf{u_b}) \times \Omega
+\gamma^{\prime\prime} \mathbf{g}
\end{equation}
where we have introduced the characteristic, relaxation or  response time of the bubble
\begin{equation}\label{eq_taub}
\tau_b= \frac{C_M d_b^2}{12 \nu \mathcal{K}(Re)},
\end{equation}
as well as the ratios  $\gamma=(1+C_M)/C_M$, $\gamma^\prime=C_L/C_M$ and $\gamma^{\prime\prime}=-1/C_M$ that correspond to the mass accelerated by the fluid, the mass involved in the lift effect and the mass accelerated by gravity, respectively, compared to the mass accelerated by the bubble motion. For spherical bubbles  $\gamma=3$,  $\gamma^\prime\approx 1$ for $\Rey \gg 1$ and $\gamma^{\prime\prime}=-2$. Note that {these coefficients differ from the ones considered} for heavy particles of density $\rho_p$,  $\gamma=\rho (1+C_M)/(\rho_p+C_M \rho)$, $\gamma^\prime=\rho C_L/(\rho_p+C_M \rho)$ and $\gamma^{\prime\prime}=(\rho_p - \rho)/(\rho_p+C_M \rho)$, resulting in different behaviors.}

Equations \ref{eq_traj_2_1}-\ref{eq_traj_2_4} (or \ref{eq_traj_tau}) with the appropriate  expressions for the drag $\Rey$-correction, $\mathcal{K}(\Rey)$, the added mass coefficient $C_M$ and the lift coefficient $C_L(\Rey, Sr)$ are commonly used to investigate turbulent dispersed bubbly flows \cite{Mazzitelli2003, Sugiyama2008, Chouippe2014, Mathai2016, Muniz2020, Ruth2021}. {Numerical studies based on Lagrangian tracking of bubbles are based on the point-force assumption (point bubble) that formally  requires $d_b \ll \eta$, where  $\eta$ is the Kolmogorov length scale. When this condition is not satisfied, Eqs. \ref{eq_traj_2_1}- \ref{eq_traj_2_4} need to include Fax\'en terms both in the drag force (Eq. \ref{eq_traj_2_2}) and in the inertia-added mass force (Eq. \ref{eq_traj_2_4}) as discussed by \citet{Homann2010}.}

\subsection{Viscous relaxation time and Stokes number}

\begin{figure}[h]
\includegraphics[width=6.5in]{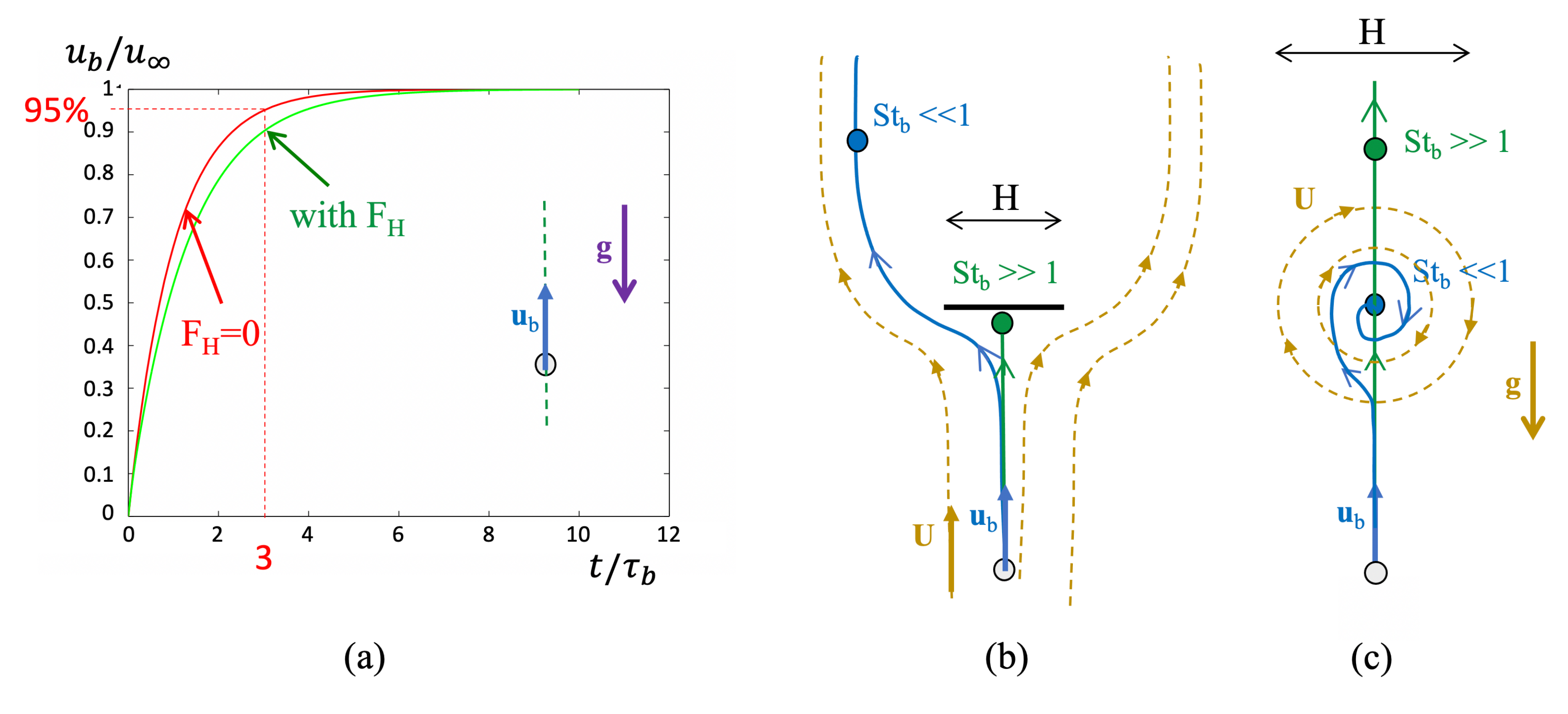}
\caption{(a) Bubble velocity evolution when released from rest. {The green line shows the effect of the history force when integrated in Eq. \ref{eq_traj_3_tau}}. (b) Bubble trajectory when varying the Stokes number $St_b$ when transported by a vertical flow. (c) Similar to (b) but for a bubble interacting with a vortex structure.}
\label{fig_bubble_St}
\end{figure}

Consider a spherical bubble placed at the bottom of a tank filled with a fluid at rest ($\mathbf{U} =0$), with no initial velocity. As discussed above, the bubble reaches its terminal velocity $u_\infty$ when the buoyancy force is balanced by the drag force. Let us consider the transient evolution of the bubble motion before reaching the terminal velocity.
Considering $\mathbf{U}=0$ in {Eq. \ref{eq_traj_tau}, grouping all the terms in $\mathbf{u_b}$ on the left side and considering} $C_M=1/2$, the equation can be simplified as:
\begin{equation}\label{eq_traj_3_tau}
\frac{d \mathbf{u_b}}{dt} + \frac{\mathbf{u_b}}{\tau_b} = -2 \mathbf{g} 
\end{equation}
{where {we see that the viscous relaxation time $\tau_b= {d_b^2}/{24 \nu \mathcal{K}(Re)}$  determines}} the transient evolution of the velocity. Note that the initial acceleration of the bubble is twice that of gravity. 
This  equation can be readily solved when $\tau_b$ is constant which is the case when $\Rey \rightarrow 0$ ($\mathcal{K}=1$) or when $\Rey \rightarrow \infty$ ($\mathcal{K}=3$). In such cases,  the evolution of the velocity is 
\begin{equation}
\mathbf{u_b}=\mathbf{u_\infty} \left[1- \exp \left(-\frac{t}{\tau_b}\right) \right]
\end{equation}
with {$\mathbf{u_\infty} = - 2 \tau_b  \mathbf{g}  = -d_b^2 \mathbf{g} / 12 \nu \mathcal{K}(Re)$}. The evolution of $\mathbf{u_b}$  with time is shown {in} Fig. \ref{fig_bubble_St}(a). The bubble reaches its terminal velocity {(95\%)} at about $t=3 \tau_b$.

The  bubble relaxation time is $\tau_b= 4 \times 10^{-4}$ s for 
a bubble with a diameter $d_b=0.1$ mm in water ($u_\infty \approx 0.01$ m/s and $Re\approx 1$). For a bubble ten times larger, $d_b=1$ mm, also in water {($u_\infty \approx 0.27$ m/s and $Re\approx 270$)}, the relaxation time increases to {$\tau_b= 1.4 \times 10^{-2}$ s}.  As a consequence,  the transient phase of the bubble motion can be neglected in most situations compared to the bubble resident time in the system. For example, for a tank of height of $h=1$ m, the resident time,  $T \approx h/u_\infty$, would be  100 s and 3 s, respectively, for the two bubble sizes considered above.

In a situation when, the liquid is moving at a uniform velocity $\mathbf{U}$, the equation of motion is
\begin{equation}\label{eq_traj_4}
\frac{d \mathbf{u_b}}{dt} + \frac{\mathbf{u_b}}{\tau_b} = -2 \mathbf{g} + \frac{\mathbf{U}}{\tau_b}.
\end{equation}
The bubble system is now subject to external forcing from two contributions: the gravity $\mathbf{g}$ and the liquid velocity $\mathbf{U}$, with the same relaxation time $\tau_b$. The final bubble velocity is then $\mathbf{u_b}=\mathbf{u_\infty} + \mathbf{U}$ and the transient phase is also described by $(1- \exp \left(-{t}/{\tau_b}\right))$. The bubble  relaxation time $\tau_b$ provides a characteristic time for which a bubble can respond to any external forcing or change in its environment (gravity and liquid motion). It is thus  used to describe how  bubbles react to various flow situation. Figures \ref{fig_bubble_St}(b) and (c) illustrate two possible situations. In Fig. \ref{fig_bubble_St}(b), the bubble is rising in a vertical upward flow so that $U \gg u_\infty$. The flow is moving around a horizontal plate.  In Fig. \ref{fig_bubble_St}(c), the bubble interacts with a vortex as it rises. In both cases two possible situations can be observed: the bubble follows the flow streamlines (as it moves around the plate in (b) or gets trapped by the vortex in (c)) or it is not influenced by the flow (it impacts the wall in (b) and  crosses  the vortex in (c)). The answer can be determined by comparing the bubble relaxation time to the characteristic time of the flow $\tau_f$ in each situation. In both cases, the characteristic flow time can be  $\tau_f \approx H/U$. A dimensionless number, the Stokes number, can thus be defined as
\begin{equation}\label{eq_traj_5}
\St_b = \frac{\tau_b}{\tau_f}.
\end{equation}
When $\text{St}_b\ll 1$ the bubble is fast to adapt to any fluid modification and is able to follow the flow {like a tracer}, while when $\St_b\gg 1$ the bubble is slow to react and follows its original trajectory, thus impacting the wall in  Fig. \ref{fig_bubble_St}(b) or moving across the vortex in Fig. \ref{fig_bubble_St}(c).  The Stokes number is, for instance, used to describe how bubbles are influenced by turbulence  {as detailed in Section \ref{Section_bubble_turb}}.

\subsection{History force}

Figure \ref{fig_bubble_St}(a) shows the bubble rising velocity {with and without considering the history force $\mathbf{F_H}$} in Eq. \ref{eq_traj_2_1}-\ref{eq_traj_2_4} {(and in Eq. \ref{eq_traj_3_tau})}. The  history force $\mathbf{F_H}$ corrects the steady drag force; that is, when the viscous effects {do not have enough time} to establish the flow field around the bubble when compared to the time scale of the flow. The history force can be calculated {under Stokes flow conditions ($\Rey \ll 1)$} as \cite{Gorodtsov1975, Yang1991}
\begin{equation}
\mathbf{F_H}= 8 \pi \mu R \int_{0}^{t}{   \exp \left[9 \frac{t-t^{\prime}}{t_{\nu}}\right] \text{erfc}\left[3 \sqrt{\frac{t-t^{\prime}}{t_{\nu}}}\right] \left(\frac{d \mathbf{U}}{d t^{\prime}} - \frac{d \mathbf{u_b}}{d t^{\prime}} \right)  d t^{\prime}}
 \label{eq_histoire}
\end{equation}
where $t_{\nu}= R^2/\nu$ is the characteristic diffusion time of momentum in the liquid. 
{The kernel $K(t-t^{\prime})$ for a spherical bubble evolves as $K(t-t^{\prime}) \approx \exp \left[9 (t-t^{\prime})/{t_{\nu}}\right] \text{erfc}\left[3 \sqrt{{(t-t^{\prime})}/{t_{\nu}}}\right]$ resulting in a finite contribution to the total force at short times, in contrast with the Basset-Boussinesq  kernel for solid sphere which diverges as $\sqrt{t_{\nu}/(t-t^{\prime})}$, resulting in less important history effects for spherical bubbles when compared to solid spheres}. In general, $F_H$ can be neglected for the case of a bubble when compared to the steady drag force \cite{magnaudet1995}. The contribution of the history force is expected to be maximum at small Reynolds number. As shown in Fig.  \ref{fig_bubble_St}(a), when integrated to the equation of motion for a rising bubble, the history force does not have a noticeable effect  on the transient motion. However, in  flows oscillating at high frequency, the history force should be added in the trajectory equation.
{Some experiments have evidenced the role of the history force. 
For example, the history force has been shown to contribute to the slow migration of bubbles rising in a vortex \cite{Candelier2005}. 
For oscillating, collapsing or growing bubbles in motion, the history force can be important. The correction induced by the volume variation has been derived by \citet{Magnaudet1998} and the importance of the history force in such condition  was reported in experiments on 
microbubbles propelled by acoustic radiation force \cite{Garbin2009} and in experiments on sonoluminescing bubbles trapped in standing sound waves \cite{Toegel2006}}.

\section{Bubble dynamics in turbulence}
\label{Section_bubble_turb}

\begin{figure}[h]
\includegraphics[width=6.5in]{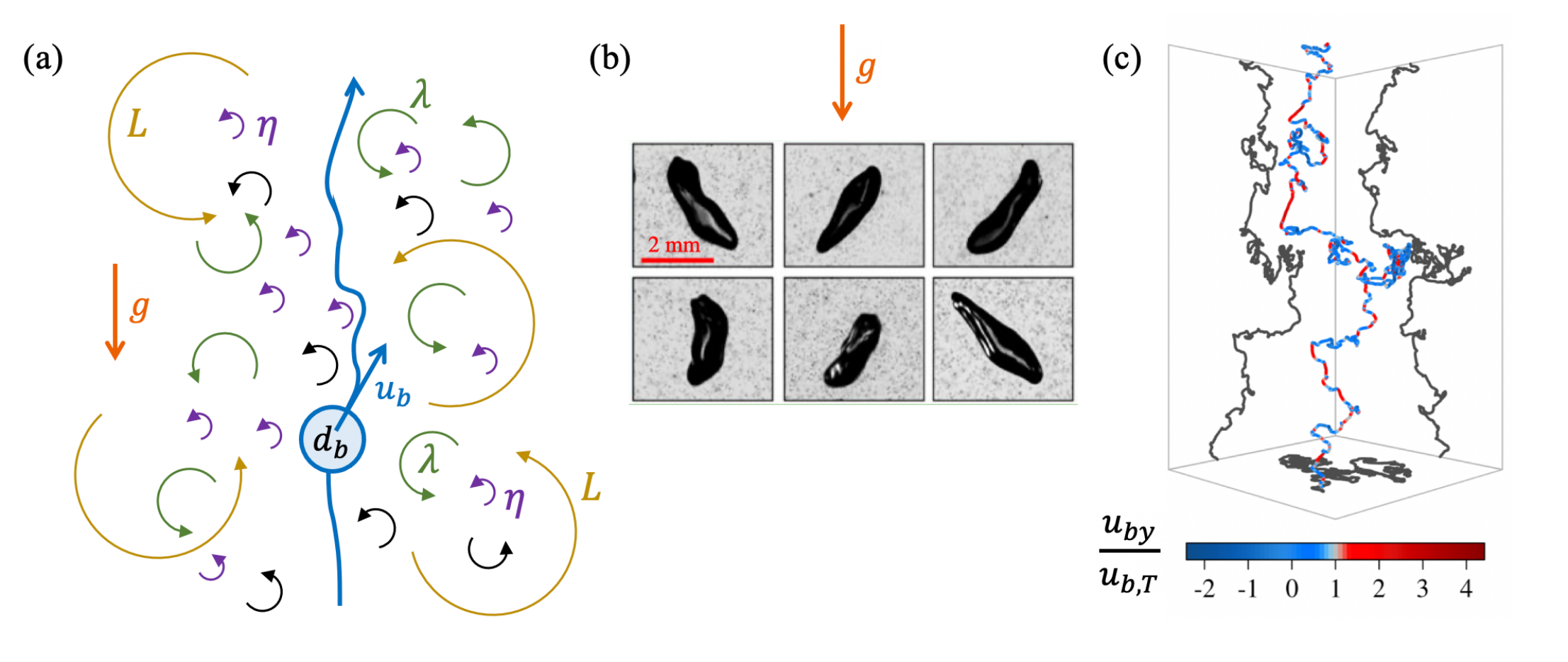}
\caption{A bubble rising in a turbulent field. (a) The turbulent field Consists of  eddies of various length scales (adapted from  \cite{Mathai2020}).  The bubble's  size $d_b$ is in the inertial subrange ($\eta < d_b < L$) where $\eta$ is the Kolomogorov length scale and $L$ is the integral length scale. $\lambda$ is the Taylor microscale. (b) Large bubble shape on six views  during its rise in an intense homogeneous and isotropic turbulence $u^\prime= 0.25$m/s, $L=3.2$cm and $\eta=38\mu$m \cite{Salibindla2020}. (c) Trajectory of a bubble, obtained by direct numerical simulation, rising in a homogeneous isotropic turbulence for $\beta=0.9$, $\eta/d_b = 0.098$, $\lambda/d_b = 1.0$, and  $L/d_b$ = 2.1. The color code indicates the instantaneous bubble vertical velocity $u_{by}$ normalized by the terminal velocity of the same bubble, $u_\infty$, in a quiescent liquid \cite{Loisy2017}.}
\label{fig_bubble_turbulence_2}
\end{figure}

A bubble  moving  through a turbulent flow interacts  with vortices (or eddies) of various length, time, velocity and acceleration scales resulting in significant bubble deformation and chaotic trajectories, as illustrated in Fig. \ref{fig_bubble_turbulence_2}. There are several  length scales relevant to the interaction of a bubble with turbulence:  $\eta$ the dissipative Kolmogorov length scale, $\lambda$, the intermediate Taylor microscale and $L$ the integral length scale of the flow. The Kolmogorov length  and  time scales are  determined by $\eta = (\nu^3/\varepsilon)^{1/4}$ and $\tau_\eta= (\nu/\varepsilon)^{1/2}$, respectively, where $\varepsilon$ is the energy dissipation rate. In the following discussion, $\ell$ represents a characteristic size of a turbulent eddy. In general, in the inertial subrange we have that $\eta \ll \ell \ll \lambda$. We also introduce  the fluctuating velocity intensity, $u^\prime$, which is the root mean square value of the free stream velocity. Using  $u^\prime$ we can define the ratio $\beta =u^\prime/u_{\infty}$ that compares $u^\prime$ with the bubble rising velocity $u_{\infty}$ without turbulence. 

\begin{figure}[h]
\includegraphics[width=5in]{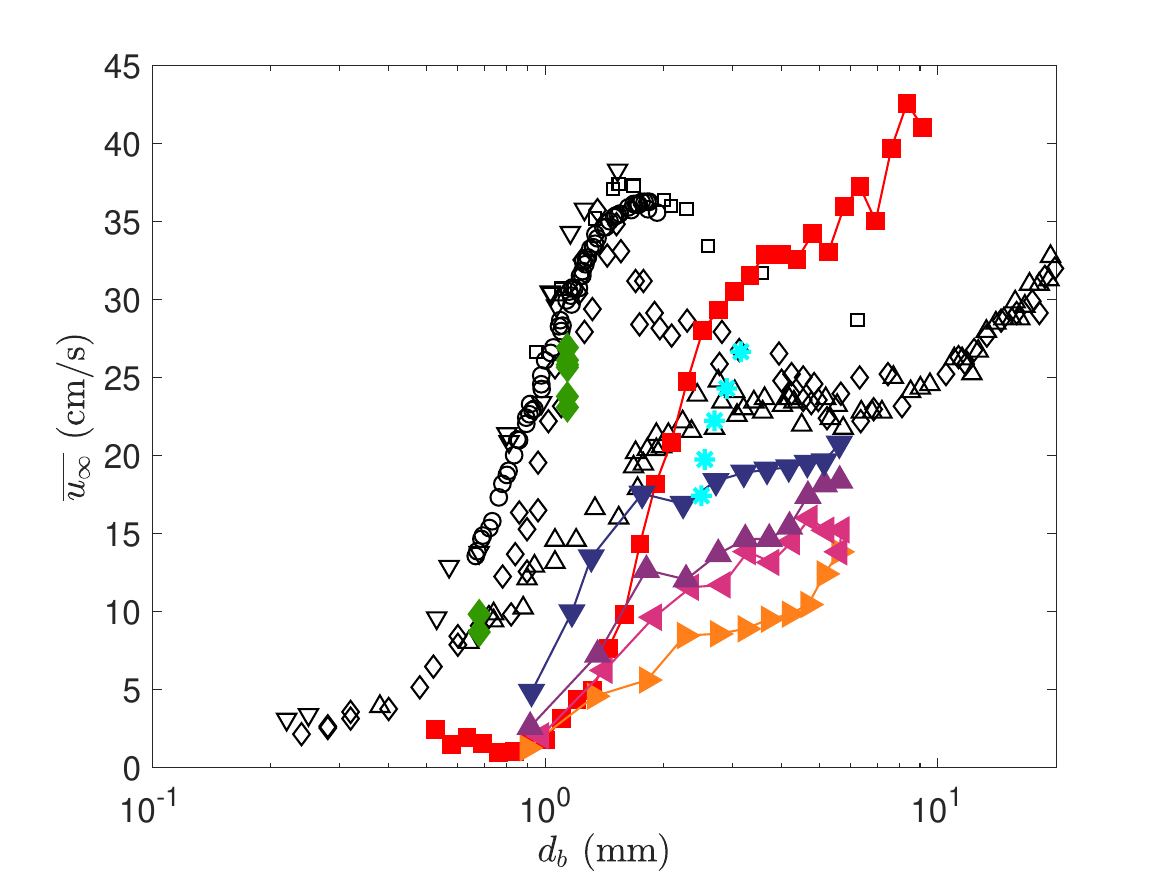}
\caption{Mean terminal velocity $\overline{u_\infty}$ of  air bubbles rising in water for different levels of turbulence $u^\prime$ as a function of their equivalent diameter $d_b$. The empty black markers show the experiments conducted in a quiescent water, shown in Fig. \ref{fig_1_new} {($u'=0$, no turbulence)}. The filled and color markers show experimental data from the literature. Clean bubbles: ({\textcolor[rgb]{0.0, 0.6, 0.0}{$\blacklozenge$}}) 0.03 m/s $< u^\prime \le$ 0.06 m/s \cite{Poorte2002}. Contaminated bubbles: ({\color{cyan}{$\ast$}}) \cite{Prakash2012}; (\R{$\blacksquare$}) $u^\prime=0.25$ m/s \cite{Salibindla2020};
({\textcolor[rgb]{0.15, 0.24, 0.52}{$\filledmedtriangledown$}}) 0.02 m/s $< u^\prime \leq$ 0.05 m/s,
({\textcolor[rgb]{0.4, 0.0, 0.3}{$\filledmedtriangleup$}}) 0.05 m/s $< u^\prime \leq$ 0.10 m/s, 
({\textcolor[rgb]{0.8, 0.0, 0.4}{$\filledmedtriangleleft$}})  0.10 m/s $< u^\prime \leq$ 0.15 m/s and  
({\textcolor[rgb]{1,0.5,0}{$\filledmedtriangleright$}})  0.15 m/s $< u^\prime \leq$ 0.20 m/s \cite{Ruth2021}. }
\label{fig_1_turb}
\end{figure}

The evolution of the mean rising velocity  $\overline{u_\infty}$ of an air bubble  in water is reported in Fig. \ref{fig_1_turb} as a function of its diameter $d_b$ for different levels of turbulence  $u^\prime$. We can compare these speeds with the free bubble rising velocity in a quiescent fluid, see Fig. \ref{fig_1_new}. Clearly, the turbulence has a significant impact on the bubble mean rising velocity revealing a strong interaction with the turbulence. The trends also seem differ for various  experimental investigations. For bubbles with a clean interface and  $ u^\prime$  ranging from 0.03 to 0.06 m/s, \citet{Poorte2002} report a decrease of the bubble mean rise velocity up to 35\% compared with the quiescent conditions. For $ u^\prime$ ranging from 0.02 to 0.2 m/s,  \citet{Ruth2021} report a significant decrease in the mean rising velocity with turbulence and  $\overline{u_\infty}$ is always found to be smaller than  the rise in quiescent water. This trend is also reported by  \citet{Salibindla2020} for bubbles smaller than $d_b=2$ mm,  whereas for larger diameters, the authors observed that bubbles actually rise faster than in quiescent water. Most of the experiments report a strong effect for small bubbles, typically less than 1 mm. Such bubbles appear to be significantly slowed down in the turbulence, their rising velocity being reduced by more than one order of magnitude. 
 
We first describe the different mechanisms frequently used to explain  bubble behavior when  rising in turbulence. {This discussion is limited to the case where the turbulence is not strong enough to cause bubble rupture, as discussed in Section \ref{Sec_bubble_rupture}.} Depending on the dominant mechanisms, different behavior can be observed resulting in bubble capture, reduction or increase of the rising velocity.
We then discuss the main results concerning small and large bubbles relative to the smallest turbulent scale $\eta$. We focus on the mean bubble rise velocity $\overline{u_\infty}$. Information on bubble velocity and acceleration fluctuations can be found in \citet{Zhang2019, Loisy2017}. The probability density function (PDF) of horizontal velocity fluctuations has been reported to be Gaussian, while for the vertical fluctuations the distribution shows small departure from Gaussianity \cite{Poorte2002, Prakash2012}. The PDFs of bubble acceleration are highly non-Gaussian and exhibit large tails. We also refer the reader to  \citet{Mathai2020} for a recent review on bubbly and buoyant particle-laden turbulent flows.

\subsection{{Main mechanisms of interaction with turbulence}}
\begin{figure}[h]
\includegraphics[width=6.5in]{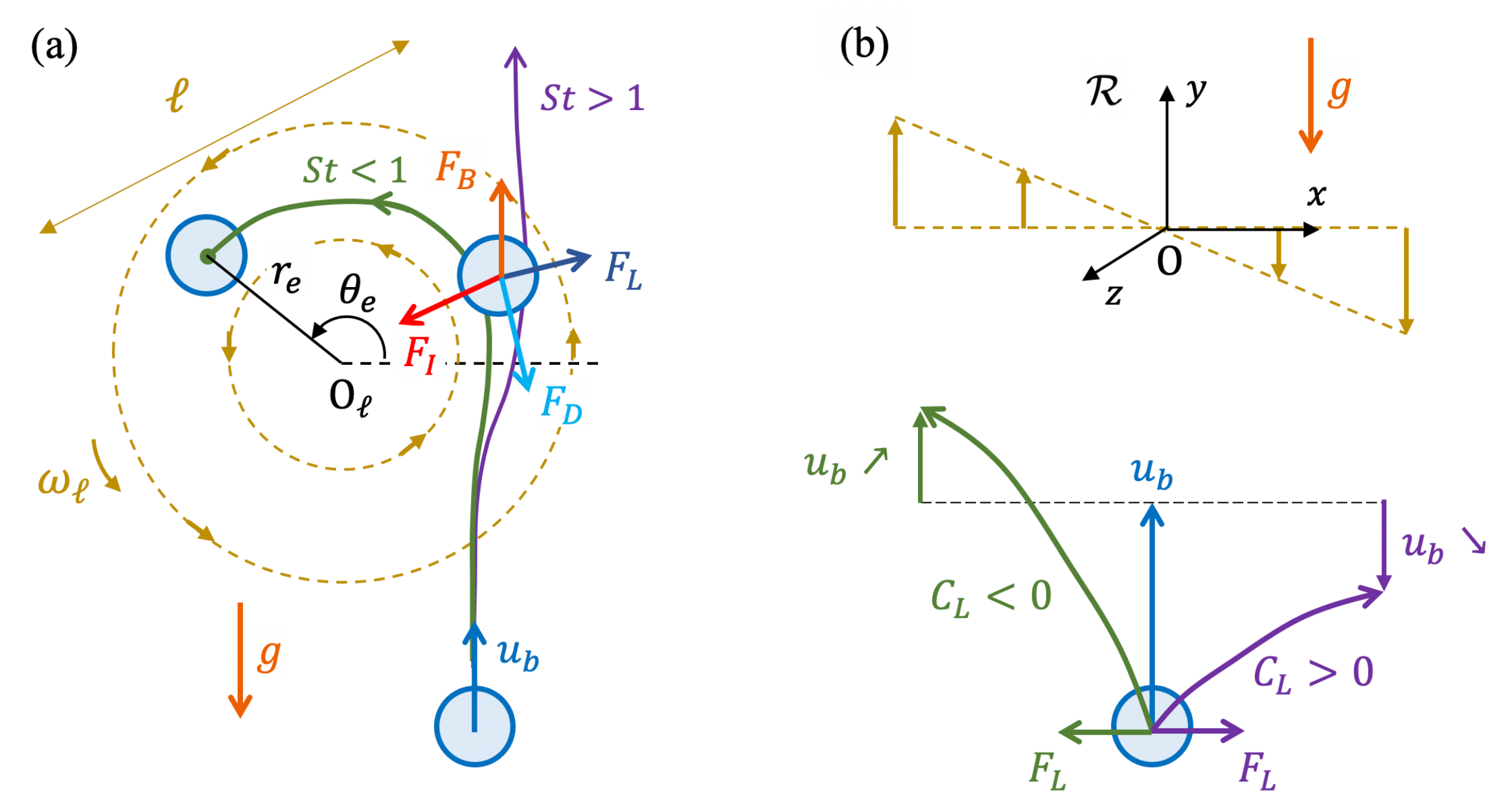}
\caption{Bubble interaction with turbulence. (a) Bubble interaction in a vortex: a schematic representation of a turbulence eddy, adapted from  \citet{Mathai2020}. (b) Bubble rising through a velocity gradient, adapted from  \citet{Spelt1997}. }
\label{fig_bubble_turbulence_0}
\end{figure}

Two mechanisms  are often considered to explain how the bubble motion   is  affected by the turbulence. These mechanisms are illustrated in   Fig. \ref{fig_bubble_turbulence_0}: bubbles can be trapped inside vortices \cite{Wang1993} or drifted toward downflow regions due to lift force effects \cite{Spelt1997}.

We first consider a bubble rising through a steady vortex of size $\ell$ in the vertical $(x,y)$-plane with a constant angular velocity $\omega_\ell$, as shown in Fig. \ref{fig_bubble_turbulence_0}(a). Cylindrical coordinates $(\mathbf{e_r}, \mathbf{e_\theta}, \mathbf{e_z})$, {with its origin at the vortex center O$_\ell$}, are used for convenience to express the forces. The velocity field is given by $\mathbf{U} =  \omega_\ell r \mathbf{e_\theta}$ and the vorticity field  is $\mathbf{\Omega} = 2 \omega_\ell \mathbf{e_z}$. According to Eqs. \ref{eq_traj_2_1}-\ref{eq_traj_2_4}, during its interaction with the vortex the bubble, moving at velocity $\mathbf{u_b} = u_{br}\mathbf{e_r} +u_{b\theta} \mathbf{e_\theta}$, is subject to the drag force $F_D = 2 \pi \mu d_b \, \mathcal{K}(Re)  \left[  - u_{br}\mathbf{e_r} + ( \omega_\ell r - u_{b\theta} ) \mathbf{e_\theta} \right] $, the inertia force $F_I = -\rho \vartheta_b (1+C_M) \omega_\ell^2 r \mathbf{e_r}$, the lift force  $F_L = 2 \rho \vartheta_b C_L \omega_\ell \left[ ( \omega_\ell r - u_{b\theta} ) \mathbf{e_r} + u_{br} \mathbf{e_\theta} \right]$ and the buoyancy force $\mathbf{g} = - g \mathbf{e_y}$.  Considering an equilibrium position $(r_e,  \theta_e)$ at which ($u_{br} =u_{b\theta}=0$), the force balance results in
\begin{equation}\label{turb_eq}
\sin \theta_e =  \left[1+C_M - 2 C_L\right]  \frac{\omega_\ell^2 r_e}{g} ; \quad \cos \theta_e = - \frac{\omega_\ell r_e}{u_\infty} 
\end{equation}
This relation indicates that $\cos \theta_e$ is always negative, thus, the equilibrium position is located  in the  left half of the vortex depicted in Fig. \ref{fig_bubble_turbulence_0}(a) where the vertical component of the fluid velocity is negative. The sign of $\sin \theta_e$ depends on the sign of $1+C_M - 2 C_L$. Typically for a spherical bubble at $Re>100$,  $C_L \approx C_M =0.5$, so that $1+C_M - 2 C_L \approx 0.5 $ is positive and, therefore, the stable position is then in the top left quadrant. 

Expression (\ref{turb_eq}) indicates that: (i) a bubble can be trapped in a vortex and (ii) once trapped, the bubble can experience a downward fluid velocity component thus reducing its vertical rise. 
As shown in the figure, both the inertia-added mass force and the lift effect are of importance. The inertia term is the driving force toward the center of the vortex. Note that in the absence of gravity and lift effects, bubbles are systematically trapped at the vortex center corresponding to a stable position at $r_e=0$. 

In general, an equilibrium radius $r_e$ can be found for any situation but in practice, bubbles are interacting with vortex of finite size $\ell$. Their time of interaction with the vortex should be larger than {both the vortex life time and} the bubble relaxation time $\tau_b$ to make the capture possible.
This provides the conditions for the vortex to capture a bubble of a certain size: ${\left[1+C_M - 2 C_L\right] \omega_\ell^2}  \ell/{g} \le 1$ and   ${\omega_\ell  \ell }/{u_\infty}\le 1$. A life time condition ($\tau_\ell \ge \tau_b$) can be estimated using  the eddy turnover time $\tau_\ell=1/\omega_\ell$.
{For eddies of characteristic size $\ell$ in the inertial subrange, $\omega_\ell\approx (\varepsilon/\ell^2)^{1/3}$}. 
From these conditions, we can define a Stokes number $\St_\ell = \tau_b \omega_\ell$ and a Froude number $\Fr_\ell= \omega_\ell^2 \ell/2 g$ with the following conditions for the capture of the bubble, {or at least} a significant interaction between the bubble and the turbulence eddy of size $\ell$:
\begin{equation}\label{turb_capt}
Fr_\ell  \le  St_\ell \le 1
\end{equation}

The second mechanism explainging the bubble motion in turbulence is related to the bubble drift when they rise through a velocity gradient as illustrated in Fig. \ref{fig_bubble_turbulence_0}(b).  \citet{Spelt1997} introduced this idea to explain the reduction of bubble rise and the important role of the lift force in bubble interaction with turbulence. This argument has been recently reused by \citet{Salibindla2020} to explain bubble rise velocity increase when the deformation is enough to induce lift reversal. 
In the scenario depicted in Fig. \ref{fig_bubble_turbulence_0}(b), consider a bubble rising at its terminal velocity $\mathbf{u_b}= u_\infty \mathbf{e_y}$ through a vertical shear flow $\mathbf{U}= - \omega x \mathbf{e_y}$. The bubble experiences a horizontal lift force $F_L= \rho \vartheta_b C_L \omega u_\infty  \mathbf{e_x}$ imposed by the vorticity field  $\mathbf{\Omega}= - \omega \mathbf{e_z}$.
For $C_L >0$ the bubble drifts toward the downward velocities, thus reducing its vertical rise, while when $C_L <0$ the bubble is pushed toward the upward velocities thus increasing its rising speed.

\subsection{{Small bubbles in turbulence}}
\begin{figure}[h]
\includegraphics[width=6.2in]{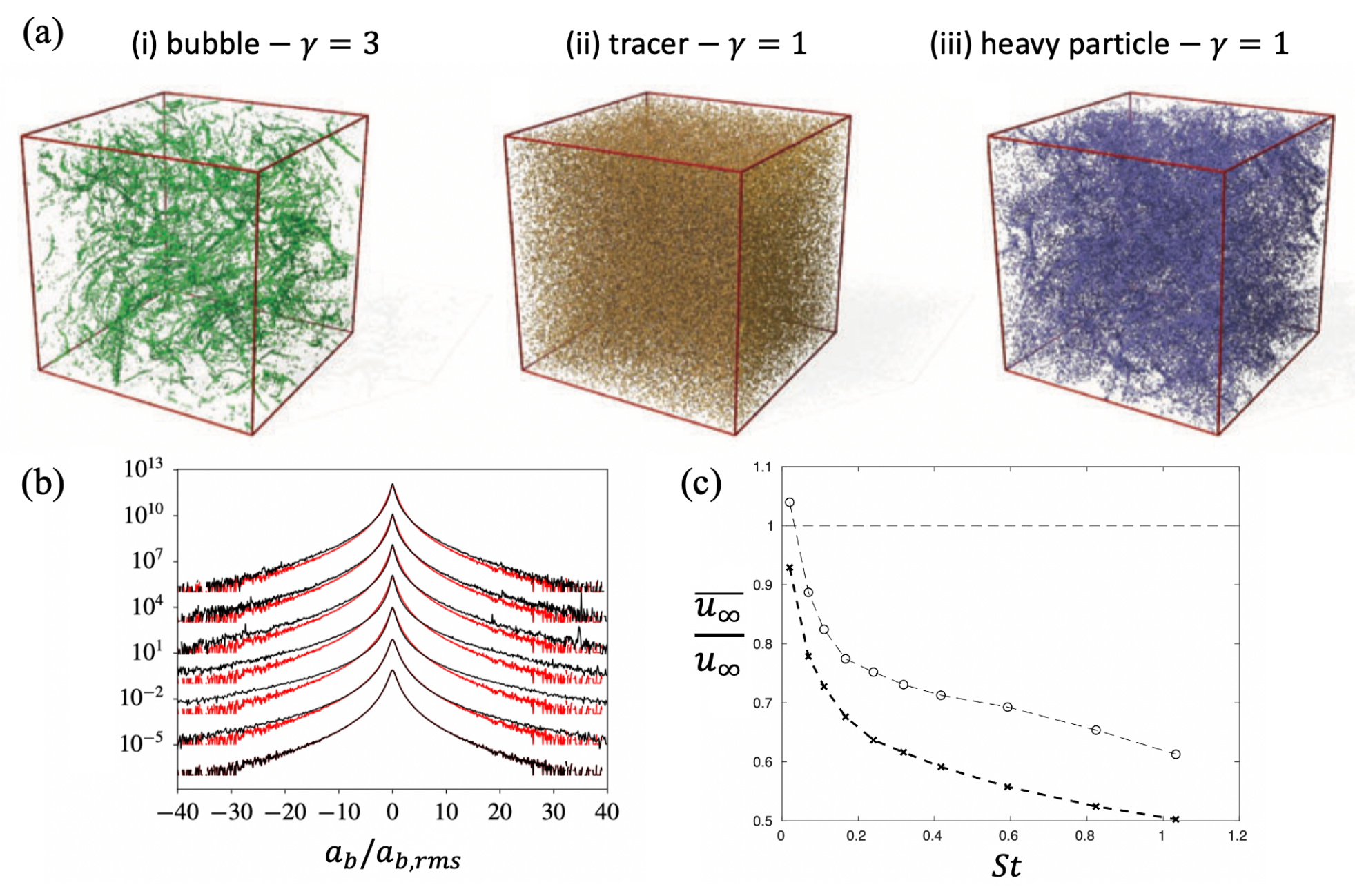}
\caption{Interaction of small bubbles with turbulence. (a) Snapshots of particle distributions in a turbulent flow field at $\St = 0.6$ for (i) $ \gamma= 3$ (bubbles), (ii) $\gamma= 1$ (tracers), and (iii) $\gamma = 0$ (heavy particles) \cite{Toschi2009}. (b) The PDF of the bubble acceleration normalized by its root-mean-square value in black for $\St$ = 0.02, 0.074, 0.20, 0.45, 1.01, 1.55 and 2.07 (shifted upward by two decades from each other for clarity, respectively) and comparison with the PDF of the acceleration of fluid tracers in red \cite{zhang2019model} in a homogeneous and isotropic turbulence at $\Rey_\lambda = 100$. (c) The average value of the rising velocity normalized by the terminal velocity \cite{Zhang2019} with predictions from Eq. \ref{Eq_kolmo} without lift force ($\circ$) and with lift force ($\times$).}
\label{fig_bubble_turbulence1}
\end{figure}

First we consider small bubbles compared to the turbulent eddies, i.e.,  when bubble size is smaller than the Kolmogorov scale of turbulence ($d_b < \eta$). This case has received significant attention in the literature  \cite{Mathai2020}. In such a situation, bubbles interact with all the scales of the flow and Eqs. \ref{eq_traj_2_1}-\ref{eq_traj_2_4} are usually normalized by using  the Kolmogorov units of length $\eta$ and time $\tau_\eta$. Considering $\mathbf{g} = - g \mathbf{e_y}$ Eq. \ref{eq_traj_2_1}-\ref{eq_traj_2_4} can be written as:
\begin{equation}\label{Eq_kolmo}
\mathbf{a_b^*}= \frac{d \mathbf{u_b^*}}{dt^*} =  \frac{1}{\St_\eta} (\mathbf{U^*} - \mathbf{u_b^*})   + \gamma  \left(\frac{\partial \mathbf{U^*}}{\partial t^*}  +  \mathbf{U^*}.\nabla^* \mathbf{U^*} \right)_{\mathbf{x_b^*}}  +  \gamma^\prime  (\mathbf{U^*} - \mathbf{u_b^*}) \times \Omega^* + \frac{1}{\Fr_\eta} \mathbf{e_y} 
\end{equation}
where the superscript $^*$ denotes the new dimensionless quantities. The parameters $\gamma=(1+C_M)/C_M$ and $\gamma^\prime = C_L/C_M$ have been introduced with Eq. \ref{eq_traj_tau}.
The Stokes number is defined here as $\St_\eta=  \tau_b/ \tau_\eta =  d_b^2/24 \nu \tau_\eta \mathcal{K}(Re)$ which compares the bubble relaxation time $\tau_b$ (Eq. \ref{eq_taub}) to the Kolmogorov time $\tau_\eta$. The Stokes number defined in this manner can be used to consider Reynolds number effects  because it includes the drag factor $\mathcal{K}(Re)$. Similarly, it can also account for bubble surface contamination  \cite{zhang2021} using  relation  (\ref{eq:drag_mei_K}) for clean spherical bubbles or relation (\ref{eq:drag_contaminated_K}) for contaminated bubbles. 
The Froude number $\Fr_\eta = a_\eta/(2 g)$ compares the fluid acceleration at the Kolmogorov scale $a_\eta=\eta/\tau_\eta^2$ 
 to the Archimedean acceleration. Note that replacing $\eta$ with $\ell$ we recover the  Stokes number $\St_\ell$ and the Froude number  $\Fr_\ell$ introduced in the previous section, (see Eq. \ref{turb_capt}).

From Eq. \ref{Eq_kolmo} it becomes clear that bubbles follow the fluid trajectories ($ \mathbf{u_b} \approx \mathbf{U}$), in the limit when $\St_\eta \rightarrow 0$. For a non-zero Stokes number, bubbles depart from fluid streamlines and distribute non-homogeneously due to their inertia (induced by their added mass), as illustrated in Fig. \ref{fig_bubble_turbulence1}(a).
Bubbles preferentially concentrate in regions of high vorticity and interact with vortices, as discussed previously. Even for low Stokes number, bubbles experience the effect of turbulence more intensely \cite{zhang2019model}. This behavior is very different from the case of heavy solid particles  ($\gamma = 3 \rho /( \rho + 2 \rho_p) < 1$, $\rho_p$ being the density of the particle) that are expelled from rotating regions due to their inertia \cite{Toschi2009} as shown in Fig. \ref{fig_bubble_turbulence1}(a). 

Also, from  Eq. \ref{Eq_kolmo}, we can conclude that the effect of gravity is negligible on the bubble motion as long as $\St_\eta/\Fr_\eta \ll1$.  Due to the combined effect of inertia and buoyancy, the bubble acceleration variance  $\overline{a^2_b}$ deviates from the fluid acceleration variance  $\overline{a^2_f}$ as \cite{Mathai2016, Mathai2020}:
\begin{equation}\label{Eq_ab}
 \frac{\overline{a^2_b}}{\overline{a^2_f}}  -1 \sim  \left(  \frac{\St_\eta}{\Fr_\eta} \right)^2
\end{equation}
Thus, both increasing $\St_\eta$ (inertia) and $1/\Fr_\eta$ (buoyancy), bubbles spend less time in turbulent eddies and can rise vertically through the turbulent flow.
The PDF of the  bubble acceleration is shown in Fig. \ref{fig_bubble_turbulence1}(b) for the limit  $\St_\eta/\Fr_\eta \rightarrow 0$. The stretched tails in the non-Gaussian PDF indicate the occurrence of very intense acceleration events which are typical of bubble behavior in turbulence. The main difference with fluid tracer acceleration is observed  for intermediate Stokes numbers ($\St_\eta \approx 0.5$).

The role of the lift force (the second to last term on the right hand side of Eq. \ref{Eq_kolmo}) has been identified to reduce the rising velocity of bubbles in turbulence \cite{mazzitelli2003b}. {This is illustrated in Fig. \ref{fig_bubble_turbulence1}(c), where $\overline{u_\infty}/u_{\infty}$ is reported as a function of $\St_\eta$: the terminal velocity is significantly reduced when considering the lift force in Eq. \ref{Eq_kolmo} \cite{Zhang2019}.} Two mechanisms are usually invoked as discussed in the previous section. First, the lift force acts to counteract  the inertia/added mass effect (first term in left hand side of Eq. \ref{Eq_kolmo}) that drives  inertial bubbles to the cores of the vortices. Second, the lift force causes the bubbles to preferentially drift toward down-flow regions, ($C_L$ being positive for small spherical bubbles). The resulting reduction of the bubble rising velocity for  $\beta =u^\prime/u_{\infty} \ll 1$ is expressed  as  \cite{Spelt1997, Poorte2002}: 
\begin{equation}\label{Eq_spelt}
 \frac{\overline{u_\infty}}{u_{\infty}}  -1 \sim  \beta^2.
\end{equation}

\subsection{{Large bubbles in turbulence}}
\begin{figure}[h]
\includegraphics[width=6.3in]{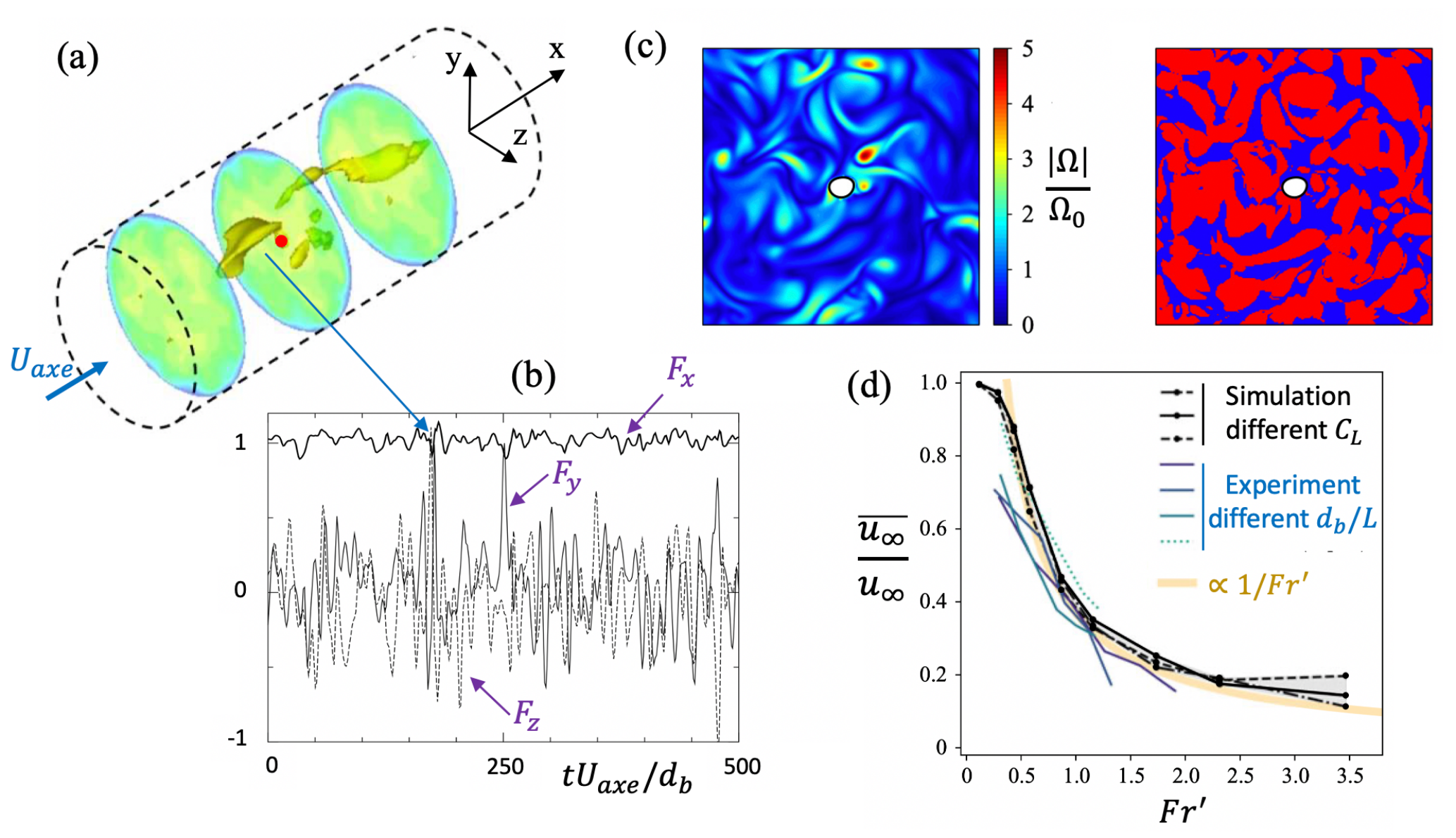}
\caption{Interaction of large bubbles with turbulence. (a) A spherical bubble fixed in the center of a turbulent pipe flow. {The bubble Reynolds number based on the bubble diameter and the centerline velocity $U_{axe}$ in the pipe is $\Rey=500$ and the turbulent pipe Reynolds number based on the pipe diameter and bulk velocity is $\Rey_{pipe}=6000$}. Turbulent eddies visualized by iso-vorticity contour are responsible for the peak in the transverse force seen in (b), which shows  the time evolution of the instantaneous force components normalized by the drag force on the bubble in the corresponding laminar flow \cite{merle2005}. (c) Bubble rising in a turbulent flow \cite{Loisy2017}: (left) normalized vorticity magnitude $\Omega$, and (right)  strain-dominated (shown in blue) and vorticity-dominated (shown in red) regions. (d) Evolution of the normalized rising velocity of a bubble in a turbulent field $\overline{u_\infty}/u_\infty$ as a function of the turbulent Froude number $ Fr^\prime = u^\prime/\sqrt{g d_b}$, adapted from  \cite{Ruth2021}.}
\label{fig_bubble turbulence}
\end{figure}

The motion of bubbles with size $d_b$ in the inertial subrange ($\eta < d_b < L$) has received less attention than point bubbles both in experiments \cite{Salibindla2020, Ruth2021} or in direct numerical simulation where the bubble surface and all turbulence scales are resolved  \cite{merle2005, Loisy2017}. The predictions from Eq. \ref{Eq_kolmo} are expected to hold for turbulent flows as long as the bubble size is much smaller than the Kolmogorov length scale. However, it is often used  to describe bubble motion for $d_b>\eta$. This can be justified by considering numerical simulations conducted for spherical bubbles {at large Reynolds number $\Rey \gg 1$} and bubble sizes that are larger than the Kolmogorov scale but smaller than the integral length scale. {Indeed, the simulations by \citet{merle2005}} which consider $\eta/d_b=0.12$, $\lambda/d_b=1.35$, $L/d_b=3.1$ for a fixed bubble in a turbulent flow ($\Rey=500$), show that  Eq. (\ref{Eq_kolmo}) is still able to reproduce the instantaneous force experienced by the bubble, particularly the lift force whose magnitude can be significantly increased when the bubble is interacting with eddies of similar size as illustrated in Fig. \ref{fig_bubble turbulence}(a-b).

This supports the idea that  bubbles with size in the inertial subrange ($\eta \ll d_b \ll L$) interact more intensely with eddies of size $d_b$.

Based on this argument, a  relevant turbulent velocity scale for bubble dynamics is the eddy velocity scale at the bubble's size: $u_d=u_\varepsilon \sim (\varepsilon d_b)^{1/3}$ \cite{Salibindla2020}.

The evolution of the mean rising bubble velocity in turbulence $\overline{u_\infty}$ reported in Fig. \ref{fig_1_turb} shows different behaviors, especially  for large bubbles. \citet{Salibindla2020} report bubble rising at a speed larger than their terminal velocity in a quiescent fluid and they provide an explanation based on the lift induced mechanism as illustrated in Fig. \ref{fig_bubble_turbulence_0}. A high turbulence level is considered ($u^\prime=0.25$m/s,  $\beta=O(1)$) and bubbles interact more intensely  with  eddies of size $\ell \approx d_b$,  resulting in more frequent and intense induced lift effect events as illustrated in Fig. \ref{fig_bubble turbulence}.  The bubble deformation is large enough to induce lift reversal resulting in more frequent situations where the bubble drifts toward upward velocities, as illustrated in Fig. \ref{fig_bubble_turbulence_0}(b).

\citet{Ruth2021}  found that for $u^\prime$ ranging from 0.02 to 0.2 m/s, the mean rising speed $\overline{u_\infty}$ decreases as $ u^\prime$ increases. They provide an explanation based on the mean behavior of  the non-linear drag force, since they consider contaminated bubbles. This behavior differs from that of clean bubbles which, at first order, experience a linear viscous drag force. 
Let us consider a drag force in the form 
$\mathbf{F_D}  \propto  \| \mathbf{U} - \mathbf{u_b} \|^\psi  (\mathbf{U} - \mathbf{u_b})$
where the parameter $\psi$ can take different values depending on the Reynolds number and the bubble surface contamination:  $\psi=0$ for a clean spherical bubble or contaminated bubble  for $Re<1$; $\psi=0.687$ for a contaminated spherical bubble for $1 \ll Re<800$; and $\psi=1$ for inertial drag force ($C_D=$constant). 
{Considering only a balance between the drag and the buoyancy, $\mathbf{F_B}$, to simplify the discussion, we have $\mathbf{F_B} + \overline{\mathbf{F_D}} \approx 0$ and $\mathbf{F_B} + \mathbf{F_D} \approx 0$ with and without turbulence, respectively. Since $\mathbf{F_B}$ is constant for a given bubble size, we have $\overline{\mathbf{F_D}} \approx \mathbf{F_D}$. It is therefore possible to directly compare $\overline{u_\infty}$ and $u_\infty$.}

We now introduce the Reynolds decomposition for both the bubble and the fluid velocities considering,  for simplicity, that the mean flow satisfies $\overline{\mathbf{U}}=0$ and the mean bubble velocity is along the vertical direction $\mathbf{e_y}$ ($\overline{u_{bx}}=\overline{u_{bz}}=0$) to express the relative velocity between the fluid and the bubble as
\begin{equation}\label{turb_drag}
 \mathbf{U} -  \mathbf{u_b} = - \overline{u_\infty}  \mathbf{e_y} +  \mathbf{u^\prime} -  \mathbf{u_b^\prime}.
\end{equation}

For a linear drag force ($\psi=0$), the magnitude of the drag force varies with the bubble rising velocity as $ \overline{{F_D}} \propto  \overline{u_\infty}$ and ${F_D} \propto  u_\infty$, which results in  $\overline{u_\infty} \approx u_{\infty}$. Considering the other forces, such as the inertia and lift forces and a linear drag ($\mathbf{F_D} =6 \pi \mu d_b  (\mathbf{U} - \mathbf{u_b})$), the expression of $\overline{u_\infty}$ was calculated by   \citet{Spelt1997} leading to Eq.\ref{Eq_spelt} which shows a decrease of the rising velocity with respect to the rise in a quiescent fluid attributed to the induced lift drift toward downward velocities, as discussed in the previous section and illustrated in Fig. \ref{fig_bubble_turbulence_0}. 

For a non-linear relation between drag and velocity ($\psi > 0$), the mean drag will be impacted by the root mean square (r.m.s.)  of the velocity fluctuations  $\overline{\mathbf{ \left| \mathbf{u^\prime} -  \mathbf{u_b^\prime} \right|^2}}$. 
Assuming $\psi =1$ for simplicity, {the magnitudes of drag force are then $\overline{{F_D}} \propto  \overline{u_\infty}^2 + 
\overline{\mathbf{ \left| \mathbf{u^\prime} -  \mathbf{u_b^\prime}\right|^2}}$ and $ F_D \propto {u_\infty}^2$, so that  the relationship between the rising velocities is now} $u_{\infty}^2 \approx \overline{u_\infty}^2 + 
\overline{\mathbf{ \left| \mathbf{u^\prime} -  \mathbf{u_b^\prime} \right|^2}}$. This clearly indicates a decrease in the rising velocity {$\overline{u_\infty}$ when compared to $u_{\infty}$}. Based on such argument \citet{Ruth2021}  have derived relations for $\overline{u_\infty}$ for both small and large turbulence fluctuations. In the limit of large fluctuations ($\beta \gg1$), they found that 
\begin{equation}\label{Eq_Ruth}
 \frac{\overline{u_\infty}}{u_{\infty}}  \sim  \frac{1}{Fr^\prime}
\end{equation}
where $ Fr^\prime = u^\prime/\sqrt{g d_b}$ is a Froude number defined  here by considering the turbulence fluctuation $u^\prime$ and the bubble diameter. They also proposed a numerical approach based on Lagrangian tracking of point bubbles in homogeneous and isotropic turbulence  by solving Eqs. \ref{eq_traj_2_1}-\ref{eq_traj_2_4}  by imposing $\mathcal{K}(\Rey)=C_D \Rey/48$ and $C_D=1$ for the unsteady drag force. Their experiments and their simulations  confirm the evolution of $\overline{u_\infty}/u_{\infty}$ with $Fr^\prime$ given by Eq. \ref{Eq_Ruth}, as shown in Fig. \ref{fig_bubble turbulence}(d). Their simulations however did not show an effect of the lift force (values of $C_L$=-0.25, 1 and 0.25 were considered) on the decrease of the rising velocity. This is in contradiction with the conclusions from other studies \cite{Loisy2017, Salibindla2020} for large bubbles rising in high levels of turbulence. 

{To summarize, a comprehensive understanding of the dynamics of large bubbles in intense turbulence remains elusive. A detailed outline of the open questions and future research directions is offered in the conclusions section.\\}

\section{Concluding remarks on open questions}

In this paper, we have discussed several key aspects of bubble motion. In any application or natural phenomena where bubbles are present, {some of which were highlighted  in the introduction,} it is crucial to understand how and where they move, their speeds, shapes and how will they interact with walls, other bubbles or with fluids with different rheologies. This task can be accomplished if we understand the physical mechanisms that influence the motion of bubbles.

By addressing the seemingly simple question, what is the terminal speed of an air bubble in water, we build the foundation of the information conveyed in this paper: drag forces balance buoyancy to determine the terminal speed, viscous, inertial and gravitational forces balance surface tension to determine the bubble shape, bubble shape influence drag, surfactants affect both drag and shape, non-uniform flows affect the bubble speed and trajectory, walls and other bubbles affect the bubble speed, as well as fluids with different rheology. All these additional forces affect the bubble motion, but a good understanding and modelling of them allows us to propose a dynamic equation that is currently used to provide predictive solutions. {Also, while composing this narrative, we clearly identify the {physical mechanisms involved and the} relevant dimensionless parameters that influence the bubble dynamics; identifying such {effects and} numbers, their meaning and relevance, is important to read and understand the literature on the subject.}

All the factors that affect the bubble motion, listed above, involve complexities and subtleties. Even at the bubble scale, the motion is affected by the details of the flow, deformation and possible contamination. In most applications, all these effects appear simultaneously 
making it difficult to interpret observations
because each effect could not be easily disentangled from the others and because coupling among effects can arise. {The challenge in conducting research in this area is to find relatively simple systems with which an idea can be methodically tested, while keeping the number of parameters that affect the bubble motion small.}

This review provides a solid foundation for understanding bubble dynamics. It surveys the classical results from the literature, while incorporating explanations of physical phenomena based on recent investigations. In our opinion, the subject of bubble dynamics has reached a good level of maturity. We believe that many of the challenges in bubble-based applications and natural phenomena can be addressed considering some of the physical descriptions discussed in this review. We note, however, that most of the current understanding is still limited to relatively simple flow configurations and, most importantly, to Newtonian liquids. 
{Writing this review helped us to identify relevant areas that, in our opinion, need attention: bubble dynamics in turbulent flows, the effect of non-Newtonian liquids and the effect of {surface contamination in presence of}  electrolytes. Many modern applications, especially for biological flows, will imply bubbles moving in turbulent fluids with a non-Newtonian nature and/or salt-rich environments for which the understanding of bubble dynamics is in its infancy. }

{For bubbles moving in turbulent flows, we conclude that understanding how large bubbles (larger than the Kolmogorov scale but smaller than the integral scale) respond to intense turbulent fields remains an open question. In particular, future works will have to clarify the interaction mechanism between turbulence and bubble deformation and contamination. A full picture of their respective impact of these factors, in particular, on the bubble rising velocity is needed. Under what conditions one can expect the mean bubble rise velocity, $\overline{u_\infty}$, in turbulence be larger than bubble rise velocity in a stagnant fluid, ${u_\infty}$? Addressing this question will require considering the coupled effects of bubble deformation and surface contamination. As discussed earlier, contaminants influence bubble deformation, and deformation affects surfactant coverage. This mutual influence can alter the interfacial vorticity, potentially modifying how the bubble is driven by drag and lift forces within the turbulent field. This effect, and its response to turbulence, could be significant in understanding aggregation and dispersion in bubbly flows.}

{The understanding of bubble dynamics in non-Newtonian fluids is much less mature than the one currently available for Newtonian fluids. Although there are many experiments reported in the literature, many of which are not recent,  our current understanding remains fragmented and needs further consolidation through focused research. With computational tools now highly advanced, it is time to combine numerical and experimental investigations to clarify key aspects of bubble dynamics. First, it would be highly desirable to obtain a version of the Clift-Grace-Weber diagram (see Fig. \ref{fig_1_new}.b) \citep{clift1978} for the case of non-Newtonian fluids. Accomplishing this task would require experiments or simulations that gradually vary the elasticity of the fluid, while keeping the viscosity constant. This approach would isolate the effects of elasticity, enabling a precise assessment of its influence on bubble shape and terminal velocity. Then, the effects of shear-dependent viscosity could be investigated, as an additional complexity to the problem.}
{A path similar to that followed for Newtonian fluids should be followed to gain an understanding of the nature of hydrodynamic interactions in non-Newtonian fluids. \citet{Zenit2018} gave a good summary of the main issues in this area. There are recent investigations that aim to understand single bubble behavior \cite{fraggedakis2016,bothe2022} and pair interactions \citep{ravisankar2022,kordalis2023} in non-Newtonian fluids that could serve as guides for future research in the area. The effect of contaminants on the motion of bubbles in non-Newtonian media has been long recognized as important \citep{astarita1965}, but a clear picture of the physical understanding is lacking. To our knowledge, studies of bubbles in turbulent fields in non-Newtonian fields do not exist. This area clearly has relevance in the context of bioreactors and drag reduction.}

{Finally,  we have recognized that the effect of electrolytes in several aspects of gas bubble dynamics remains poorly understood or unresolved. Electrolytes are substances, soluble in water, which can have natural positive or negative electrical charges. These charges can have a significant effect in the production of hydrogen via electrolysis: the bubble nucleation, coalescence and detachment from an electrode surface, as shown in Fig. \ref{fig_examples}(c), impact the generated bubble swarm. Since Hydrogen is our best option to replace fossil fuels, it is extremely important to study how bubble dynamics can help improve the efficiency to produce hydrogen in a green and sustainable way using this method \citep{angulo2020}.}
{All additives to air-water systems, such as surfactants, alcohols, and other substances including electrolytes, are usually classified as contaminants. The effects of contaminants were discussed in this review, but certainly not extensively. When used in small quantities, electrolytes inhibit bubble coalescence \cite{lessard1971,zenit2001,Orvalho2009, Orvalho2021} due to short range repulsion forces induced by the electrical charges. However, a clear understanding of the effect of electrolytes at large concentration, such as those as encountered in electrolyzers, is still needed. Recent experiments of air bubbles in low-concentration electrolyte solutions \citep{Quinn2014, Hessenkemper2020} report that the  relationship between rise velocity and bubble size is not significantly affected by presence of electrolytes, independent of solute type. However, they found that electrolyte can have a significant effect on how surfactant molecules move and accumulate on the bubble surface, possibly leading to changes in shape, drag and interactions. For the specific case of hydrogen bubbles,  \citet{Mandalahalli2023} found that  the rise characteristics (velocity and shape)  at different concentrations of a mixture electrolytes, can be explained by their effect on the liquid properties (density, viscosity, and surface tension). Due to the limited amount of information, it is not yet possible to conclude if these findings can be applied generally to other gas-liquid combinations. Other complexities that need to be addressed for the particular case of hydrogen bubbles is their potential for reactivity which make them susceptible to react contaminants suspended in water; hydrogen is also known to be consumed by bacteria in natural environments \citep{barz2010,lappan2023}. Future investigations will need to account for these additional complexities, which have not been discussed in the present paper.}

With the many challenges in bubble dynamics described here, we remain excited about the discoveries that lie ahead. We sincerely hope that this review will serve as a guide and inspiration for future researchers in the filed of bubble dynamics to pursue some of these studies in the years to come.

\begin{acknowledgments}
We thank {our students and colleagues} for many helpful discussions and suggestions during the preparation of this manuscript. We are grateful to M. Ravisankar for conducting some additional experiments to obtain bubble shape images shown in the paper. 

\end{acknowledgments}

\bibliography{biblio}

\end{document}